\documentclass[longauth]{aa} 

\newcommand\hmmax{0}
\newcommand\bmmax{0}

\usepackage{graphicx}
\usepackage{txfonts}
\usepackage[colorlinks=true,
            linkcolor=blue,
            urlcolor=blue,
            citecolor=blue]{hyperref}
\usepackage{booktabs}
\usepackage[flushleft]{threeparttable}
\usepackage{upgreek}
\usepackage{textcomp}
\usepackage{bm}
\makeatletter
\renewcommand*\aa@pageof{, page \thepage{} of \pageref*{LastPage}}
\makeatother

\def\ed{{\sc{EddySed}}\xspace}

\def\f1{f_{\rm I}}
\def\mj{{M}$_{\textrm{\tiny Jup}}$\xspace}
\def\rj{${R}_{\textrm{\tiny Jup}}$\xspace}

\newcommand{\klip}{{\tt KLIP}\xspace}

\def\teff{$T_{\rm eff}$}
\def\Tint{\ensuremath{T_{\rm int}}\xspace}
\def\Tbot{\ensuremath{T_{\rm bot}}\xspace}
\def\kzz{\ensuremath{K_{\rm zz}}\xspace}
\def\fsed{\ensuremath{f_{\rm sed}}\xspace}
\def\h2o{H$_{2}$O\xspace}
\def\ch4{CH$_{4}$\xspace}
\def\nh3{NH$_{3}$\xspace}
\def\co2{CO$_{2}$\xspace}
\def\ph3{PH$_{3}$\xspace}

\def\Zpl{\ensuremath{Z_{\rm pl}}\xspace}
\def\Zstar{\ensuremath{Z_{\star}}\xspace}

\def\beq{\begin{equation}}
\def\eeq{\end{equation}}

\def\t2{\tau_{\rm II}}

\def\sigmas0{\Sigma_{\rm s,0}}

\def\({\left(}
\def\){\right)}
\def\<{\left<}
\def\>{\right>}

\defcitealias{ackerman_precipitating_2001}{AM01}
\defcitealias{molliere_retrieving_2020}{M20}
\defcitealias{zhang_elemental_2023}{Z23}
\defcitealias{guillot_radiative_2010}{G10}

\begin{document} 

   \title{Four-of-a-kind? Comprehensive atmospheric characterisation of the HR~8799 planets with VLTI/GRAVITY}
    \titlerunning{HR8799 Atmospheres}
    \authorrunning{Nasedkin et al.}
   \subtitle{}

   \author{E.~Nasedkin\inst{\ref{mpia}}\fnmsep\thanks{Corresponding Author, \href{mailto:nasedkin@mpia.de}{nasedkin@mpia.de}}
   \and P.~Molli\`ere\inst{\ref{mpia}}
 \and S.~Lacour\inst{\ref{lesia},\ref{esog}}
 \and M.~Nowak\inst{\ref{cam}}
 \and L.~Kreidberg\inst{\ref{mpia}}
 \and T.~Stolker\inst{\ref{leiden}}
 \and J.~J.~Wang\inst{\ref{northwestern}}
 \and W.~O.~Balmer\inst{\ref{jhupa},\ref{stsci}}
 \and J.~Kammerer\inst{\ref{esog}}
 \and J.~Shangguan\inst{\ref{mpe}}
 \and R.~Abuter\inst{\ref{esog}}
 \and A.~Amorim\inst{\ref{lisboa},\ref{centra}}
 \and R.~Asensio-Torres\inst{\ref{mpia}}
 \and M.~Benisty\inst{\ref{ipag}}
 \and J.-P.~Berger\inst{\ref{ipag}}
 \and H.~Beust\inst{\ref{ipag}}
 \and S.~Blunt\inst{\ref{northwestern}}
 \and A.~Boccaletti\inst{\ref{lesia}}
 \and M.~Bonnefoy\inst{\ref{ipag}}
 \and H.~Bonnet\inst{\ref{esog}}
 \and M.~S.~Bordoni\inst{\ref{mpe}}
 \and G.~Bourdarot\inst{\ref{mpe}}
 \and W.~Brandner\inst{\ref{mpia}}
 \and F.~Cantalloube\inst{\ref{lam}}
 \and P.~Caselli \inst{\ref{mpe}}
 \and B.~Charnay\inst{\ref{lesia}}
 \and G.~Chauvin\inst{\ref{cotedazur}}
 \and A.~Chavez\inst{\ref{northwestern}}
 \and E.~Choquet\inst{\ref{lam}}
 \and V.~Christiaens\inst{\ref{liege}}
 \and Y.~Cl\'enet\inst{\ref{lesia}}
 \and V.~Coud\'e~du~Foresto\inst{\ref{lesia}}
 \and A.~Cridland\inst{\ref{leiden}}
 \and R.~Davies\inst{\ref{mpe}}
 \and R.~Dembet\inst{\ref{lesia}}
 \and J.~Dexter\inst{\ref{boulder}}
 \and A.~Drescher\inst{\ref{mpe}}
 \and G.~Duvert\inst{\ref{ipag}}
 \and A.~Eckart\inst{\ref{cologne},\ref{bonn}}
 \and F.~Eisenhauer\inst{\ref{mpe}}
 \and N.~M.~F\"orster Schreiber\inst{\ref{mpe}}
 \and P.~Garcia\inst{\ref{centra},\ref{porto}}
 \and R.~Garcia~Lopez\inst{\ref{dublin},\ref{mpia}}
 \and E.~Gendron\inst{\ref{lesia}}
 \and R.~Genzel\inst{\ref{mpe},\ref{ucb}}
 \and S.~Gillessen\inst{\ref{mpe}}
 \and J.~H.~Girard\inst{\ref{stsci}}
 \and S.~Grant\inst{\ref{mpe}}
 \and X.~Haubois\inst{\ref{esoc}}
 \and G.~Hei\ss el\inst{\ref{actesa},\ref{lesia}}
 \and Th.~Henning\inst{\ref{mpia}}
 \and S.~Hinkley\inst{\ref{exeter}}
 \and S.~Hippler\inst{\ref{mpia}}
 \and M.~Houll\'e\inst{\ref{cotedazur}}
 \and Z.~Hubert\inst{\ref{ipag}}
 \and L.~Jocou\inst{\ref{ipag}}
 \and M.~Keppler\inst{\ref{arizona}}
 \and P.~Kervella\inst{\ref{lesia}}
 \and N.~T.~Kurtovic\inst{\ref{mpe}}
 \and A.-M.~Lagrange\inst{\ref{ipag},\ref{lesia}}
 \and V.~Lapeyr\`ere\inst{\ref{lesia}}
 \and J.-B.~Le~Bouquin\inst{\ref{ipag}}
 \and D.~Lutz\inst{\ref{mpe}}
 \and A.-L.~Maire\inst{\ref{ipag}}
 \and F.~Mang\inst{\ref{mpe}}
 \and G.-D.~Marleau\inst{\ref{duisburg},\ref{tuebingen},\ref{bern},\ref{mpia}}
 \and A.~M\'erand\inst{\ref{esog}}
 \and J.~D.~Monnier\inst{\ref{umich}}
 \and C.~Mordasini\inst{\ref{bern}}
 \and T.~Ott\inst{\ref{mpe}}
 \and G.~P.~P.~L.~Otten\inst{\ref{sinica}}
 \and C.~Paladini\inst{\ref{esoc}}
 \and T.~Paumard\inst{\ref{lesia}}
 \and K.~Perraut\inst{\ref{ipag}}
 \and G.~Perrin\inst{\ref{lesia}}
 \and O.~Pfuhl\inst{\ref{esog}}
 \and N.~Pourr\'e\inst{\ref{ipag}}
 \and L.~Pueyo\inst{\ref{stsci}}
 \and D.~C.~Ribeiro\inst{\ref{mpe}}
 \and E.~Rickman\inst{\ref{esa}}
 \and J.B.~Ruffio\inst{\ref{sandiego}}
 \and Z.~Rustamkulov\inst{\ref{jhueps}}
 \and T.~Shimizu \inst{\ref{mpe}}
 \and D.~Sing\inst{\ref{jhupa},\ref{jhueps}}
 \and J.~Stadler\inst{\ref{mpa},\ref{origins}}
 \and O.~Straub\inst{\ref{origins}}
 \and C.~Straubmeier\inst{\ref{cologne}}
 \and E.~Sturm\inst{\ref{mpe}}
 \and L.~J.~Tacconi\inst{\ref{mpe}}
 \and E.~F.~van~Dishoeck\inst{\ref{leiden},\ref{mpe}}
 \and A.~Vigan\inst{\ref{lam}}
 \and F.~Vincent\inst{\ref{lesia}}
 \and S.~D.~von~Fellenberg\inst{\ref{bonn}}
 \and F.~Widmann\inst{\ref{mpe}}
 \and T.~O.~Winterhalder\inst{\ref{esog}}
 \and J.~Woillez\inst{\ref{esog}}
 \and \c{S}.~Yaz\i{}c\i{}\inst{\ref{mpe}}
 \and  the~GRAVITY~Collaboration}
\institute{ 
   Max-Planck-Institut f\"ur Astronomie, K\"onigstuhl 17, 69117 Heidelberg, Germany
\label{mpia}      \and
   LESIA, Observatoire de Paris, PSL, CNRS, Sorbonne Universit\'e, Universit\'e de Paris, 5 place Janssen, 92195 Meudon, France
\label{lesia}      \and
   European Southern Observatory, Karl-Schwarzschild-Stra\ss{}e 2, 85748 Garching, Germany
\label{esog}      \and
   Institute of Astronomy, University of Cambridge, Madingley Road, Cambridge CB3 0HA, United Kingdom
\label{cam}      \and
   Leiden Observatory, Leiden University, P.O. Box 9513, 2300 RA Leiden, The Netherlands
\label{leiden}      \and
   Center for Interdisciplinary Exploration and Research in Astrophysics (CIERA) and Department of Physics and Astronomy, Northwestern University, Evanston, IL 60208, USA
\label{northwestern}      \and
   Department of Physics \&\ Astronomy, Johns Hopkins University, 3400 N. Charles Street, Baltimore, MD 21218, USA
\label{jhupa}      \and
   Space Telescope Science Institute, 3700 San Martin Drive, Baltimore, MD 21218, USA
 \label{stsci}  \and
   Max-Planck-Institut f\"ur Extraterrestrische Physik, Giessenbachstra\ss e~1, 85748 Garching, Germany
\label{mpe}      \and
   Universidade de Lisboa - Faculdade de Ci\^encias, Campo Grande, 1749-016 Lisboa, Portugal
\label{lisboa}      \and
   CENTRA - Centro de Astrof\'{i}sica e Gravita\c{c}\~{a}o, IST, Universidade de Lisboa, 1049-001 Lisboa, Portugal
\label{centra}      \and
   Univ. Grenoble Alpes, CNRS, IPAG, 38000 Grenoble, France
\label{ipag}      \and
   Aix Marseille Univ, CNRS, CNES, LAM, Marseille, France
\label{lam}      \and
   Universit\'e C\^{o}te d'Azur, Observatoire de la C\^{o}te d'Azur, CNRS, Laboratoire Lagrange, France
\label{cotedazur}      \and
  STAR Institute, Universit\'e de Li\`ege, All\'ee du Six Ao\^ut 19c, 4000 Li\`ege, Belgium
\label{liege}      \and
   Department of Astrophysical \&\ Planetary Sciences, JILA, Duane Physics Bldg., 2000 Colorado Ave, University of Colorado, Boulder, CO 80309, USA
\label{boulder}      \and
   1.\ Institute of Physics, University of Cologne, Z\"ulpicher Stra\ss e 77, 50937 Cologne, Germany
\label{cologne}      \and
   Max-Planck-Institut f\"ur Radioastronomie, Auf dem H\"ugel 69, 53121 Bonn, Germany
\label{bonn}      \and
   Universidade do Porto, Faculdade de Engenharia, Rua Dr.~Roberto Frias, 4200-465 Porto, Portugal
\label{porto}      \and
   School of Physics, University College Dublin, Belfield, Dublin 4, Ireland
\label{dublin}      \and
   Departments of Physics and Astronomy, Le Conte Hall, University of California, Berkeley, CA 94720, USA
\label{ucb}      \and
   European Southern Observatory, Casilla 19001, Santiago 19, Chile
\label{esoc}      \and
   Advanced Concepts Team, European Space Agency, TEC-SF, ESTEC, Keplerlaan 1, NL-2201, AZ Noordwijk, The Netherlands
\label{actesa}      \and
Department of Astronomy and Steward Observatory, University of Arizona, Tucson, AZ, USA 
\label{arizona}     \and
   University of Exeter, Physics Building, Stocker Road, Exeter EX4 4QL, United Kingdom
\label{exeter}      \and
   Fakult\"at f\"ur Physik, Universit\"at Duisburg--Essen, Lotharstra\ss{}e 1, 47057 Duisburg, Germany
\label{duisburg}      \and
   Instit\"ut f\"ur Astronomie und Astrophysik, Universit\"at T\"ubingen, Auf der Morgenstelle 10, 72076 T\"ubingen, Germany
\label{tuebingen}      \and
   Physikalisches Institut, Universit\"at Bern, Gesellschaftsstr.~6, 3012 Bern, Switzerland
\label{bern}      \and
   Astronomy Department, University of Michigan, Ann Arbor, MI 48109 USA
\label{umich}      \and
   Academia Sinica, Institute of Astronomy and Astrophysics, 11F Astronomy-Mathematics Building, NTU/AS campus, No. 1, Section 4, Roosevelt Rd., Taipei 10617, Taiwan
\label{sinica}      \and
   European Space Agency,  
   ESA Office, Space Telescope Science Institute, 3700 San Martin Drive, Baltimore, MD 21218, USA
\label{esa}   \and
    Department of Astronomy \& Astrophysics,  University of California, San Diego, La Jolla, CA 92093, USA
 \label{sandiego}   \and
   Department of Earth \& Planetary Sciences, Johns Hopkins University, Baltimore, MD 21218, USA
\label{jhueps}      \and
   Max-Planck-Institut f\"ur Astrophysik, Karl-Schwarzschild-Str. 1, 85741 Garching, Germany
\label{mpa}      \and
   Excellence Cluster ORIGINS, Boltzmannstra\ss{}e 2, 85748 Garching bei München, Germany
\label{origins}    
}

   \date{Received 2024-01-24; accepted 2024-04-25}

 
  \abstract{
  With four companions at separations from 16 to 71 au, HR 8799 is a unique target for direct imaging, presenting an opportunity for a comparative study of exoplanets with a shared formation history. 
  Combining new VLTI/GRAVITY observations obtained within the ExoGRAVITY program with archival data, we performed a systematic atmospheric characterisation across all four planets. 
  We explored different levels of model flexibility to understand the temperature structure, chemistry, and clouds of each planet using both {\tt petitRADTRANS} atmospheric retrievals and fits to self-consistent radiative--convective equilibrium models. 
  Using Bayesian model averaging to combine multiple retrievals (a total of 89 across all four planets), we find that the HR 8799 planets are highly enriched in metals, with [M/H] $\gtrsim$1, and have stellar to superstellar atmospheric C/O ratios.
  The C/O ratio increases with increasing separation from $0.55^{+0.12}_{-0.10}$ for d to $0.78^{+0.03}_{-0.04}$ for b, with the exception of the innermost planet, which has a C/O ratio of $0.87\pm0.03$. 
  Such high metallicities are unexpected for these massive planets, and challenge planet-formation models.
  By retrieving a quench pressure and using a disequilibrium chemistry model, we derive vertical mixing strengths compatible with predictions for high-metallicity, self-luminous atmospheres.
  Bayesian evidence comparisons strongly favour the presence of HCN in HR 8799 c and e, as well as \ch4 in HR 8799 c, with detections at $>5\sigma$ confidence.
  All of the planets are cloudy, with no evidence of patchiness. 
  The clouds of c, d, and e are best fit by silicate clouds lying above a deep iron cloud layer, while the clouds of the cooler HR 8799 b are more likely composed of Na$_{2}$S.
  With well-defined atmospheric properties, future exploration of this system is well positioned to unveil further details of these planets, extending our understanding of the composition, structure, and formation history of these siblings.}
   

   \maketitle
%

\section{Introduction}\label{sec:intro}
Directly imaged exoplanets provide an ideal laboratory for understanding the formation and evolution of planetary systems.
These young systems provide unique insight into widely separated companions ($\gtrsim 10$~au): by directly measuring their emission spectra, we can peer into regions of their atmospheres inaccessible through other techniques.
While spectroscopically similar to their brown dwarf cousins, these young, low-surface-gravity exoplanets display unique spectral shapes and colours \citep{faherty_spectral_2018}, indicative of differences in their chemistry, clouds, and formation history \citep{marley_patchy_2010,marley_masses_2012,charnay_self-consistent_2018}.

The \object{HR 8799} system is a benchmark target for directly imaged exoplanets, containing four planets \citep{marois_direct_2008,marois_images_2010}, an inner debris disc \citep{boccaletti_imaginghr8799_2023}, and an outer Kuiper-belt-like disc \citep{su_debris_2009}.
This is among the best-studied systems of exoplanets, with a wide range of photometric and spectroscopic data.
The spectroscopic data cover the near-infrared region (1--4~$\upmu$m) at varying spectral resolution, while the photometric data extend out to 15~$\upmu$m  with the recent addition of JWST/MIRI observations \citep{boccaletti_imaginghr8799_2023}.
Most of these studies, together with extensive modelling work, have tried to answer the following main questions:
\begin{enumerate}
    \item How did the HR 8799 system form?
    \item What are the dynamics of the system? Is it stable, and how do the planets and disc interact?
    \item What are the atmospheres of each planet made of, and how have they evolved through time?
\end{enumerate}
In the present work, we attempt to directly answer  question (3), which has implications for question (1). Using Bayesian atmospheric retrievals \citep[e.g.][]{madhusudhan_exoplanetary_2019} as well as fits to 1D self-consistent models, we infer the atmospheric properties of each of the four planets.
To date, the only comprehensive retrieval study of all four of the HR 8799 planets was by \cite{lavie_helios-retrieval_2017}.
New, high-precision K-band spectra obtained with the VLTI/GRAVITY as part of the ExoGRAVITY program \citep{gravity_collaboration_first_2019, lacour_exogravity_2020}, together with updated atmospheric models and opacity databases, provide motivation and the means to perform a systematic reanalysis of this system.

With effective temperatures in the range of 1000--1400~K, the HR 8799 planets sit in the middle of the L/T transition \citep{kirkpatrick_dwarfs_1999}.
This spectral transition is marked by changes in chemistry between L- and T-type objects, from CO-dominated carbon chemistry in the hotter objects to methane chemistry as the temperature falls below $\sim$1300~K.
While this transition is well established for brown dwarfs, detections of \ch4 in exoplanets remain elusive: there have been tentative detections in the atmosphere of \object{HR 8799 b} \citep{barman_simultaneous_2015, ruffio_deep_2021}, but the only convincing detections have come from JWST observations of cool, transiting exoplanets \citep{bell_methane_2023, madhusudhan_carbon_2023} and the coldest directly imaged exoplanets, such as 51 Eridani b \citep{brown-sevilla_revisiting_2023,whiteford_retrieval_2023}.
This `missing methane' is thought to be driven by convective upwelling in the atmospheres, dredging material from deeper, hotter regions of the atmosphere where CO is more favoured by equilibrium chemistry \citep{fegley_atmospheric_1996}.
Precise constraints on the abundance of both CO and \ch4 would allow better constraints on this vertical mixing, which is typically parameterised by the vertical diffusion coefficient \kzz.

The sharp change in colour in the L/T transition is thought to be caused by the sinking of silicate clouds through the atmosphere, as the cloud base shifts deeper as the effective temperature decreases \citep{burrows_chemical_1999}.
Once the cloud base sinks below the photosphere, the impact of the cloud opacity is increasingly removed from the spectrum, causing the blueward shift characteristic of T dwarfs as the effective temperature falls below 1300~K.
Following the mid-infrared observations with \textit{Spitzer}/IRS \citep{cushing_spitzer_2006, cushing_atmospheric_2008}, \cite{suarez_ultracool_2022_LT} identified a trend in the silicate absorption feature at 9~$\upmu$m as a function of temperature.
The strength of this absorption feature was found to correlate positively with the near-infrared colour for L dwarfs, which is often used as a proxy for cloudiness.
The HR 8799 planets lie comfortably below the temperature at which silicate clouds are expected to occur entirely below the photosphere, yet their red colour and near-infrared spectral shape are thought to be clear hallmarks of thick silicate cloud coverage \citep{molliere_retrieving_2020}.
However, \cite{line_uniform_2015} and \cite{suarez_ultracool_2023_grains} find that these clouds are not only sensitive to temperature, but also to surface gravity, which plays a role in determining the size and therefore settling speed of the aerosol particles.
As young, giant exoplanets still retain significant heat from formation, their atmospheres remain inflated due to low surface gravity, which will in turn result in cloud properties that are unique to this class of object; observations of VHS~1256~b \citep{miles_jwst_2023} remain the only spectroscopic observations of a silicate feature in a directly imaged planet.
\cite{burningham_cloud_2021} and  \cite{vos_patchy_2023} use atmospheric retrievals to identify the detailed structure and composition of the clouds, providing for the first time evidence to support the use of particular cloud compositions in these substellar atmospheres.

The mechanism through which four super-Jupiter planets can form in a single system is unclear. 
The presence of both an inner and an outer debris disc implies that the planets formed within a circumstellar disc; that is to say they did not form like stars. 
However, it is still unclear whether these objects formed through gravitational instability (GI) \citep{bodenheimer_calculations_1974,adams_eccentric_1989} or via core accretion \citep{pollack_formation_1996}.
Evolutionary models \citep{saumon_evolution_2008} suggest that the current temperatures of the planets suggests hot-start boundary conditions for their evolution, which is more typically associated with GI \citep[but also see][]{mordasini_characterization_2017}. 
GI models, such as that of \cite{helled_metallicity_2010}, find that the amount of heavy elements accreted by the planets should be small, implying nearly stellar compositions for all four planets.
Likewise, current composition estimates suggest that the planets share a C/O ratio with their host star \citep{hoch_assessing_2023}, but may be slightly enriched in metals, leading to tension with the predictions of the GI models. 

In addition to understanding the formation mechanism, \cite{molliere_interpreting_2022} present a framework through which we can infer the conditions of the formation environment from measured atmospheric parameters.
However, these authors, and many others \citep[e.g.][]{eistrup_molecular_2018,cridland_connecting_2019_refractoryc-o,cridland_connecting_2020_c-o_n-o,turrini_tracing_2021,pacetti_chemical_2022}, demonstrate that this is not a straightforward task.
The \cite{oberg_effects_2011} model links the planet C/O ratio to the location of formation relative to snow lines in the disc. 
This model provides a simplified view through which we can understand the impact of disc conditions on the outcomes of planet formation, but the complex and time-evolving physics and chemistry of discs and forming planets make solving the inverse problem challenging.
Nevertheless, the best hope for linking the atmospheric properties back to the protoplanetary disc is to infer robust atmospheric elemental abundances and link these to interior models to determine the bulk planetary composition \citep{guillot_interiors_2005, fortney_self-consistent_2011_SolarSystem}, thereby determining what disc conditions could lead to the diversity of planet-formation outcomes.

While new data and modelling techniques are beneficial, interpreting such model--data comparisons for exoplanet spectra is far from trivial. 
Multiple techniques must be studied simultaneously to paint a consistent portrait of these worlds. 
Biases in inferred planet parameters are a common challenge in direct-imaging analyses: fits to emission spectra often find unphysically small radii that are inconsistent with evolutionary tracks \citep{marley_masses_2012}. 
Retrievals using free molecular abundances tend to find higher C/O ratios than when disequilibrium chemistry models are considered \citep{lavie_helios-retrieval_2017, wang_chemical_2020}, possibly due to additional oxygen sequestered in refractory clouds \citep{fonte_oxygen_2023}.
The inferred effective temperatures (\teff) of each planet can vary by hundreds of kelvin, within a region of parameter space where the chemical timescales can vary by orders of magnitude over tens of kelvin \citep{zahnle_methane_2014}.
Complicating matters further are the known discrepancies between spectral measurements \citep[such as between SPHERE and GPI in the H-band; see][]{molliere_retrieving_2020}, leading to uncertainties in both the shape and overall flux calibration of the spectra that are not reflected in the formal uncertainties.
Attempting to address this problem, \cite{nixon_methods_2023} demonstrate the use of Bayesian model averaging (BMA), which can be used to combine the posterior distributions of multiple models, allowing some degree of model uncertainty to be formally incorporated into the inferred parameter uncertainties.
Finally, \cite{greco_measurement_2016} and \cite{nasedkin_impacts_2023} demonstrate the importance of properly accounting for the covariance in low-resolution IFS data --- a thorough treatment of IFS data is necessary to ensure meaningful and unbiased posterior probability distributions.


Many of the questions of chemistry and formation will be addressed through the use of the various instruments aboard JWST.
This telescope will open new observational windows, extending out to the mid-infrared, and allow new characterisation methods, such as molecular mapping of the system \citep{patapis_direct_2022}.
Simultaneous measurements of CO and \ch4 features between $3$ and $5\,\upmu$m will allow constraints to be placed on the vertical mixing in the atmosphere, and more precise estimates of the C/O and metallicity.
Nevertheless, ground-based observations remain crucial: the innermost companion will remain challenging to measure spectroscopically without a coronagraph; across most of its spectra, \object{HR 8799 e} is below the $2\times10^{-5}$ contrast threshold at 300 mas obtained in \cite{ruffio_highcontrast_2023}.

The present study provides a comprehensive examination of the atmospheres of the HR 8799 companions, making use of new, high-$S/N$ observations obtained with VLTI/GRAVITY, together with a combination of retrieval methods and self-consistent modelling.
We present further context and background information on the HR 8799 system in Section \ref{sec:context}.
The data used in this work are described in Section \ref{sec:obs}, while the details of the {\tt petitRADTRANS} ({\tt pRT}) forward model are described in Section \ref{sec:modelling}, with the self-consistent models introduced in Section \ref{sec:selfconsistent}.
The results of the atmospheric retrievals and self-consistent grid fits are presented in Section \ref{sec:retrievals_all}.
We discuss the limitations of this study, additional work to validate our results, and the implications of our findings in Section \ref{sec:discussion}.
The appendices contain details of the data and data analysis (\ref{ap:photometry}, \ref{ap:algocomp}), model validation (\ref{ap:validation}), implementation details (\ref{ap:hansen}), and tables of the complete set of retrieval results (\ref{ap:retrievalresults}).

\section{The planetary system of HR 8799}\label{sec:context}
\begin{figure}
    \centering
    \includegraphics[width=\linewidth]{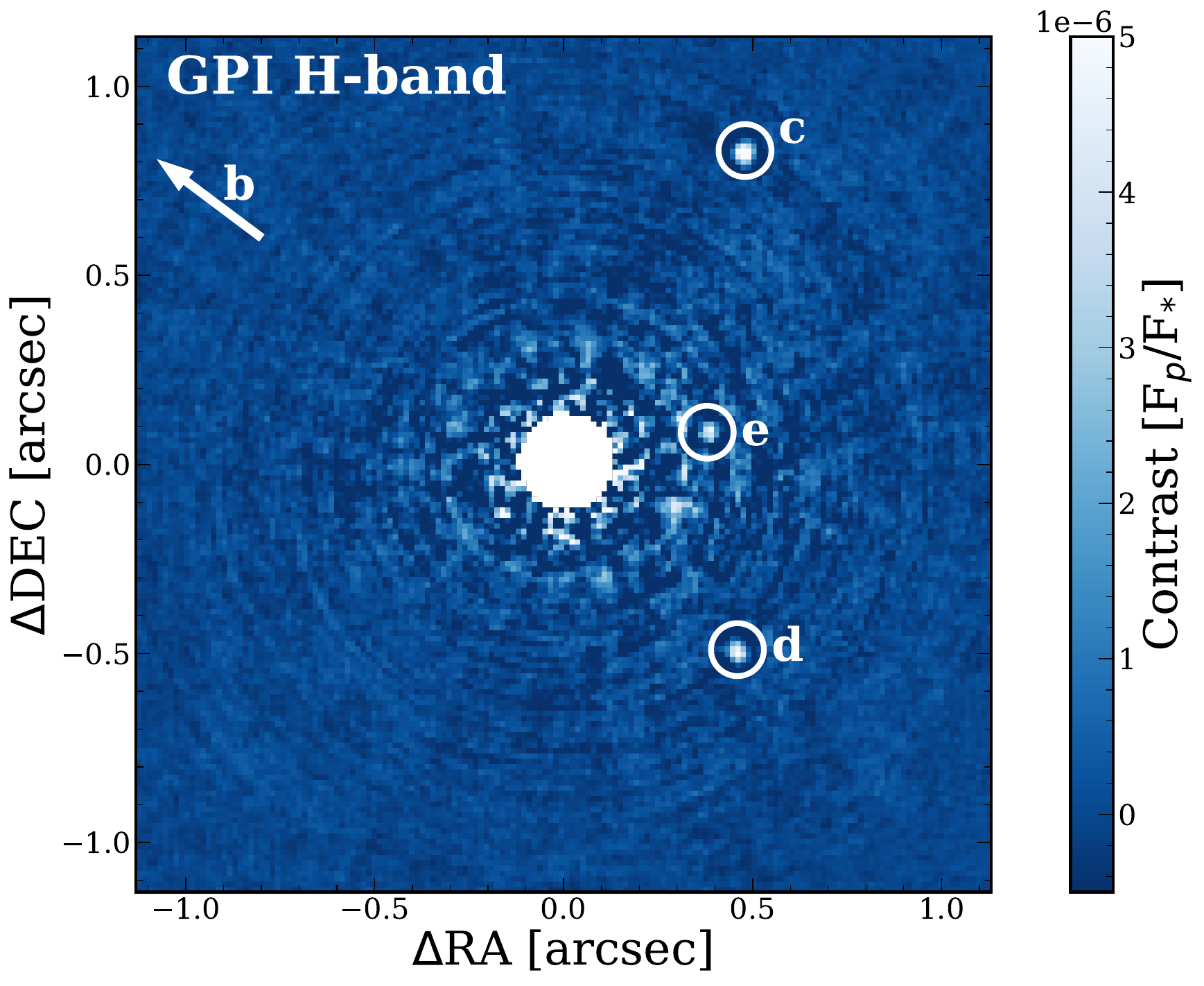}
    \caption{ HR 8799 planets as imaged in the H-band with the Gemini/GPI IFU, originally published in \cite{greenbaum_gpi_2018}. The IFU cube was processed using {\tt KLIP}, and the image is mean combined along the spectral axis. HR 8799 b is outside the field of view of GPI.}
    \label{fig:hr8799image}
\end{figure}
While HR~8799 is one of the most well studied exoplanetary systems (as seen in Figure \ref{fig:hr8799image}), there remains significant disagreement in the literature both with respect to the spectroscopic measurements and the inferred planet properties.
In general, these super jupiters all host very cloudy atmospheres, with significant impacts of disequilibrium chemistry. 
The spectra of these planets show characteristics of low surface gravity and have been classified at the L/T transition, though the variability typical of these objects has not yet been observed in the companions.
The composition of the companions is generally found to be moderately enriched compared to the host star, and while \h2o and CO are the dominant absorbers, the C/O ratio estimates vary significantly between models.
The measurements of the bulk planet properties discussed in this section, together with the results of this work are compiled in Tables \ref{tab:literature_values_b}, \ref{tab:literature_values_c}, \ref{tab:literature_values_d}, and \ref{tab:literature_values_e} for planets b, c, d, and e respectively.

\subsection{HR 8799 A: The host star}\label{sec:hoststar}
HR 8799 A is an A5V \citep{gehren_model_1977,cannon_henrydraper_1993} to F0+VkA5mA5 \citep{gray_hr8799stellartype_2003} type star, host to four detected planets and an inner and outer debris disc \citep{su_debris_2009}. 
It was one of the first identified $\gamma$ Doradas pulsators \citep{kaye_gammadoradus_1999, zerbi_gammador_1999}, and has been classified as a $\lambda$ Bo\"{o}tis star \citep{sadakane_lambdaboo_2006, moya_lambdaboo_2010} due to its depletion of heavy elements in the atmosphere.
Due to this depletion, the [Fe/H] of HR 8799 A is measured to be subsolar, with measurements ranging from $-0.47\pm0.10$ \citep{gray_hr_1999,sadakane_lambdaboo_2006} to between $-0.32\pm0.1$ and $-0.12\pm0.1$, depending on the inclination angle \citep{moya_lambdaboo_2010}.
TESS photometry allowed the measurement of this inclination angle, finding a core rotation period of $\approx$0.7 days, which combined with $v\sin i$ and stellar radius measurements would result in a preliminary stellar inclination of $\approx$28° \citep{sepulveda_20_2023}, and would favour the higher metallicity case presented by \cite{moya_lambdaboo_2010}.
Using high resolution spectroscopy from the LBT/PEPSI and HARPS instruments, \cite{wang_chemical_2020} found a very subsolar iron metallicity of $-0.52\pm0.08$, but found the relative carbon (C/H = $0.11\pm0.12$) and oxygen (O/H = $0.12\pm0.14$) abundances to be consistent with solar as is characteristic of $\lambda$ Bo\"{o}tis stars.
The derived C/O ratio from their measurements was $0.54^{+0.12}_{-0.09}$.
For this work, we use a {\tt BT-Nextgen} stellar model as fitted in \cite{nasedkin_impacts_2023} with best-fit parameters of \teff = 7200 K, log g = 3.0, and [Fe/H] = 0.0, slightly cooler than the models used in previous studies \citep{zurlo_first_2016,greenbaum_gpi_2018}.
However, this temperature is in line with \cite{sepulveda_dynamical_2022}, though based on their dynamical mass estimate of the host star and the radius measurement of \cite{baines_chara_2012} they find a higher surface gravity of 4.28.

Most indicators place the HR 8799 system between 25~and 60~Myr in age.
Using the debris disc as evidence, \cite{zuckerman_young_2004} and \cite{rhee_characterization_2007} estimate an age of 30 Myr.
\cite{zuckerman_tucanahorologium_2011} identified HR 8799 as `a likely member of the $\sim$30 Myr old Columba Association', thus providing an age and an association of stars with a shared formation history to which we can compare HR~8799 and its companions.
However, using asteroseismology \cite{moya_age_2010} found that an age of $\sim$1 Gyr or greater is also compatible with their measurements. 
While we continue to use the standard $\sim$30 Myr age for the system, we acknowledge that there remains some uncertainty in the age and activity of the host star.

\begin{table*}[t]
\centering
\begin{threeparttable}
    \centering
    \begin{footnotesize}
    \caption{Summary of literature and derived planet properties for HR 8799 b.}
    \label{tab:literature_values_b}
    \begin{tabular}{llllllllll}
    \toprule
    \textbf{Planet} & \textbf{Ref.}& \textbf{Clouds} & \textbf{M} & \textbf{log g} & $\mathbf{T_{\rm{\textbf{eff}}}}$ & $\mathbf{R}$ & \textbf{[M/H]} & \textbf{C/O} & $\mathbf{\log L_{\textbf{bol}}/L_{\odot}}$ \\
     & & &$\left[M_{\rm Jup}\right]$ & [cgs] & $\left[K\right]$ & $\left[R_{\rm Jup}\right]$ & & \\
    \midrule
    b   & B11a & Slab & 0.1$-$3.3 & $3.5\pm0.5$ & $1100\pm100$ & $0.63-0.92$ & \ldots & \ldots & $-5.1\pm0.1$\\ 
        & C11  & Thick&5$-$15    & $4-4.5$     & $800-1000$   & \ldots           & \ldots & \ldots & \ldots\\
        & G11  & Slab &1.8       & $4$         & $1100$       & 0.69        & \ldots & \ldots & \ldots\\
        & M11  & Power law &2$-$12    & $3.5-4.3$   & $750-850$    & \ldots           & \ldots & \ldots & \ldots\\
        & M12  & \citetalias{ackerman_precipitating_2001} & 26        & $4.75$      & $1000$       & 1.11        & 0 & \ldots & $-4.95\pm0.06$\\[1.5pt]
        & L13 & Slab  & $16^{+5}_{-4}$ & $5.0^{+0.1}_{-0.2}$  & $900^{+30}_{-60}$  & $0.66^{+0.07}_{-0.04}$ & \ldots & 0.96 & $-5.1\pm0.1$\\[1.5pt]
        & B15 & Slab  & \ldots  & 3.5 & 1000 & \ldots & \ldots & $0.55-0.7$ & \ldots\\
        & B16 & ER4   & \ldots & $3.4-3.8$ & $1100-1200$ & $0.6-0.7$ & 0.5 & \ldots &\ldots\\
        & L17 & Mie   & \ldots & $4\pm0.1$ & $^{a}320\pm20$ & $1.08\pm0.02$ & \ldots & 0.92$\pm0.01$ & \ldots\\[1.5pt]
        & W21$^{b}$ & BT-Settl & \ldots & $4.8^{+0.4}_{-0.8}$ & $1423.3^{+212.6}_{-278.4}$ & \ldots & \ldots & \ldots & \ldots \\[3pt]
        & R21$^{c}$ & Slab & \ldots &$3.1^{+0.03}_{-0.03}$ & $1180^{+14}_{-14}$ & \ldots & 0 & $0.578^{+0.004}_{-0.005}$ & \ldots \\
        \cmidrule{2-10}
        & Best $A\cap B^{d}$  & \citetalias{ackerman_precipitating_2001} & $6.0_{-0.3}^{+0.3}$ & $4.10_{-0.04}^{+0.03}$ & $942_{-13}^{+12}$ & $1.11_{-0.03}^{+0.03}$ & $0.96_{-0.08}^{+0.08}$ & $0.78_{-0.04}^{+0.03}$ & $-5.08_{-0.04}^{+0.04}$\\[3pt] 
        & BMA $A\cap B^{e}$  & \citetalias{ackerman_precipitating_2001} & $6.0^{+0.4}_{-0.3}$ & $4.10^{+0.03}_{-0.04}$ & $942^{+12}_{-15}$ & $1.10^{+0.03}_{-0.03}$  & $0.96^{+0.08}_{-0.08}$ & $0.78^{+0.03}_{-0.04}$ & $-5.08_{-0.04}^{+0.04}$\\[3pt]
        & $A\cap B^{f}$  & \citetalias{ackerman_precipitating_2001} & $6.0^{+0.3}_{-0.3}$ & $4.10^{+0.06}_{-0.06}$ & $936^{+22}_{-34}$ & $1.11^{+0.08}_{-0.08}$ & $1.1^{+0.1}_{-0.2}$ & $0.73^{+0.04}_{-0.04}$ & $-5.08^{+0.06}_{-0.06}$\\[3pt]
        & Grids    & Various & $4.6-6.0$ & $3.5-4.5$ &  $850-1100$ & $0.73-1.2$ & $>0.7$ & $0.3-0.55$ & $-5.14$ to $-5.28$ \\
\bottomrule
    \end{tabular}
    \begin{tablenotes}
    \small
    \item \textbf{Notes on clouds:} both the `Thick' and `Slab' clouds are based on \cite{burrows_landt_2006}, and are vertically extended throughout the atmosphere above a base pressure, with a decaying mass fraction, though the slab clouds account for a greater range of aerosol opacities. The `Power law' clouds parameterise the vertical extent and position using a power law, and fix the base pressure to the location at 2300 K. `\citetalias{ackerman_precipitating_2001}' clouds balance the cloud sedimentation and vertical mixing to determine the particle size, and use the sedimentation fraction to determine the vertical extent. `ER4' is the Exo-Rem4 model from \cite{bonnefoy_first_2016}. `Mie' clouds do not use physical optical constants, but directly fit for mie scattering parameters. The `{\tt BT-Settl}' cloud model is based on radiation hydrodynamical simulations that solve for the diffusion and mixing of aerosol particles \citep{allard_models_2012}.
    \item \textbf{References:} B11a: \cite{barman_clouds_2011}; C11: \cite{currie_combined_2011}; G11: \cite{galicher_m-band_2011}; M11: \cite{madhusudhan_model_2011}; M12: \cite{marley_masses_2012}; L13: \cite{lee_atmospheric_2013}; B15: \cite{barman_simultaneous_2015}; B16: \cite{bonnefoy_first_2016}; L17: \cite{lavie_helios-retrieval_2017}; W21: \cite{wang_detectionHiResHR8799_2021}; R21: \cite{ruffio_deep_2021}.
    \item $^{a}$ Only \Tint, a model parameter, is reported.
    \item $^{b}$ W21 used high resolution spectroscopy, and did not infer masses or radii, using masses of $7.2\pm0.7$ \mj for the inner three planets and $5.8\pm0.5$ \mj for HR~8799~b. A radius of $1.2\pm0.1$ \rj was used for all planets.
    \item $^{c}$ R21 uncertainties were limited by the coarseness and boundaries of their model grid.
    \item $^{d}$ Single best retrieval parameters.
    \item $^{e}$ Bayesian model averaged parameters from group $A\cap B$.
    \item $^{f}$ Unweighted average parameters from group $A\cap B$.
    \end{tablenotes}
    \end{footnotesize}
\end{threeparttable}
\end{table*}

\subsection{Photometric studies}
The HR~8799 system has been the subject of extensive photometric characterisation, from the red-optical out to the mid-infrared with JWST/MIRI.
The outer three companions were originally detected in \cite{marois_direct_2008}, with HR~8799~e following in \cite{marois_images_2010}.
Many photometric studies  \citep[e.g.][]{lafreniere_hstnicmos_2009,fukagawa_h-band_2009,metchev_pre-discovery_2009,currie_combined_2011,bergfors_vltnaco_2011,galicher_m-band_2011,soummer_orbital_2011,skemer_first_2012,currie_direct_2012,esposito_lbt_2013,skemer_directly_2014,currie_deep_2014,maire_leech_2015,rajan_characterizing_2015,petit_dit_de_la_roche_new_2019,biller_high-contrast_2021,boccaletti_imaginghr8799_2023} have identified the companions as L/T transition objects, with near-infrared colours compatible with more extended clouds than L-dwarfs of similar temperatures.
This is generally explained as a result of the young age and low surface gravity, where the lower gravity allows the condensate particles to remain aloft above the photosphere at lower temperatures.
Even in the earliest studies, disequilibrium chemistry was used as an explanation for the drop in the continuum flux due to CO absorption at 4~$\upmu$m \citep{currie_combined_2011, janson_spatially_2010}.

In addition to the four known companions there have been many searches for a fifth companion, interior to HR~8799~e.
\cite{thompson_deep_2023} used long time baseline astrometry and deep L' imaging with Keck/NIRC2 to search for this hypothesised companion, finding that an additional companion fits both the astrometry and photometry better than a four planet solution, but does not result in a significant detection.
For now, we  only examine the four confirmed companions further.

\cite{bonnefoy_first_2016} explores the implications of the near-infrared photometry for all four of the companions, comparing them to spectrally similar field objects from the SpeX PRISM library.
Empirically, the HR~8799 planets are much more red in colour than field dwarfs of similar spectral type.
They also show that using the dereddening coefficients for corundum (Al$_{2}$O$_{3}$), iron (Fe), enstatite (MgSiO$_{3}$), and forsterite (Mg$_{2}$SiO$_{4}$) from \cite{marocco_extremely_2014}, the colours of the companions more closely match those of field dwarfs.
However, they cannot quantitatively distinguish the chemical composition of the clouds, which requires mid-infrared spectroscopic observations of condensate absorption features \citep{burningham_cloud_2021, miles_jwst_2023}.

\cite{marley_masses_2012} and \cite{bonnefoy_first_2016} use estimates of the surface gravity and radius from spectroscopic fits to constrain the overall mass and luminosity of the planets, which they in turn compared to planetary evolution models, such as those of \cite{baraffe_evolutionary_2003} and \cite{saumon_evolution_2008}.
With self-consistent, radiative equilibrium models, the planet radius is often difficult to fit, with the radius underestimated by over 30\%\ compared to expectations from the evolutionary models (e.g. \citealt{bonnefoy_first_2016}).

\begin{table*}[ht]
\centering
\begin{threeparttable}
    \centering
    \begin{footnotesize}
    \caption{Summary of literature and derived planet properties for \object{HR 8799 c}}
    \label{tab:literature_values_c}
    \begin{tabular}{llllllllll}
    \toprule
    \textbf{Planet} & \textbf{Ref.}& \textbf{Clouds} & \textbf{M} & \textbf{log g} & $\mathbf{T_{\rm{\textbf{eff}}}}$ & $\mathbf{R}$ & \textbf{[M/H]} & \textbf{C/O} & $\mathbf{\log L_{\rm \textbf{bol}}/L_{\odot}}$ \\
     & & &$\left[M_{\rm Jup}\right]$ & [cgs] & $\left[K\right]$ & $\left[R_{\rm Jup}\right]$ & & \\
    \midrule
    c   & C11 & Thick      & 7$-$17.5    & $4-4.5$      & $1000-1200$   & \ldots      & 0       & \ldots & $-4.7\pm0.1$\\
        & G11 &  Slab      & 1.1         & $3.5$         & $1200$       & 0.97        & \ldots  & \ldots  & $-4.7\pm0.1$\\
        & M11 &  Power law & $3-11$      & $4.0-4.3$     & $950-1025$   & \ldots      & \ldots  & \ldots & $-4.7\pm0.1$\\
        & M12 & \citetalias{ackerman_precipitating_2001}       & $8-11$      & $4.1\pm0.1$  & $950\pm60$    & $1.32-1.39$ & 0  & \ldots & $-4.90\pm0.1$\\
        & K13 & \ldots     & $3-7$       & $3.5-4.0$    & $1100\pm100$   & $1-1.5$    & \ldots  & $0.65\pm0.1$  & \ldots\\
        & B16 & ER4     & \ldots & $3.8-3.9$ & 1200 & 1.0 & 0.5 & \ldots & \ldots \\
        & L17 & Mie     & \ldots & $4.5\pm0.1$ & $^{a}960\pm20$ & $1.25\pm0.02$ & \ldots & 0.55$\pm0.01$ & \ldots\\
        & G18 & Various & \ldots & $3.5-4.0$ & $1100-1350$ & $0.7-1.2$ & \ldots & \ldots &$-4.58$ to $-4.82$ \\[1.5pt]
        & W20 & Deck    & \ldots & $3.97^{+0.03}_{-0.03}$ & $1054^{+7}_{-5}$ & $1.47\pm0.02$ & \ldots & $0.58^{+0.06}_{-0.06}$ & $-4.59\pm0.004$ \\[3pt]
        & W20$^{b}$ & Deck & \ldots & $3.95^{+0.04}_{-0.12}$ & $1102\pm2$ & 1.20 & \ldots & $0.39^{+0.06}_{-0.06}$ & $-4.69\pm0.0002$\\[3pt]
        & W21$^{c}$ & BT-Settl & \ldots & $5.4^{+0.1}_{-0.2}$ & $1474.4^{+24.4}_{-36.3}$ & \ldots & \ldots & \ldots & \ldots \\[3pt]
        & R21$^{d}$ & Slab & \ldots & $3.63^{+0.03}_{-0.02}$ & $1200^{+*}_{-14}$  & \ldots & 0 & $0.562^{+0.004}_{-0.005}$ & \ldots\\[3pt]
        & W23 & \citetalias{ackerman_precipitating_2001} & \ldots &  $4.17^{+0.41}_{-0.47}$ & $^{e}1421^{+92}_{-72}$ &  $1.01^{+0.09}_{-0.08}$ & $^{f}0.51^{+0.40}_{-0.43}$ & $0.67^{+0.12}_{-0.15}$  & \ldots \\
    \cmidrule{2-10}
        & Best $A\cap B$  & \citetalias{ackerman_precipitating_2001} & $8.5_{-0.4}^{+0.4}$ & $4.26_{-0.03}^{+0.02}$ & $1158_{-12}^{+11}$ & $1.10_{-0.01}^{+0.01}$ &$1.27_{-0.06}^{+0.06}$ & $0.66_{-0.01}^{+0.01}$& $-4.71_{-0.02}^{+0.02}$\\[3pt] 
        & BMA $A\cap B$ & \citetalias{ackerman_precipitating_2001} & $8.5^{+0.4}_{-0.5}$ & $4.26^{+0.02}_{-0.03}$ & $1159^{+11}_{-12}$ & $1.10^{+0.01}_{-0.01}$ & $1.27^{+0.05}_{-0.06}$ & $0.66^{+0.01}_{-0.01}$ & $-4.71^{+0.02}_{-0.02}$\\[3pt]
        & $A\cap B$  & \citetalias{ackerman_precipitating_2001} & $8.6^{+0.4}_{-0.4}$ & $4.25^{+0.04}_{-0.14}$ & $1159^{+24}_{-76}$ & $1.10^{+0.23}_{-0.03}$ & $1.2^{+0.1}_{-0.1}$ & $0.63^{+0.05}_{-0.02}$ & $-4.70^{+0.03}_{-0.03}$\\[3pt]
        & Grids    & Various & 1.24$-$10.3 & 3.5$-$4.5  & 1100$-$1200 &$0.8-1.31$  & $>1.0$  &  $0.3-0.8$ & $-4.65$ to $-4.72$\\
\bottomrule
    \end{tabular}
    \begin{tablenotes}
    \small
    \item \textbf{References:} C11: \cite{currie_combined_2011}; G11: \cite{galicher_m-band_2011}; M11: \cite{madhusudhan_model_2011}; M12: \cite{marley_masses_2012}; L13: \cite{lee_atmospheric_2013}; K13: \cite{konopacky_detection_2013}; B16: \cite{bonnefoy_first_2016}; L17: \cite{lavie_helios-retrieval_2017}; G18: \cite{greenbaum_gpi_2018}; W20: \cite{wang_chemical_2020}; W21: \cite{wang_detectionHiResHR8799_2021}; R21: \cite{ruffio_deep_2021}; W23:\cite{wang_retrieving_2023_HR8799c}.
    \item $^{a}$ Only \Tint, a model parameter, is reported.
    \item $^{b}$ W20 compared using strong and weak radius priors to enforce physicality.
    \item $^{c}$ W21 used high resolution spectroscopy, and did not infer masses or radii, using masses of $7.2\pm0.7$ \mj for the inner three planets and $5.8\pm0.5$ \mj for HR~8799~b. A radius of $1.2\pm0.1$ \rj was used for all planets.
    \item $^{d}$ R21 uncertainties were limited by the coarseness and boundaries of their model grid.
    \item $^{e}$ W23 report the temperature at 3.3 bar rather than the effective temperature.
    \item $^{f}$ The metallicity of W23 was found by averaging the retrieved C/H and O/H ratios.
    \end{tablenotes}
    \end{footnotesize}
\end{threeparttable}
\end{table*}

\subsubsection{Variability}
Young brown dwarfs are known to be highly variable \citep{radigan_strong_2014,vos_search_2019,vos_patchy_2023}.
L/T transition objects display stronger photometric variability -- up to 30\%\ -- , though this amplitude is rare outside of the transition regime \citep{radigan_independent_2014}.
However, as we view the HR~8799 system nearly pole on, it is difficult to see the effects of rotational variation, in addition to the technical challenges of observing variability with high-contrast imaging instruments.
\cite{apai_high-cadence_2016} and \cite{biller_high-contrast_2021} have placed upper limits on the photometric variability of the two outermost HR~8799 planets: 10\% for b and 25\% for c.
\cite{wang_atmospheric_2022_CHARIS} used the Subaru/CHARIS instrument to attempt to monitor H-band variability  in HR 8799 c and d, placing upper limits of 10\% and 30\% respectively.
The atmospheric turbulence, stellar contamination, and significant post-processing required to measure the innermost companion has so far prevented measurements of variability for HR 8799 e.

\subsubsection{Orbital dynamics}
Within the context of directly imaged exoplanets, the HR~8799 companions orbit relatively near to their host star, from a projected separation of 16~au for e out to 71~au for b.
Astrometric monitoring has allowed for the precise characterisation of the orbits of the companions, demonstrated in such studies as \cite{wang_dynamical_2018,brandt_hr8799b_2021} and \cite{thompson_deep_2023}.
Using such orbital fitting techniques, \cite{zurlo_orbital_2022} inferred dynamical masses for each of the companions. 
While they explore a range of models, we use the fit assuming the planets are in a near-resonant 8:4:2:1 configuration and a host star mass of 1.47 $M_{\odot}$.
This model produced results typical of the range of models explored; the mass estimates for each companion are: $e=7.6\pm0.9$ \mj, $d=9.2\pm0.1$ \mj, $c=7.7\pm0.7$ \mj, and $b=5.8\pm0.4$ \mj. 
These dynamical mass estimates, as well as those of \citet{brandt_hr8799b_2021},  are broadly consistent with mass estimates from evolutionary models, assuming hot start conditions \citep{marley_masses_2012}.
Further astrometric analysis of the GRAVITY data will be examined in a forthcoming paper from Chavez et al.\ (in prep).

\begin{table*}[t]
\centering
\begin{threeparttable}
    \centering
    \begin{footnotesize}
    \caption{Summary of literature and derived planet properties for \object{HR 8799 d}}
    \label{tab:literature_values_d}
    \begin{tabular}{llllllllll}
    \toprule
    \textbf{Planet} & \textbf{Ref.}& \textbf{Clouds} & \textbf{M} & \textbf{log g} & $\mathbf{T_{\rm{\textbf{eff}}}}$ & $\mathbf{R}$ & \textbf{[M/H]} & \textbf{C/O} & $\mathbf{\log L_{\rm \textbf{bol}}/L_{\odot}}$ \\
     & & &$\left[M_{\rm Jup}\right]$ & [cgs] & $\left[K\right]$ & $\left[R_{\rm Jup}\right]$ & & \\
    \midrule
    d   & C11 & Thick& 5$-$17.5    & $3.75-4.5$       & $1000-1200$   & \ldots       & 0       & \ldots & $-4.7\pm0.1$\\
        & G11 & Slab& 6            & $4.0$            & $1100$        & 1.25        & \ldots  & \ldots & \ldots\\
        & M11 & Power law& $3-11$   & $3.5-4.2$       & $850-1000$    & \ldots       & \ldots  & \ldots & \ldots\\
        & M12 & \citetalias{ackerman_precipitating_2001}& $8-11$        & $4.1\pm0.1$     & $1000\pm75$   & $1.33-1.41$  & 0  & \ldots & $-4.80\pm0.09$\\
        & B16 & ER4  & \ldots & $4.4-4.5$ & $1200-1300$ & $0.9-1.1$ & 0.5 & \ldots & \ldots\\
        & L17 & Mie  & \ldots & $4.2\pm0.2$ & $^{a}1420\pm10$ & $0.96\pm0.05$ & \ldots & 0 &\ldots \\
        & G18 & Various & \ldots & $3.5-4.0$ & $1100-1600$ & $0.65-1.4$ & \ldots & \ldots &$-4.58$ to $-4.82$ \\
        & W21$^{b}$ & BT-Settl & \ldots & $5.1^{+0.3}_{-0.4}$ & $1558.8^{+50.9}_{-91.4}$ & \ldots & \ldots & \ldots & \ldots \\[3pt]
        & R21$^{c}$ & Slab & $3.7^{+0.03}_{-0.03}$ & $1200^{+*}_{-14}$ & \ldots & \ldots & 0 & $0.551^{+0.004}_{-0.005}$ & \ldots\\
    \cmidrule{2-10}
        & Best $A\cap B$ & \citetalias{ackerman_precipitating_2001} & $9.19_{-0.07}^{+0.08}$  & $4.18_{-0.03}^{+0.04}$ &  $1177_{-21}^{+21}$ & $1.26_{-0.06}^{+0.05}$ & $1.2_{-0.1}^{+0.2}$ & $0.61_{-0.04}^{+0.03}$ & $-4.63_{-0.04}^{+0.04}$ \\[3pt] 
        & BMA $A\cap B$ & \citetalias{ackerman_precipitating_2001} & $9.19^{+0.08}_{-0.07}$ & $4.18^{+0.06}_{-0.04}$ & $1179^{+31}_{-28}$ & $1.26^{+0.06}_{-0.08}$ & $1.2^{+0.2}_{-0.2}$ & $0.60^{+0.04}_{-0.06}$ & $-4.62^{+0.05}_{-0.04}$\\[3pt]
        & $A\cap B$  & \citetalias{ackerman_precipitating_2001} & $9.20^{+0.09}_{-0.08}$ & $4.19^{+0.07}_{-0.04}$ & $1179^{+38}_{-36}$ & $1.25^{+0.06}_{-0.09}$  & $1.2^{+0.4}_{-0.3}$ & $0.55^{+0.12}_{-0.10}$ & $-4.61^{+0.05}_{-0.05}$\\[3pt]
        & Grids    & Various &8.2$-$9.9 & 3.5$-$4.5 & 1200$-$1300 & 0.97$-$1.21 & $>0.0$ & $0.2-0.55$  & $-4.59$ to $-4.65$\\
\bottomrule
    \end{tabular}
    \begin{tablenotes}
    \small
    \item \textbf{References:} C11: \cite{currie_combined_2011}; G11: \cite{galicher_m-band_2011}; M11: \cite{madhusudhan_model_2011}; M12: \cite{marley_masses_2012}; B16: \cite{bonnefoy_first_2016}; L17: \cite{lavie_helios-retrieval_2017}; G18: \cite{greenbaum_gpi_2018}; W21: \cite{wang_detectionHiResHR8799_2021}; R21: \cite{ruffio_deep_2021}.
    \item $^{a}$ Only \Tint, a model parameter, is reported.

    \item $^{b}$ W21 used high resolution spectroscopy, and did not infer masses or radii, using masses of $7.2\pm0.7$ \mj for the inner three planets and $5.8\pm0.5$ \mj for HR~8799~b. A radius of $1.2\pm0.1$ \rj was used for all planets.
    \item $^{c}$ R21 uncertainties were limited by the coarseness and boundaries of their model grid.
    \end{tablenotes}
    \end{footnotesize}
\end{threeparttable}
\end{table*}

\subsection{Spectroscopic characterisation}
In addition to the multitude of photometric observing campaigns, the spectroscopic characterisation of the HR 8799 planets has traced the development of dedicated exoplanet instrumentation, from long-slit spectrographs \citep{janson_spatially_2010} to high-contrast integral field spectrographs (IFS) \citep{ingraham_gemini_2014, zurlo_first_2016} to fibre-fed high resolution spectrometers \citep{wang_detectionHiResHR8799_2021}.
These observations cover a broad swath of wavelength ranges and spectral resolving powers, leading to often conflicting photometric calibration and inferred atmospheric parameters.
In particular the H-band spectra as observed with SPHERE \citep{zurlo_first_2016}, GPI \citep{greenbaum_gpi_2018} and CHARIS \citep{wang_atmospheric_2022_CHARIS} display different flux peaks and different H-band shapes.
As several atmospheric parameters such as $\log g$ and the water abundance are strongly impacted by the shape of this band, they have remained challenging to measure. 

Many results for individual planets have been presented in the literature.
HR 8799 b was first explored in \cite{bowler_near-infrared_2010} with Keck/OSIRIS, where they identify an L5-T2 spectral type, moderate levels of cloudiness and potential impacts of disequilibrium chemistry. 
\cite{barman_clouds_2011} added additional H-band OSIRIS observations, and inferred the low temperature, low surface gravity, and low \ch4 abundance of HR 8799 b through the triangular shape of the H-band feature.
They also suggest that higher metallicity grids, up to 10$\times$ solar, may be able to better fit the data and provide more plausible radii than their solar metallicity models.
This data was augmented with additional wavelength coverage in the K band in \cite{barman_simultaneous_2015}, where they claim simultaneous detections of \h2o, CO, and tentatively \ch4. 
\cite{oppenheimer_reconnaissance_2013} obtained low resolution spectra for all four planets in the Y, J and H bands using the Project 1640 instrumentation suite at the Palomar Hale Telescopes, and thus providing the only additional measurement for HR 8799 b.
However these spectra are very low $S/N$, and are significantly discrepant from subsequent measurements.

\cite{janson_spatially_2010} were the first to spectroscopically explore HR 8799 c, using the VLT/NACO L-band spectrometer.
While they were limited in the available $S/N$, they still discussed the impact of disequilibrium chemistry on the overall shape of the spectrum, finding that there was strong CO absorption beyond 4~$\upmu$m.
These L-band observations were later succeeded by LBT/ALES observations \citep{doelman_l-band_2022,liu_applying_2023}, where low resolution spectra at moderate $S/N$ were obtained for the c, d, and e planets.
\cite{konopacky_detection_2013} presented the first conclusive evidence of CO and water absorption lines in a directly imaged exoplanet through K-band observations of HR 8799 c using Keck/OSIRIS, measuring the C/O ratio to be slightly above the stellar value at $0.65^{+0.10}_{-0.05}$.
The Gemini/GPI instrument provided the first spectra obtained using a coronagraphic instrument in \cite{ingraham_gemini_2014}, measuring both the c and d planets in the H and K bands. 
This was followed up with additional post-processing using {\tt KLIP} \citep{soummer_detection_2012,pueyo_detection_2016}
in \cite{greenbaum_gpi_2018}, where HR 8799 e was also detected. 
All three of the planets were found to best match mid-to-late L-type spectra, with HR 8799 c being most consistent with an L6 dwarf.
Consistent with the photometric models, they found HR 8799 c to have a temperature between 1100~K and 1300K, with a $\log g$ around 4.0.
As with the photometry, the self-consistent models they used to infer the planet properties struggled to obtain radius estimates consistent with predictions of evolutionary models.
\cite{wang_detectionHiResHR8799_2021} use high resolution spectroscopy to measure the rotation of c, d, and e, finding an upper limit of 14~km/s for c, and measurements of $10.1^{+2.8}_{-2.7}$~km/s for d and $15.0^{+2.3}_{-2.6}$~km/s for e.
\cite{wang_retrieving_2023_HR8799c} combine several of these datasets and perform a retrieval analysis to constrain the composition of HR 8799 c, finding [C/H] = $0.55^{+0.36}_{-0.39}$, [O/H] = $0.47^{+0.31}_{-0.32}$, and C/O = $0.67^{+0.12}_{-0.15}$.
These results depended strongly on the details of the forward model used in the retrieval, and the [C/H] parameter could vary from 0.55 to 0.95, while the [O/H] from 0.47 to 0.80, though they all represent significant enrichment compared to the host star abundances. 
These results are also significantly discrepant from those of \cite{wang_chemical_2020}, who found elemental abundance ratios for HR 8799 c of [C/H] = $0.16^{+0.12}_{-0.13}$, [O/H] = $0.13^{+0.08}_{-0.08}$, and C/O = $0.58^{+0.06}_{-0.06}$, though they also found that enforcing strong mass priors led to both the metallicities and C/O ratio being subsolar. 
\cite{ruffio_radial_2019,ruffio_deep_2021} and \cite{wang_detectionHiResHR8799_2021} explore HR 8799 c using moderate and high resolution spectroscopy respectively.
Both works characterise the dynamics of the planets, with \cite{ruffio_deep_2021} measuring the radial velocities for planets b, c, and d, finding them to be $-9.1\pm0.4$~km/s, $-11.1\pm0.4$~km/s, and $-11.6\pm0.8$~km/s respectively, placing important constraints on the allowed orbits for the planets.
They also confirm the presence of water and CO, but are unable to significantly detect \ch4, which was consistent with the results of \cite{wang_detecting_2018}.

\begin{table*}[t]
\centering
\begin{threeparttable}
    \centering
    \begin{footnotesize}
    \caption{Summary of literature and derived planet properties for HR 8799 e}
    \label{tab:literature_values_e}
    \begin{tabular}{llllllllll}
    \toprule
    \textbf{Planet} & \textbf{Ref.}& \textbf{Clouds} & \textbf{M} & \textbf{log g} & $\mathbf{T_{\rm{\textbf{eff}}}}$ & $\mathbf{R}$ & \textbf{[M/H]} & \textbf{C/O} & $\mathbf{\log L_{\rm \textbf{bol}}/L_{\odot}}$ \\
     & & &$\left[M_{\rm Jup}\right]$ & [cgs] & $\left[K\right]$ & $\left[R_{\rm Jup}\right]$ & & \\
    \midrule    e   & B16 & ER4  & \ldots         & $3.7-4.1$ & $1200-1300$ & $0.9-1.0$ & 0.5 & \ldots & \ldots\\
        & L17 & Mie  & \ldots         & $3.8\pm0.3$ & $^{a}1230\pm30$ & $1.2\pm0.1$ & \ldots & 0 & \ldots\\
        & G18 & Various & \ldots      & $3.5-4.0$ & $1100-1650$ & $0.6-1.4$ & \ldots & \ldots &$-4.58 - -4.75$ \\[1.5pt]
        & M20 & \citetalias{ackerman_precipitating_2001} & $4.81^{+8.78}_{-3.33}$ & $4.00^{+0.46}_{-0.52}$  & $1154^{+49}_{-48}$ & $1.12^{+0.09}_{-0.09}$  & $0.48^{+0.25}_{-0.29}$  & $0.60^{+0.07}_{-0.08}$ & \ldots\\[3pt]
        & W21$^{b}$ & BT-Settl & \ldots & $3.7^{+0.3}_{-0.1}$ & $1345.6^{+57.0}_{-53.3}$ & \ldots & \ldots & \ldots & \ldots \\
    \cmidrule{2-10}
        & Best $A\cap B$  & \citetalias{ackerman_precipitating_2001} & $7.5_{-0.6}^{+0.6}$  & $4.20_{-0.06}^{+0.06}$ & $1172_{-27}^{+29}$ & $1.14_{-0.05}^{+0.05}$ & $1.9_{-0.1}^{+0.1}$ & $0.88_{-0.02}^{+0.02}$&  $-4.71_{-0.06}^{+0.05}$\\[3pt] 
        & BMA $A\cap B$ & \citetalias{ackerman_precipitating_2001} & $7.5^{+0.7}_{-0.7}$ & $4.20^{+0.06}_{-0.06}$ & $1161^{+33}_{-34}$ & $1.12^{+0.05}_{-0.05}$ & $1.9^{+0.1}_{-0.2}$ & $0.88^{+0.02}_{-0.02}$ & $-4.72^{+0.06}_{-0.06}$\\[3pt]
        & $A\cap B$  & \citetalias{ackerman_precipitating_2001} & $7.5^{+0.7}_{-0.7}$& $4.3^{+0.1}_{-0.1}$ & $1198^{+41}_{-77}$ & $1.05^{+0.15}_{-0.08}$ & $1.8^{+0.3}_{-0.4}$ & $0.84^{+0.06}_{-0.07}$ & $-4.71^{+0.07}_{-0.08}$\\[3pt]
        & Grids    & Various & 1.07$-$8.8 & 3.5$-$4.5 & 1100$-$1400 & 0.75$-$1.24 & $>1.0$ & $>0.55$ & $-4.70- -4.78$ \\
    \bottomrule
    \end{tabular}
    \begin{tablenotes}
    \small
    \item \textbf{References:} B16: \cite{bonnefoy_first_2016}; L17: \cite{lavie_helios-retrieval_2017}; G18: \cite{greenbaum_gpi_2018}; M20: \cite{molliere_retrieving_2020}; W21: \cite{wang_detectionHiResHR8799_2021}; 
    \item $^{a}$ Only \Tint, a model parameter, is reported.
    \item $^{b}$ W21 used high resolution spectroscopy, and did not infer masses or radii, using masses of $7.2\pm0.7$ \mj for the inner three planets and $5.8\pm0.5$ \mj for HR~8799~b. A radius of $1.2\pm0.1$ \rj was used for all planets.
    \end{tablenotes}
    \end{footnotesize}
\end{threeparttable}
\end{table*}

The first reliable spectroscopic measurements of HR 8799 d and e were published by \cite{zurlo_first_2016}. 
These were obtained using the VLT/SPHERE instrument, and were the first YJH band observations of the inner two planets, and remain the highest quality observations in this band.
Together with the modelling work in \cite{bonnefoy_first_2016}, they classify both planets as L6-L8 dusty dwarfs, and confirm that only thick cloud models based on the Exo-REM self-consistent modelling code provide reasonable fits to the data, finding effective temperatures of 1200~K, $\log g$ in the range of 3.0-4.5, and metallicities of 0.5 for both planets.
Compared to previous modelling work of \cite{madhusudhan_model_2011} and \cite{barman_clouds_2011}, the Exo-REM models provided better fits to the data, due to improvements in the opacity databases, cloud treatments, and the inclusion of disequilibrium chemistry.
Subsequent SPHERE observations, such as those in \cite{wahhaj_search_2021} have maintained consistent spectral shapes with these earlier observations.
The \cite{gravity_collaboration_first_2019} performed the first interferometric observations of an exoplanet, measuring the K-band spectrum of HR~8799~e. 
HR 8799 e was detected as well, and they performed atmospheric analyses on all three planets using the full spectra at 1--2.5~$\upmu$m.
They found that the spectrum of HR 8799 d has a substantially different shape than the other two planets, but that all three shared supersolar metallicities and effective temperatures around 1100~K.

\subsection{Retrieval studies}
Atmospheric retrievals \citep[e.g.][]{madhusudhan_temperature_2009, benneke_retrieval_2012, waldmann_tau-rex_2015, burningham_retrieval_2017, molliere_petitradtrans_2019} are widely used to solve the inverse atmosphere problem, inferring planet properties such as the thermal structure, chemical composition, and cloud properties from spectroscopic observations.
The HR 8799 planets are among the first directly imaged planets to have retrieval methods applied to their spectra.
\cite{lee_atmospheric_2013} performed the first retrieval study of HR 8799 b, using the spectrum published of \cite{barman_clouds_2011}.
This pilot study explored various levels of cloudiness, particle sizes, and compositions, finding that the planet is likely cloudy, with relatively large particle sizes (1.5$\upmu$m) and a supersolar metallicity.
They note the long-standing degeneracies between the cloud level and the planet radius, making it difficult to distinguish between different levels of cloudiness in the models.
The first systematic characterisation of all four planets was performed in \cite{lavie_helios-retrieval_2017} using the {\tt HELIOS-Retrieval} package, with the key goal of constraining the composition of all four planets using the data of \cite{zurlo_first_2016}.
After fitting for molecular abundances, they infer the elemental C/H and O/H ratios for each planet, finding oxygen enrichment for b, c, and e, and carbon enrichment for b and c. 
They find a strongly superstellar C/O ratio for b of 0.9, a stellar value for c, but were unable to constrain the ratio for the inner two planets.
While previous works on HR 8799 e were limited due to a lack of high $S/N$ K-band data, \cite{molliere_retrieving_2020} made use of the GRAVITY spectrum obtained in \cite{gravity_collaboration_first_2019}, together with the SPHERE data of \cite{zurlo_first_2016} and the GPI data of \cite{greenbaum_gpi_2018}. 
Using the {\tt pRT} retrieval framework and a novel temperature profile, they inferred a highly cloudy atmosphere, implementing clouds with multiple scattering.
They infer modest enrichement of [M/H]=$0.48^{+0.25}_{-0.29}$ and a C/O ratio of $0.60^{+0.07}_{-0.08}$.
Finally, \cite{wang_chemical_2020} and \cite{wang_retrieving_2023_HR8799c} both perform {\tt pRT} retrieval studies of HR 8799 c. 
The latter study is unique in including high resolution data in the retrieval framework, allowing precise measurements of the elemental abundance ratios, finding modest enrichment of both carbon and oxygen.

\subsection{Self-consistent atmospheric  modelling}
Motivated by the considerable volume of observations, extensive theoretical modelling work has been performed to better understand the physics of the atmospheres of the HR 8799 planets and similar substellar objects.
Brown dwarf atmospheres saw extensive 1D modelling efforts \citep[e.g.][]{chabrier_evolutionary_2000, allard_limiting_2001, burrows_landt_2006, saumon_evolution_2008}, driven largely by the need to trace the evolution of these continuously cooling objects over time.
These studies demonstrated the necessity of accounting for silicate clouds in the atmospheres of L/T dwarfs, used to explain the red colour of these objects in the near infrared.
Applied specifically to the young, low-gravity companions, \cite{madhusudhan_model_2011} developed one of the first models specifically for the HR~8799 companions to constrain their mass and age. 
They identify forsterite and iron as being the important contributors to the clouds, and infer planetary ages between 10~and 150~Myr, consistent with stellar measurements.
\cite{marley_masses_2012} provides a deep review of the state of modelling of the atmospheres of the HR 8799 planets, further developing the model of \cite{saumon_evolution_2008}.
They find masses and ages for the planets consistent with the stellar properties, and that the companions share approximately consistent properties with L/T dwarfs of similar effective temperatures and surface gravities.
Using the cloud model of \citet[][hereafter \citetalias{ackerman_precipitating_2001}]{ackerman_precipitating_2001}, they infer an \fsed parameter of 2, implying that the clouds are moderately extended throughout the atmosphere.
More recent self consistent models such as petitCODE \citep{molliere_model_2015}, ATMO \citep{tremblin_fingering_2015, phillips_new_2020, petrus_x-shyne_2023}, Exo-REM \citep{charnay_self-consistent_2018}, and Sonora (\cite{marley_sonora_2021, karalidi_sonora_2021}; Morley et al.\ in prep) have been developed specifically to understand the thermal structure and clouds of directly imaged planets.
There remain degeneracies between reddening and damping of spectral features via continuum opacity sources and through reductions in the temperature gradient, hypothesised to be due to diabatic convection \citep{tremblin_thermocompositional_2019}.

\cite{zahnle_methane_2014} provide an in-depth exploration of the impacts of disequilibrium chemistry on cool, self-luminous atmospheres, providing predictions for the CO, \ch4, and NH$_{3}$ abundances as a function of vertical mixing and effective temperature, identifying the key transition between \ch4 and CO dominated chemistry at around 1100~K.
\cite{moses_composition_2016} uses a disequilibrium model including photochemistry to predict the chemical composition for a range of surface gravities and effective temperatures, and provides column abundance predictions for HR 8799 b, finding that the CO abundance should dominate over \ch4, assuming a solar composition.
\cite{soni_effect_2023} extend this to superstellar metallicities and vertical mixing strengths, using the constraints on the CO and \ch4 abundances from \cite{barman_simultaneous_2015} to infer a vertical mixing strength of $\log \kzz \in [7,10]$ for the $10\times$ solar metallicity case.
To better understand the planet structure, \cite{thorngren_mass-metallicity_2016} derive a mass-metallicity relationship.
As the mass of the object increases, the metallicity tends to decrease, consistent with predictions of core accretion formation, as heavier objects accrete and retain more H$_{2}$ and He relative to a lower mass object.
From their relationship, they predict that a 6 $M_{\rm Jup}$ planet should have a $\Zpl/\Zstar$  ratio of between 3 to 5 (in a 68\% confidence interval).

\begin{figure*}[t]
    \includegraphics[width=\linewidth]{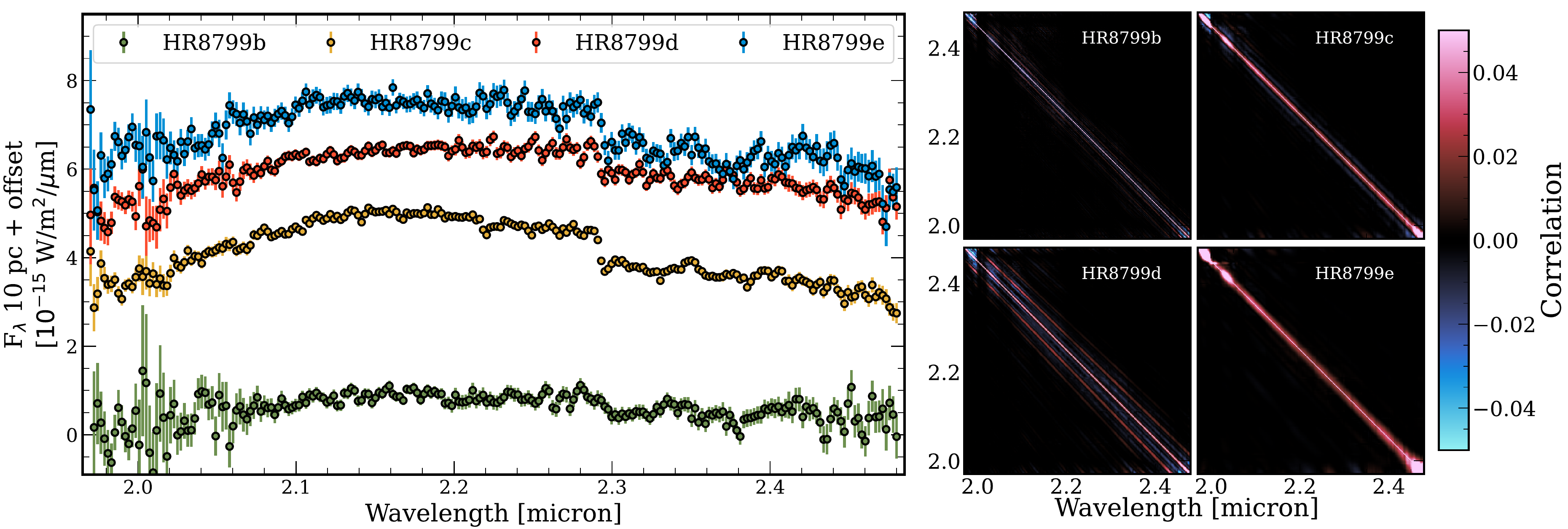}
    \caption{Flux-calibrated VLTI/GRAVITY K-band spectra for each of the HR8799 planets, normalised to 10 pc. Each spectrum above that of HR 8799 b has an additional 1.5$\times10^{-15}$ W/m$^{2}$/$\upmu$m offset. The panels on the right show the empirical correlation matrices for each of the four planets. The colour bar is scaled to highlight weak correlations in the GRAVITY data.}
    \label{fig:gravityspectra}
\end{figure*}

In addition to the 1D modelling efforts, global circulation models (GCMs) of self-luminous, substellar objects, such as those of \cite{showman_atmospheric_2013, tan_atmospheric_2021, tan_atmospheric_2021-1} have been developed.
These 3D models allow for the exploration of atmospheric dynamics, longitudinal variations, and time variability.
Recent observations are beginning to validate these 3D models: \cite{suarez_ultracool_2023_latitudes} finds that brown dwarfs are cloudier when viewing the equator, which is consistent with the cloudiness predictions of rapidly rotating brown dwarfs in \cite{tan_atmospheric_2021-1}.
Likewise, the prediction of patchy clouds in the photosphere region leading to variability \citep{showman_atmospheric_2013} seems to match the observations of high variability in low-gravity atmospheres with silicate clouds \citep{vos_patchy_2023}.

\subsection{Formation}
With four massive planets on wide orbits, HR 8799 provides a unique system with which to test formation scenarios.
In general, these fall under the categories of either gravitational instability models \citep[e.g.][]{perri_nebulainstability_1974,cameron_diskinstability_1978,adams_eccentric_1989,laughlin_gi_1996,boss_giant_1997}, where the planets form via the direct collapse of the gas into a substellar object, or core accretion \citep{pollack_formation_1996,bodenheimer_insituformation_2000}, where a dense core of heavy material grows slowly until it is massive enough to experience runaway accretion and gather an extended hydrogen-helium envelope.
GI models tend to produce larger planets on wider orbits with solar compositions, while core accretion scenarios form closer-in planets on more circular orbits, with the possibility of greater metal enrichment.
\cite{dodson-robinson_formation_2009} tested both of these scenarios for HR 8799, finding that while core accretion may better explain the near-orbital resonances of the system, it struggled to form planets on such wide orbits (beyond 30~au), and could not rule out the possibility of direct gravitational collapse.
Similarly, \cite{nero_did_2009} find that while HR 8799 b may have formed through gravitational instability, it is unlikely that disc fragmentation could have formed the inner three companions.

In addition to constraints from the mass and location of the companions, the present-day planet composition provides insight into the formation and evolution history.
The template for this was developed in \cite{oberg_effects_2011}, demonstrating how the C/O ratio in the gas and dust varies as a function of position in the disc, which would in turn impact the outcome of the formation process.
\cite{eistrup_molecular_2018} extended this model to include time evolution, and \cite{molliere_interpreting_2022} presented a framework to link the measured planet properties to the disc environment in a Bayesian framework, which allows testing the effect of various formation assumptions.
However, due to the uncertainty in the atmospheric measurements, combined with the many outstanding questions in formation modelling, this link remains tenuous.

The different formation scenarios can lead to dramatically different amounts of energy retained in the planet following the formation process.
So-called `hot-start' models result in young planets retaining the gravitational potential energy as internal heat, to be radiated and cooled over time \citep{marley_luminosity_2007,mordasini_characterization_2017}.
This scenario is typically associated with formation due to gravitational instability.
In cold-start scenarios, often tied to core-accretion models, this energy is radiated away by accretion shocks as the gas flows from the circumstellar disc onto the forming planet, resulting in a lower internal energy \citep{mordasini_characterization_2012, szulagyi_thermodynamics_2017}.
This is a useful, though simplified picture of planet formation.
Additional complication comes from the energetics of the accretion shock during core accretion, where different radiative efficiencies can lead to different initial entropies of the forming planet \citep{marlau_shock1_2017}. 
These shock-resolving models find typical internal energies that are an order of magnitude higher than in typical cold-start scenarios \citep{marleau_shock2_2019}, thus lying somewhere between the hot and cold start scenarios.
Over time, all of these scenarios converge to the same cooling rate, though precise mass and luminosity estimates can distinguish between the two scenarios for the first $\sim$100~Myr \citep{baraffe_evolutionary_2003,saumon_evolution_2008}.
Current measurements of planet masses, temperatures, and radii generally favour hot or warm start models, but can only definitely exclude the coldest initial conditions, such as the cold-start models of \cite{marley_luminosity_2007}.
The hot-start models of \cite{baraffe_evolutionary_2003} led to predictions of 7 $M_{\rm Jup}$ for the inner three planets, and $5 M_{\rm Jup}$ for HR 8799 b, which are approximately consistent with the current dynamical mass estimates of \cite{zurlo_orbital_2022}.
Using the hot-start model of \cite{saumon_evolution_2008}, \cite{marley_masses_2012} finds that the radii of all of the planets should be slightly larger than $1 R_{\rm jup}$, and that even assuming very cold initial conditions the planet radii should never fall below $1 R_{\rm jup}$, though this claim did not account for significantly nonsolar composition.

Finally, HR 8799 is home to both an inner and outer debris disc, imaged with \textit{Spitzer} \citep{su_debris_2009}, \textit{Herschel} \citep{matthews_resolved_2014}, JWST \citep{boccaletti_imaginghr8799_2023}, and \textit{ALMA} in the millimeter \citep{wilner_resolved_2018}.
The inner debris disc has a temperature of around 150~K and is confined to within 10~au, while the cold outer debris disc is analagous to the Kuiper belt in our own Solar System \citep{geiler_scattered_2018}, but at a much wider separation (90--300~au).
The structure of the outer disc appears to be sculpted by an additional gravitational component, though it is unclear whether this is due to HR 8799 b or an additional unseen companion \citep{contro_dynamical_2015,faramaz_detailed_2021}.
The inner disc has been detected in thermal emission \citep{su_debris_2009} and resolved using MIRI coronagraphic imaging \citep{boccaletti_imaginghr8799_2023}.
Modelling efforts have placed tentative limits of $\sim 1 M_{\rm Jup}$ on the allowed mass of companions interior to HR~8799~e \citep{gozdziewski_orbital_2018}.

\section{Observations}\label{sec:obs}
While the new GRAVITY spectra represent the best available K-band observations of the HR 8799 system, additional data are required to constrain planetary properties such as surface gravity and C/O ratio.
We combine published datasets across a wide wavelength range from a variety of sources in order to present the most complete possible picture of this system. 
Archival photometric data of the companions are also included in our analysis, the details of which can be found in appendix \ref{ap:photometry}.
Also included in appendix \ref{ap:photometry} is the stellar photometry used in fitting the {\tt BT-Nextgen} model, with which the companion contrast measurements are flux calibrated.
In this section we present a brief overview of the spectroscopic datasets included in the retrieval analysis, with the key observational parameters listed in Table \ref{tab:data}.
All of the observational data, together with the complete set of retrievals results is available on Zenodo\footnote{\url{https://zenodo.org/records/10914429}}.

\begin{figure*}[t]
    \includegraphics[width=\linewidth]{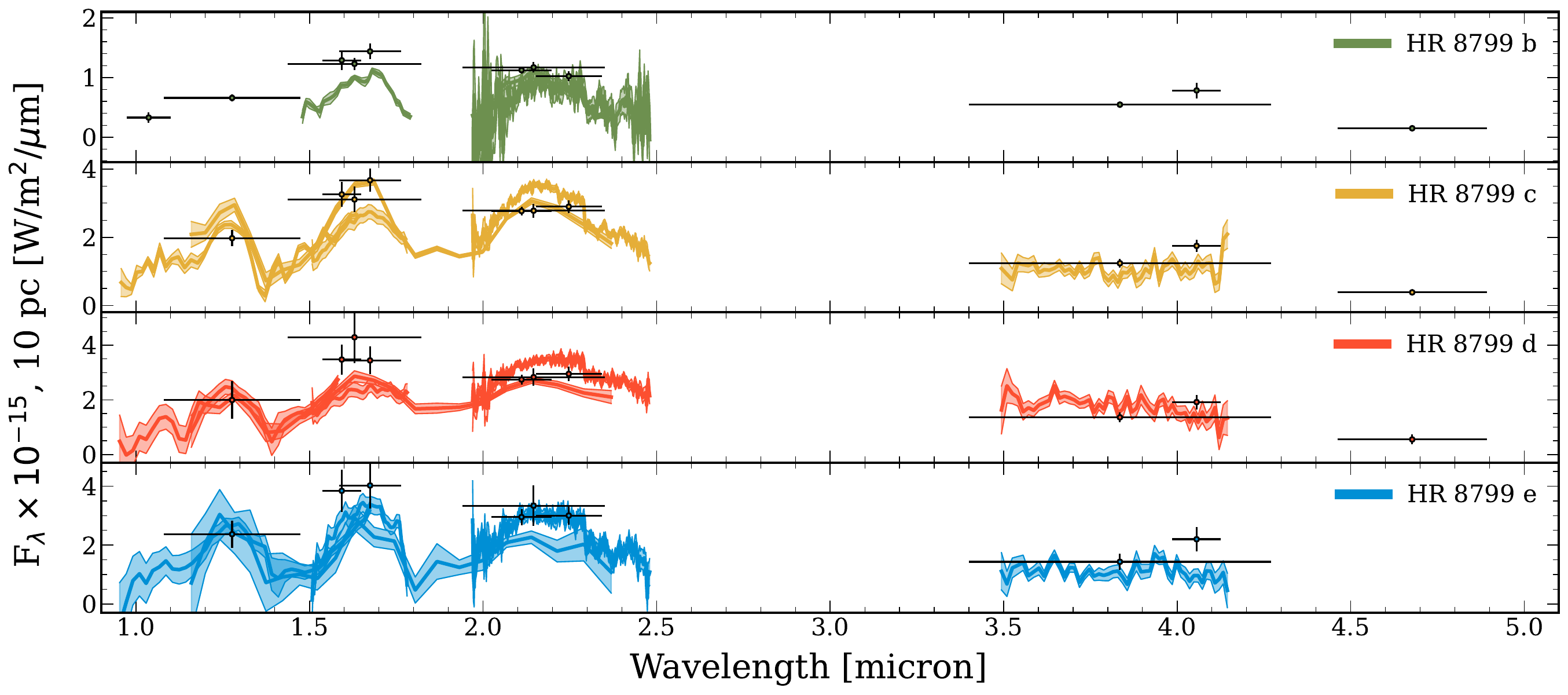}
    \caption{Data of the HR 8799 planets. The OSIRIS b and ALES d datasets have scaling factors of 1.0 and 1.2 applied, respectively. Not shown are the MIRI photometry points from \cite{boccaletti_imaginghr8799_2023}. References: OSIRIS \citep{barman_clouds_2011}; SPHERE \citep{zurlo_first_2016}; GPI \citep{greenbaum_gpi_2018}; CHARIS \citep{wang_chemical_2020,wang_atmospheric_2022_CHARIS}; ALES \citep{doelman_l-band_2022}.
    }
    \label{fig:allspectra}
\end{figure*}

\subsection{GRAVITY data}  
In Figure \ref{fig:gravityspectra} we present new VLTI/GRAVITY observations of HR~8799~e, together with the first interferometric observations of d, c, and b taken as part of the ExoGRAVITY project \citep{lacour_exogravity_2020}, under ESO program ID 1104.C-0651.
GRAVITY is a K-band spectroscopic interferometer that combines light from either the four 8~m Unit Telescopes (UTs) of the VLT, or the 1.8-m Auxiliary Telescopes \citep{gravity_collaboration_first_2017}.
With baselines of up to 134~m, GRAVITY provides unprecedented spatial resolution, allowing for the detection of companions close to their host stars and the measurement of relative astrometry with a precision of few tens of $\upmu$as.
All observations of the HR8799 system were obtained using the UTs, with the dual-field mode of GRAVITY. The medium resolution mode was used, which offers a resolution of R$\sim$500 over a nominal wavelength range of 2.0~to 2.4~$\upmu$m. 

Two different strategies were used for the observations and data-reduction. Observations of HR~8799~e at all dates, except on 2 dates (11 November 2019 and 02 July 2023) were obtained using the on-axis strategy, in which a 50/50 beam-splitter is used to separate the field to between the science and fringe-tracking channels of GRAVITY. 
In this mode, observations with the science channel pointing at the location of the planet are interleaved with observations obtained with the fibre pointed at the central star. 
The on-star observations are then used to calibrate both the interferometric phase and amplitudes. This is similar to the observations reported in \cite{nowak_direct_2020}. 
The second strategy is the dual-field/off-axis strategy, in which the roof-mirror is used to split the field. The use of the roof-mirror is required to observe planets at larger separation, because the field of view of the beam-splitter does not reach these targets. In this case, the metrology zero point is calibrated using observations of the dedicated calibrator HD~25535, and the interferometric amplitude using an on-axis observation of the central star, typically done at the end of the observation sequence. This strategy is similar to the observation of Sgr~A* by \citet{gravity_modeling_2020}.

The data-reduction was performed using the tools developed for the ExoGRAVITY large program\footnote{\url{https://gitlab.obspm.fr/mnowak/exogravity}}. The main steps of the reductions are as follows:
\begin{enumerate}
    \item All data are first reduced with the GRAVITY pipeline \citep{lapeyrere_gravity_2014}, up to the `astroreduced' data product, which keeps individual DITs separated. 
    \item For the on-axis observations, the phase reference is extracted from the on-star observations and subtracted from the on-planet observations. For the off-axis observations, this phase-reference is extracted from the observations of the binary-calibrator HD~25535. In both cases, the amplitude reference is taken using the on-star observations.
    \item The stellar light (also called stellar speckle) is subtracted from the reduced data by fitting a fourth-order polynomial in wavelength multiplied by the amplitude reference. The astrometry of the planet is then extracted from the observations.
    \item The contrast spectrum is then extracted using a model that also takes into account the residual starlight, and the planet astrometry previously extracted. 
\end{enumerate}

This procedure yields a contrast spectrum for each planet, at each observation date. The spectrum extraction, which consists entirely of linear operations on the complex coherent flux, also propagates the errors reported by the pipeline as covariance matrices. These covariance matrices allow for correlations over the wavelength channels and between the real and imaginary parts of the coherent flux. However, it should be noted that the GRAVITY pipeline does not report such covariances, and so the extraction code starts with fully diagonal covariance matrices.

For each planet, all the available spectra are then combined using a covariance-weighted combination. The final contrast spectrum $C = (c_{\lambda_1}, c_{\lambda_2}, \dots{}, c_{\lambda_{n}})^T$ and its associated covariance matrix $W$ arre given by:
\begin{align}
    W &= \left[\sum_t{{W_t}^{-1}}\right]^{-1} \\
    C &= W\cdot{}\left[\sum_t{{W_t}^{-1}\cdot{}C_t}\right],
\end{align}
where $C_t$ and $W_t$ represent the contrast spectrum and its associated covariance matrix on a given observing date $t$.

The contrast spectra are then converted to fluxes using a model of the stellar flux. For HR~8799, we used a {\tt BT-Nextgen} model fit to the near infrared photometry, the details of which are more thoroughly discussed in Section \ref{sec:hoststar} and are based on \cite{nasedkin_impacts_2023}.

The faintest companion, HR~8799~b, was detected with a mean $S/N$ of 3.4 per wavelength channel. HR~8799~c was observed with a mean $S/N$ of 27.5 per channel, while HR~8799~d and HR~8799~e had a mean $S/N\approx20$ and $S/N\approx10$ respectively.
These observations were taken over a 5 year period. With the 50 microarcsecond astrometric precision of GRAVITY, this will allow the detection of planet--planet orbital perturbances within a few years \citep{covarrubias_n-body_2022}, and we leave such analysis to future work.

\subsection{Archival data}
In addition to the new GRAVITY spectra, we also include archival data covering a broad wavelength range, presented in Fig.~\ref{fig:allspectra}.
\cite{molliere_retrieving_2020} noted that the SPHERE \citep{zurlo_first_2016} and GPI \citep{greenbaum_gpi_2018} data are inconsistent with each other in the H-band.
In order to reduce systematic variation and to account for correlations, we rereduce the data with up-to-date pipelines, and reprocesses the datasets optimally as described in \cite{nasedkin_impacts_2023} using \klip \citep{soummer_detection_2012,pueyo_detection_2016}.
However, in order to best extract the planet signal we use \klip in ADI+SDI mode, in comparison to ADI only mode as described in the previous study. 
Both the reprocessed SPHERE and GPI spectra can be found in Figs. \ref{fig:sphereextract} and \ref{fig:gpiextract}.
In total, our dataset includes nearly 400 data points for each planet: $N_{b}=297,~ N_{c}=391,~N_{d}=387,~N_{e}=388$.

\subsubsection{SPHERE}
Two sets of VLT/SPHERE \citep{beuzit_sphere_2008, beuzit_sphere_2019} data are considered in this study: the first was taken during the commissioning run of the SPHERE instrument on 12 August 2014, and was originally published in \cite{zurlo_first_2016}. 
This is still the deepest SPHERE observation of HR 8799 covering the full YJH range, but due to the orientation of the field of view does not include HR~8799~c. 
This dataset was reprocessed as in \cite{nasedkin_impacts_2023} using \klip in ADI+SDI mode, and we extract spectra and covariance matrices for both the e and the d companions.
The second SPHERE dataset was published in \cite{flasseur_paco_2020}, who processed the dataset using the {\tt PACO-ASDI} algorithm and were able to extract a spectrum for HR~8799~c in addition to d and e. 

Additional SPHERE observations, such as presented in \cite{biller_high-contrast_2021} or \cite{wahhaj_search_2021} are available. 
However, in the case of \cite{biller_high-contrast_2021} the observations of the host star used for photometric calibration that were taken before and after the science observations are of insufficient $S/N$. 
While we attempted to calibrate the companion spectra using the satellite spots, this was unreliable.
Finally, these observations only cover the Y and J bands, and lack the overlap with the GPI H-band spectrum, which is important for ensuring compatibility across instruments. Therefore we continue with only the datasets of \cite{zurlo_first_2016} and \cite{flasseur_paco_2020}.

\subsubsection{GPI}
Gemini/GPI \citep{macintosh_first_2014} observations of HR8799, originally published in \cite{greenbaum_gpi_2018}, were taken on 17 November 2013, 18 November 2013, and 19 September 2016 for the K1, K2 and H bands respectively.
These were reduced using the standard GPI reduction pipeline (version 1.4.0), and reprocessed with {\tt KLIP} using the same methods as the SPHERE data.
As the new GRAVITY observations supersede the GPI data in the K-band, we only consider the GPI H-band data for this work.

\subsubsection{CHARIS}
Subaru/CHARIS \citep{groff_charis_2015,groff_first_2017} observations of HR~8799 c, d, and e were presented in \cite{wang_chemical_2020} and \cite{wang_atmospheric_2022_CHARIS}. 
These observations cover 1.2--$2.4~\upmu$m range at low resolution. 
\cite{wang_atmospheric_2022_CHARIS} primarily examined these data for temporal variability, while here we combine the full two nights of observations in order to obtain the highest precision spectrum for each of the three planets.
We take the mean spectrum for both nights, and add the errors in quadrature, dividing by the square root of the number of observations (i.e. by $\sqrt{2}$) to obtain a spectrum for each planet.

\subsubsection{ALES}
\cite{doelman_l-band_2022} presented L-band observations of HR~8799 c, d, and e obtained using the LBT/ALES instrument \citep{skemer_first_2015}.
These supersede the VLT/NACO L-band observations of HR~8799~c of \cite{janson_spatially_2010}, and are the first L-band spectra of HR~8799~d and e.
These data also include covariance matrices, estimated using the analytic method of \cite{greco_measurement_2016}.

\subsubsection{OSIRIS}
Archival Keck/OSIRIS \citep{larkin_osiris_2006} data taken between 2009 and 2010 is included for HR8799b, as published in \citep{barman_clouds_2011}.
HR~8799~b falls outside the field of view of most high-contrast-imaging IFUs, so OSIRIS is joined only by GRAVITY in measuring the near infrared spectrum of the planet.
With an unbinned spectral resolution of R$\approx$4000, and an integration time of 2700s in the H-band and 1800s in the K-band, when binned to a spectral resolution of R$\approx$60 the OSIRIS data achieves a per-channel $S/N$ comparable to or better than that of the GRAVITY observations.
As the OSIRIS data were not taken using standard ADI observing modes, we did not attempt any rereduction or reprocessing of the archival data, apart from rescaling the flux and uncertainty by the current \textit{Gaia} distance estimate of $41.2925pm0.15$ pc \citep{gaia_edr3_2020}.

We also include the K-band spectra of \cite{konopacky_detection_2013} (Figure 2 of that work). As published, this spectrum is not flux calibrated, and so we always fit for a flux-scaling term. 
For HR~8799~b and c, these K-band spectra allow us to explore the impact of different measurements on the retrieved atmospheric parameters, and to determine if our methods can reproduce the results of earlier work.

More recent observations of the HR~8799 planets using OSIRIS have been explored in \cite{ruffio_deep_2021}, but these spectra are continuum subtracted, requiring a somewhat different modelling framework than the rest of the data considered in this work. 
As such we do not fit these data, but we do use them as an additional check on the robustness of our fits when examining the best-fit models at higher resolution.

%

\section{Atmospheric modelling}\label{sec:modelling}
The forward models of our atmospheric retrieval setup were computed using {\tt pRT} version 2.7 \citep{molliere_petitradtrans_2019}, a fast, open-source radiative transfer code with which we calculate the emission spectrum of a planetary atmosphere\footnote{\url{https://petitradtrans.readthedocs.io/}}.
Our fiducial setup was based on that of \cite{molliere_retrieving_2020}, used to retrieve the atmospheric properties of HR~8799~e, though substantial improvements to the code have been made and are detailed further in \cite{nasedkin_atmospheric_2024}.
We explore a wide range of model parameterisations, summarising the parameters and prior distributions used in Table \ref{tab:priors_retrievals}.
As we consider several thermal profile parameterisation, we compare their prior distributions separately in Table \ref{tab:thermal_priors_retrievals}.

To allow for both a data-driven and physically motivated approach, we retrieved either $\log g$ and $R_{\rm pl}$ with uniform priors or $R_{\rm pl}$ and $M_{\rm pl}$, with Gaussian priors set by the dynamical mass estimates of \cite{zurlo_orbital_2022} and broad Gaussian priors centred at 1.1 $R_{\rm Jup}$, in line with estimates from evolutionary models \citep{marley_masses_2012}.

As the computational cost of a retrieval varied greatly between the planets, it was unfeasible to run every model for every planet. 
As our primary point of comparison we explored the different temperature profile parameterisations for each planet, and ran both disequilibrium and free chemistry retrievals for each planet. 
Due to its low computational run time, we ran additional models for HR 8799 e, focusing on different cloud parameterisations.
\begin{table}[t]
    \centering
    \caption{Retrieval prior; temperature profile priors are included in Table \ref{tab:thermal_priors_retrievals}. $\mathcal{N}(\mu,\sigma)$, $\mathcal{U}$(low, high).}
    \label{tab:priors_retrievals}
    \begin{tabular}{ll}
    \toprule
       \textbf{Parameter}  & \textbf{Prior} \\
    \midrule
       $\log g$  & $\mathcal{U}\left(2.5,5.5\right)$\\[1pt]
       Radius [R$_{\rm Jup}$] & $\mathcal{U}\left(0.7,2.0\right)$\\
                              & $\mathcal{N}\left(1.1,0.1\right)$\\[1pt]
        Mass [M$_{\rm Jup}$]  & $\mathcal{N}\left(\mu_{\rm M,\;dyn},\sigma_{\rm M,\;dyn}\right)$\\[1pt]
        $\log P_{\rm Quench}$ [log bar]& $\mathcal{U}\left(-6.0,3.0\right)$\\[1pt]\ignorespaces
        [M/H] & $\mathcal{U}\left(0.5, 2.5\right)$\\[1pt]
        C/O   & $\mathcal{U}\left(0.1, 1.6\right)$\\[1pt]
        $\sigma_{\rm LN}$ & $\mathcal{U}\left(1.05, 3.0\right)$\\[1pt]
        \fsed & $\mathcal{U}\left(0.0, 10.0\right)$\\[1pt]
        $\log \kzz$ & $\mathcal{U}\left(5.0, 13.0\right)$\\[1pt]
        log Eq. Cloud Scaling & $\mathcal{U}\left(-2.5, 2.5\right)$\\[1pt]
        log Cloud Mass Fracs. & $\mathcal{U}\left(-6.5, 0.0\right)$\\[1pt]
        log Cloud $P_{\rm Base}$ & $\mathcal{U}\left(-6.0, 3.0\right)$\\[1pt]
        log Mass Fracs. & $\mathcal{U}\left(-7.0, 0.3\right)$\\[1pt]
    \bottomrule
    \end{tabular}

\end{table}

\subsection{Thermal structure}
\begin{table*}[t]
    \centering
    \begin{threeparttable}
    \centering
    \caption{Priors for temperature profiles.}
    \label{tab:thermal_priors_retrievals}
    \begin{small}
    \begin{tabular}{lclclclc}
    \toprule
       \multicolumn{2}{c}{\textbf{Spline}} & \multicolumn{2}{c}{\textbf{Guillot (\citetalias{guillot_radiative_2010})}} & \multicolumn{2}{c}{\textbf{Molli\`{e}re (\citetalias{molliere_retrieving_2020})}} &\multicolumn{2}{c}{\textbf{Zhang (\citetalias{zhang_elemental_2023})}}\\
       Name & Prior & Name & Prior & Name & Prior & Name & Prior\\
    \midrule
       $T_{0}$ & $\mathcal{U}$(0 K, 300K) & \Tint & $\mathcal{U}$(300 K, 2500K) & \Tint & $\mathcal{U}$(300 K, 2000K) &  \Tbot & $\mathcal{U}$(2000 K, 12000K) \\
       $T_{i}$ & $\mathcal{U}$(300 K, 11900 K) & $T_{\rm equ}$ & $\mathcal{U}$(10 K, 100K)& $T_{0}$ & $\mathcal{U}$(0, $T_{\rm Edd}$) &  $d\log T/d\log P_{0}$ & $\mathcal{N}(0.25,0.025)$ \\
       $\gamma$ & $\Gamma^{-1}$(1, $5\times10^{-5}$) & $\gamma$ & $\mathcal{N}(0,2)$ &  $T_{1}$ & $\mathcal{U}$(0, $T_{0}$) &  $d\log T/d\log P_{1}$ & $\mathcal{N}(0.25,0.045)$ \\
       
       \ldots& \ldots & $\log\kappa_{\rm IR}$ & $\mathcal{U}$($-4$, 1) &   $T_{2}$ & $\mathcal{U}$(0, $T_{1}$) & $d\log T/d\log P_{2}$ & $\mathcal{N}(0.26,0.05)$ \\
        \ldots&\ldots &\ldots & \ldots& $\alpha$ & $\mathcal{U}$(1, 2) &  $d\log T/d\log P_{3}$ & $\mathcal{N}(0.2,0.05)$ \\
        \ldots& \ldots&\ldots & \ldots& $\log\delta$ & $\mathcal{U}$(0, 1) &  $d\log T/d\log P_{4}$ & $\mathcal{N}(0.12,0.045)$ \\
        \ldots& \ldots&\ldots & \ldots&\ldots &\ldots &  $d\log T/d\log P_{5}$ & $\mathcal{N}(0.07,0.07)$ \\
    \bottomrule
    \end{tabular}
    \end{small}
    \begin{tablenotes}
    \small
    \item\textbf{Notes}
    \item $\mathcal{N}(\mu,\sigma)$: Gaussian prior. 
    \item $\mathcal{U}$(low, high): Uniform prior. 
    \item $\Gamma^{-1}\left(\alpha,\beta\right)$: Inverse Gamma function prior. 
    \end{tablenotes}
\end{threeparttable}
\end{table*}

We compared a set of four temperature structures in our model comparison in order to distinguish the amount of model flexibility justified by the data and the impact of the temperature structure on other retrieved atmospheric parameters.
While the thermal structure of these self-luminous objects is thought to be well-understood from 1D and 3D atmospheric models, this comparison will validate these predictions using an independent, data-driven methodology.
At the same time, using the best physical understanding of the thermal structure may help constrain other parameters with greater accuracy and precision; it is necessary to compare both approaches to ensure consistent results. 
For all different temperature profiles we computed an effective temperature after the spectrum computation, by integrating $F_{\lambda}$ over wavelength and applying the Stephan-Boltzmann law.
To do this, we integrated a low resolution spectrum from 0.8~to 250~$\upmu$m. 
The lower limit is set by the wavelength coverage of the data; the optical band is unconstrained and leads to unrealistically large uncertainty on the effective temperature.
The long wavelength limit is set by the wavelength coverage of the opacity databases.

\subsubsection{Spline profile}
To allow the data to fully determine the temperature profile of the atmosphere, we used a Piecewise Cubic Hermite Interpolating Polynomial as implemented in the \verb|scipy.interpolate.| \verb|PchipInterpolator| function. 
Following the prescription of \cite{line_uniform_2015}, we penalised curvature in the temperature profile by adding an additional term to the likelihood function,
\begin{equation}
    \log p(\mathbf{T}) = \frac{1}{2\gamma}\sum_{i=1}^{N-1}\left(T_{i+1} - 2T + T_{i-1}\right)^{2} - \log\left(2\pi\gamma\right).
\end{equation}
This is the additional penalty term, which we found by taking the sum of the discrete second derivative of the temperature profile.
An additional hyperparameter, $\gamma$, was also included, with an inverse gamma distribution prior.
If $\gamma$ is large, (disfavoured by the prior), then the data truly demands strong curvature in the profile, while if $\gamma$ is small, the data favours smoother profiles.
Following \cite{line_uniform_2015}, we set the parameters of the prior distribution on $\gamma$ based on the work of \cite{lang_bayesian_2004,Rahman_2005} and \cite{jullion_robust_2007}:
\begin{equation}
    \Gamma^{-1}\left(\gamma\right) = \frac{\beta^{\alpha}}{\Gamma\left(\alpha\right)}\left(\frac{1}{\gamma}\right)^{\alpha + 1}\exp\left(-\frac{\beta}{\gamma}\right),
\end{equation}
for fixed $\alpha$ and $\beta$ parameters given in Table \ref{tab:thermal_priors_retrievals}.
We repeated the retrievals and varied the number of nodes in the profile, which allowed us to use a Bayes factor comparison to determine the allowable level of complexity.
This also allowed us to explore how the pressure-temperature profile can compensate for the presence of clouds by reducing the temperature gradient in the photospheric region.

\subsubsection{Guillot profile} 
The \cite{guillot_radiative_2010} (G10) profile is a simple analytical model, constructed to estimate the thermal structure of irradiated planets:
\begin{multline}
    \label{eqn:modifguillot}
      T_{\rm Guillot}^4 = \frac{3\Tint^{4}}{4}\left(\frac{2}{3} + \tau\right) \\
            +\frac{3T_{\rm irr}^4}{4}\left(\frac{2}{3} + \frac{1}{\gamma\sqrt{3}} + 
            \left(\frac{\gamma}{\sqrt{3}} - \frac{1}{\gamma\sqrt{3}}\right)e^{-\gamma\tau\sqrt{3}}\right),
\end{multline}
where $T_{\rm irr} = \sqrt{2}T_{\rm equ}$ and $\tau = P\times\kappa_{\rm IR}/g$.
$T_{\rm equ}$ is the equilibrium temperature of an irradiated body, \Tint is the intrinsic temperature of the planet, and $g$ is the surface gravity.
$\kappa_{\rm IR}$ is the mean infrared opacity, and $\gamma$ is the ratio between the optical and infrared opacities.
While these parameters can be physically interpreted, we treat them as nuisance parameters that control the shape of the profile, rather than self-consistently linking them to the chemical opacities.
As the HR 8799 planets are widely separated, $T_{\rm irr}$ is small, and thus the profile reduces to an \cite{eddington_effect_1930} profile, which corresponds to keeping only the first term on the righthand side of Equation~(\ref{eqn:modifguillot}).

\subsubsection{Molli\`{e}re profile}
Introduced in \cite{molliere_retrieving_2020} (M20), this is a physically motivated temperature profile, split into three distinct regions in altitude.

The middle level of the atmosphere contains the photosphere.
In this region the temperature profile follows an Eddington profile, as in the first term of the Guillot profile in Equation \ref{eqn:modifguillot}.
However, for this profile we parameterise the opacity $\tau$ as a function of pressure ($P$): 
\begin{equation}
    \tau = \delta P^{\alpha}
\end{equation}
and retrieve parameters of $\log\delta$ and $\alpha$, together with \Tint, as in the \citetalias{guillot_radiative_2010} profile.

The upper atmosphere is defined as the region above $\tau=0.1$. 
Above this level, four pressure points are defined, equidistant in $\log P$.
The deepest pressure point, at $\tau = 0.1$ is fixed to the temperature of the Eddington profile of the middle atmospheric region, while the remaining temperature points are freely retrieved parameters, subject to the constraint that the temperature decreases with altitude \citep{kitzmann_helios-r2_2020}, as inversions are not expected in self-luminous objects.
The temperature profile is then interpolated from a cubic spline between the three points.
Combined with the Eddington profile parameters, this results in a total of 6 parameters to describe the temperature profile.

The base of the atmosphere is defined as a moist adiabat, up to the radiative-convective boundary.
This boundary occurs when the temperature gradient of the Eddington profile is Schwarzchild unstable:
\begin{equation}\label{eqn:schz_inst}
    \frac{dT}{dr} < \frac{T}{P}\frac{dP}{dr}\left(1 - \frac{1}{\gamma_{\rm ad}}\right).
\end{equation}
The moist adiabatic gradient is a function of the temperature, pressure, and chemical composition, and as such is interpolated from the disequilibrium chemistry table, discussed further in Section \ref{sec:poormansnoneq}. 
Once the atmosphere is unstable to convection, the temperature profile is forced onto the moist adiabat.

\subsubsection{Zhang profile} 
\cite{zhang_elemental_2023} (Z23) introduced a novel P-T parameterisation, incorporating the results of radiative-convective equilibrium models into the retrievals via careful prior selection. 
This is accomplished by fitting for the gradient of the temperature with respect to pressure, as opposed to directly retrieving the temperature as in the spline profile. 
The prior locations and widths of the gradients were determined by empirically measuring the temperature gradients in self-consistent radiative-convective models, thus providing a means to enforce the physics of these models in a retrieval framework.
The atmosphere between $10^{3}$ bar and $10^{-3}$ bar was divided up into six layers, equidistant in log pressure. 
The temperature at the bottom of the atmosphere (\Tbot) was freely retrieved. 
For the remaining layers, $\left. d\log T/d\log P\right|_{i}$ were retrieved as free parameters.
The temperature profile was then found by interpolating the gradient to the full pressure grid, and integrating to find the temperature at each pressure.
\begin{align}\label{eqn:grad}
    &T_{0} = T_{\rm Bot}\\
    &T_{i+1} = \exp\left(\log T_{i} + \left(\log P_{i+1} - \log P_{i}\right)\left(\frac{d\log T}{d\log P}\right)_{i}\right)
\end{align}
The atmosphere was isothermal above 10$^{-3}$ bar.

\subsection{Chemistry}\label{sec:chemistry}
Understanding the atmospheric chemistry of the HR8799 planets is one of the key goals of this work.
We compared a simplified disequilibrium chemistry model to a free chemistry retrieval with vertically constant abundances.
We primarily used opacities from the ExoMol database \citep{tennyson_exomol_2012,chubb_exomolop_2020}, and included \h2o \citep{ExoMol_H2O},  CO \citep{rothman_hitemp_2010}, \ch4 \citep{ExoMol_CH4}, \co2 \citep{ExoMol_CO2}, NH$_{3}$ \citep{ExoMol_NH3}, HCN \citep{ExoMol_HCN}, H$_{2}$S \citep{ExoMol_H2S}, PH$_{3}$ \citep{exomol_ph3}, FeH \citep{wende_feh_2010}, Na \citep{allard_new_2019}, K \citep{allard_k-h_2016}, SiO \citep{exomol_sio}, TiO \citep{exomol_tio}, and VO \citep{exomol_vo}.

\subsubsection{Free chemistry}
In the free chemistry retrievals, we assumed a vertically constant mass fraction for each species, and retrieved the log mass fraction abundance ($\log X_{i}$ for each of \h2o, CO, \ch4, \co2, HCN, H$_{2}$S, NH$_{3}$, FeH, Na, and K), subject to the constraint that the sum of the mass fractions is less than one.
Due to the lack of spectroscopic data in in the Y and J bands, we did not retrieve FeH, Na or K for HR 8799 b in the free retrievals to reduce the number of free parameters.
The hydrogen and helium mass fractions were calculated by using the solar abundances (0.766 for H$_{2}$, 0.234 for He), and multiplying by one minus the sum of the retrieved mass fractions $(1 - \Sigma_{i} X_{i})$.
The set of molecules included covers the most abundant trace species in the atmosphere, and in an atmosphere with strong vertical mixing in the photosphere the assumption of a vertically constant abundance is reasonable for \h2o, CO, and \ch4, though other species, such as FeH, have been found to have nonvertically constant abundances \citep{rowland_feh_2023}.
To measure the significance of a detection, we performed a `leave-one-out' retrieval, and calculated the Bayes factor between the complete retrieval and the retrieval when excluding a single chemical species.
This comparison was performed using the Zhang temperature profile and clouds condensing at their equilibrium saturation condition.
While the detection significant may vary with different setups, this setup is representative of typical retrievals, and ensures consistent comparisons.

To determine the bulk properties of [M/H] and C/O for the free retrievals, we converted the retrieved mass fraction abundances to volume mixing ratios. 
[M/H] is defined as the ratio the planetary elemental abundances from the measurements of \h2o, CO, \ch4, \co2, NH$_{3}$, and H$_{2}$S to the solar elemental abundances.
The C/O ratio was likewise found from the number ratio of the carbon and oxygen atoms in the same set of molecular species.

\subsubsection{Interpolated (dis)equilibrium}\label{sec:poormansnoneq}
The disequilibrium model used a grid of equilibrium chemical abundances interpolated along dimensions of pressure and temperature, as well as [M/H] and C/O, which were freely retrieved parameters.
The metallicity parameter scaled all of the elemental abundances, while the C/O scaled only the oxygen abundance.
Initial test retrievals used a prior range of [-1.5,1.5], but the high metallicity demanded by the data led to the choice of a prior range of [-0.5,2.5] for the retrievals included in this work.
The model of disequilibrium chemistry of the CO, \ch4, \h2o system was based on transport-induced quenching, resulting in a vertically constant abundance above a given pressure.
This (log) quench pressure was one of the retrieved parameters. 
The equilibrium abundances used to build the grid were computed using \verb|easyChem| \citep{molliere_observing_2017}, which minimises the Gibbs free energy for the system at a given pressure, temperature, and atomic composition. 
We included all of the species listed in Section \ref{sec:chemistry} as opacity sources, though 95 species are included in equilibrium chemical network used to determine the molecular abundances.
For the alkali metals we used the wing profiles of \cite{allard_k-h_2016,allard_new_2019}.

\subsection{Clouds}
We considered three cloud parameterisations in this work.
The first was based of the model of \cite{ackerman_precipitating_2001}, where cloud particles are lofted into the atmosphere through eddy diffusion ($\kzz$), and settle back down with a speed proportional to the parameter \fsed. 
We retrieved each of these parameters independently, together with $\sigma_{\rm LN}$, which is the width of the log-normal particle size distribution.

While these parameters determine the structure of the clouds, we also determined the cloud opacity through the use of cloud optical constants for different cloud compositions, allowing us to differentiate between compositions, grain structure (amorphous or crystalline), and whether the particle shapes are spherical or based on the distribution of hollow spheres (DHS).
In addition to the standard log-normal particle sized distribution used with \citetalias{ackerman_precipitating_2001} clouds, we incorporated the \cite{hansen_multiple_1971} particle size distribution, which has been proposed to be a more accurate representation of the particle size than a log-normal distribution.
Instead of $\sigma_{\rm ln}$, we retrieved the mean effective width $b_{\rm h}$. 
The details of how these parameters shape the distribution, together with how they were incorporated into the \citetalias{ackerman_precipitating_2001} model are described in Appendix~\ref{ap:hansen}.

We tested a range of cloud compositions, including MgSiO$_{3}$ (both crystalline and amorphous particle shapes), Mg$_{2}$SiO$_{4}$, Fe, Al$_{2}$O$_{3}$, KCl, and Na$_{2}$S, as well as several combinations of these compositions. 
Each cloud composition had a mass fraction abundance at the cloud base that could either be scaled from an equilibrium value or freely retrieved, together with a unique \fsed, which allowed for different vertical extents for different cloud compositions.

Nominally, the cloud base occurs at the intersection of the cloud condensation curve and the temperature pressure profile.
However, for our second parameterisation we included the cloud base pressure as a freely retrieved parameter, to determine if the clouds form where expected in the atmosphere.
We followed the derivation of \citetalias{ackerman_precipitating_2001} and
\cite{molliere_observing_2017} to obtain the cloud abundance throughout the atmosphere.
The abundance of the cloud species $X_{i}$ was defined at this base pressure $P_{0}$, and decreases with altitude to the power of \fsed:
\begin{equation}
    X_{i}(P) = X_{i,0}\left(\frac{P}{P_{0}}\right)^{f_{\rm sed}}
.\end{equation}
At pressures higher than $P_{0}$, the cloud is not condensed, and thus $X_{i}=0$.

Finally, we also tested a simple grey cloud deck, where the cloud top pressure was freely retrieved, and the cloud acts as a source of opacity at the base of the photosphere.

For each of these models we could retrieve a cloud patchiness fraction, $f_{\rm c}$. 
In this setup, we first calculated the usual cloudy spectrum, $\vec{S}_{\rm cd}$.
We then turned off the cloud opacity sources, and calculated another clear atmosphere spectrum, $\vec{S}_{ \rm cl}$.
We then combined the two spectra, weighted by $f_{\rm c}$:
\begin{equation}
    \vec{S} = f_{\rm c}\vec{S}_{\rm cd} + (1-f_{\rm c})\vec{S}_{\rm cl}.
\end{equation}
This approach divides the atmosphere into only clear and cloudy components. 
Other approaches, such as those of \cite{vos_patchy_2023} or \cite{mccarthy_multiple_2024} have different patchiness fractions for different cloud layers, allowing for different degrees of cloudiness.
Our approach reduces the number of parameters and is simple to implement in a retrieval framework, but future work should explore the patchiness of individual layers of clouds in the atmosphere.

\subsection{Retrieval setup}
We used the {\tt pyMultiNest} \citep{buchner_x-ray_2014} wrapper of {\tt MultiNest} \citep{feroz_multimodal_2008, feroz_importance_2013} as the basis for our nested sampling routine, as testing showed that it runs significantly faster than the {\tt UltraNest} sampler, which may provide more accurate estimates of the Bayesian evidence. 
For all retrievals we used 4000 live points to ensure dense posterior sampling and coverage of the parameter space.
We set the sampling efficiency to 0.05, and used constant efficiency mode in order to reduce computation time.
Comparisons to retrievals using 4000 live points and a sampling efficiency of 0.8 without constant efficiency showed that this choice does not bias the posterior estimates, and that the importance nested sampling evidence estimate is of sufficient precision for model comparison.
The ($\log\mathcal{Z}$) evidence tolerance was set to 0.1, ensuring precise estimates of the evidence and ensuring convergence of the retrievals. 

\subsection{Retrieval ranking}
\begin{figure}[b]
    \centering
    \includegraphics[width = 0.65\linewidth]{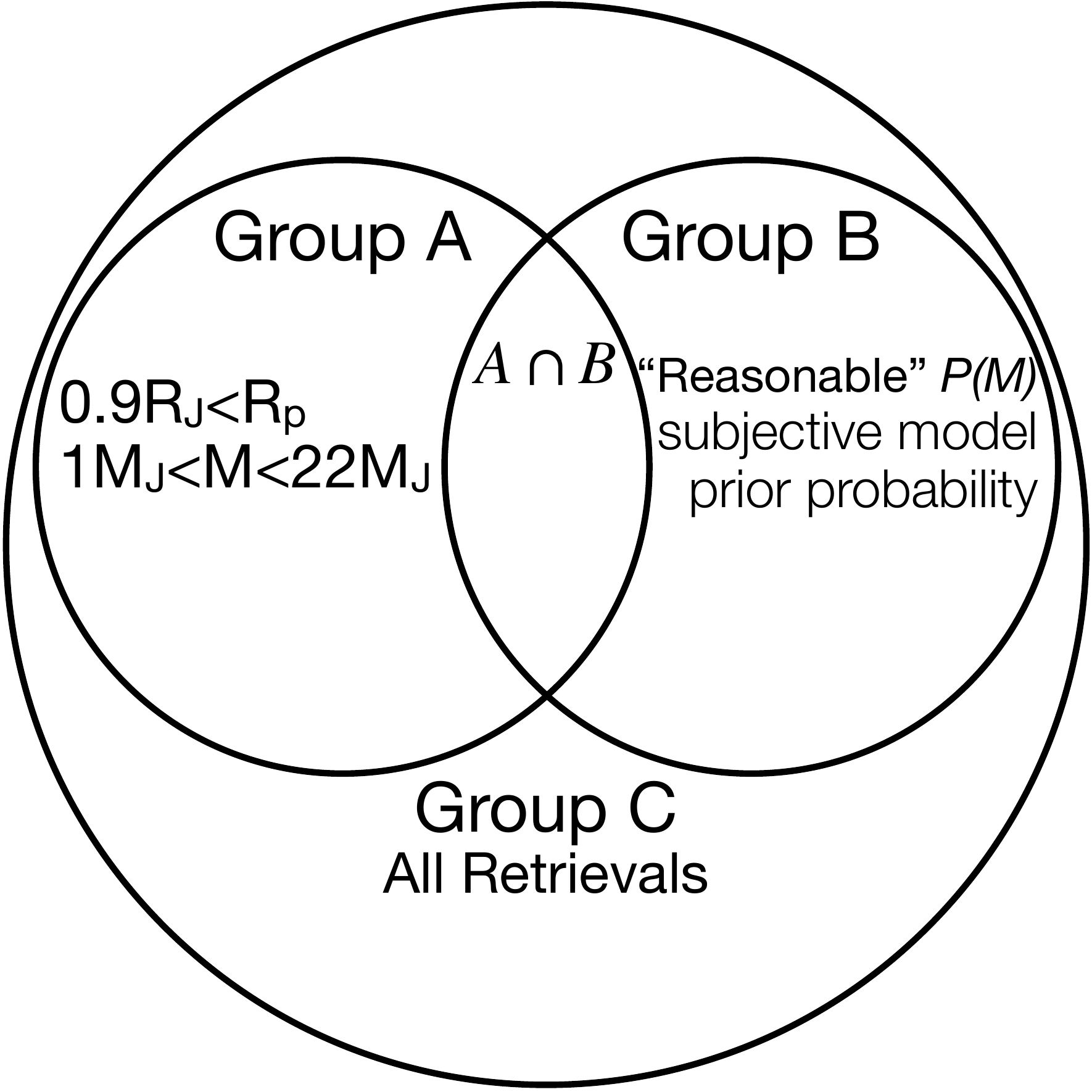}
    \caption{Illustration of how retrievals are grouped in this work. Group A retrievals are selected based on having physically plausible posterior distributions. Group B retrievals are models that are subjectively chosen to have a `reasonable' prior probability. Group C includes the entire set of retrievals run in this work.}
    \label{fig:ranking}
\end{figure}

\begin{table*}[t]
    \centering
    \caption{Self-consistent grid boundaries and step sizes.}
    \label{tab:grid_boundaries}
    \begin{tabular}{r|llllll}
    \toprule
       \textbf{Parameter}  & \textbf{ATMO} & \textbf{Diamondback} & \textbf{Exo-REM} & \textbf{petitCODE, Cool} & \textbf{petitCODE, Hot} \\
    \midrule
        $T_{\rm eff}$ [K]& [800, 3000] & [900,2400] & [400,2000] & [500,850] &[1000,1800]\\
                         & 50 & 100 & 50 & 50 & 100\\[2pt]
        $\log g$ [dex]   & [2.5, 5.5] & [3.5,5.5] & [3.0,5.0] & [3.0,5.0] & [3.5,5.5]\\
                         & 0.1 & 0.5 & 0.5 & 0.5 & 0.5\\[2pt]
        [M/H]            & [-0.6, 0.6] & [-0.5,0.5] & [-0.5,1.0]& [0.0,1.4] & [-0.3,0.3]\\
                         & 0.3 & 0.5 & 0.5 & 0.2 & 0.3\\[2pt]
        C/O              & [0.3, 0.7] & 0.458 & [0.1,0.8] & 0.55 & [0.55,0.75,0.90]\\
                         & 0.24 &\ldots & 0.05 & \ldots & \ldots\\[2pt]
        \fsed    & \ldots & [1,2,3,4,8] & \ldots & [0.5,2.0] & [1.5,4.5]\\
                         & \ldots & \ldots & \ldots  & 0.5 & 1.5\\[2pt]
        $\gamma_{\rm ad}$    & [1.01, 1.05] & \ldots & \ldots & \ldots &\ldots\\
                         & 0.02 & \ldots & \ldots & \ldots & \ldots\\        
    \bottomrule
    \end{tabular}
\end{table*}

Considering the number and range of models run, we must devise a system to systematically evaluate the quality of the retrieval.
A true Bayesian approach would be to exclusively use the Bayes factor to evaluate the model fits.
However, without well-defined prior odds for each model, we cannot quantitatively account for the prior probability of a given model.
For example, based on the current understanding of these objects, the prior probability of a clear atmosphere model should be less than that of a cloudy model, but there is no clear way of assigning an objective probability.
Instead we subjectively grouped some models into a `low odds' category.
While these are useful for validating our assumptions about the planets and testing the inclusion of different datasets, they should not contribute significantly to a final combined parameter estimate.
We further sorted the retrievals a posteriori, creating in total three tiers of retrieval results, illustrated in Fig.~\ref{fig:ranking}. 
We focused our overall analysis on a subset of retrievals that are both  plausible models, use consistent datasets, and produce physically reasonable results.

\textit{Group A}: This set of retrievals is defined as those with physically reasonable posterior values.
Based on evolutionary models, it is expected that the HR 8799 planets have radii greater than 1 R$_{\rm Jup}$.
Even high-mass, cold brown dwarfs over 1 Gyr in age are found to have minimum radii of  $\sim$0.88 R$_{\rm Jup}$, with the minimum radius increasing with increasing metallicity \citep{burrows_dependence_2011}. 
Thus we exclude from group A any retrievals with a median retrieved radius less than  0.9 R$_{\rm Jup}$.
We additionally enforce that the mass estimate should be broadly consistent with the dynamical mass estimates: the median retrieved mass must be greater than 1 M$_{\rm Jup}$ and less than 22 M$_{\rm Jup}$, approximately double the highest dynamical mass estimate of any of the planets \citep{zurlo_orbital_2022}. 
The exact positioning of these cuts does not significantly impact the results.


\textit{Group B}: this set of retrievals includes those that we consider to have a high prior model probability $P(M)$; equivalently we are assigning a model prior probability $P(M)=0$ to those models that we believe do not describe these atmospheres well.
Specifically, we exclude models with a clear atmosphere, those with with poorly parameterised temperature profiles used during validation studies (e.g. retrievals using only 2 nodes to define a spline temperature profile), and those using data inconsistent with our fiducial dataset.
Thus while the retrievals using OSIRIS data for HR 8799 c are highly ranked by the Bayes factor, we exclude them from our analysis and from the Bayesian Model Average, as the Bayes factor is only a relevant metric when comparing like datasets.
Likewise, a clear atmosphere would require a diabatic temperature profile to explain the reddening of the emission spectra, which we do not include in the retrievals and therefore the clear models are unlikely to be physically meaningful.
As the Bayes factors are weighted heavily towards the best retrievals, a weighted posterior distribution effectively reduces to that of Group A.

\textit{Group C}: the complete set of retrievals included in this work, regardless of prior or posterior likelihood. 
As the full set of retrievals includes highly unrealistic atmospheric models by design, we do not present combined posterior distributions, but only explore specific comparison retrievals used to validate different model assumptions.
Ultimately we found that all of the retrievals fall into group A, universally finding reasonable estimates for the planet masses and radii.

The best set of retrievals is the intersection of groups A and B (indicated by $A\cap B$ when used to refer to a particular retrieval), which are retrievals that have physically plausible posterior values, and whose model we believe is a reasonable representation of the atmosphere.
Tables \ref{tab:full_results_b} to \ref{tab:full_results_e} list the complete set of retrieval results, classifying the individual retrievals by group and sorting by the Bayes factor. 
We turn to \cite{kass_bayesfactors_1995} for an interpretation of the Bayes factor in terms of frequentist statistical significance.
Thus a $\Delta\log_{10}\mathcal{Z}>1$ is considered substantial evidence,  and $\Delta\log_{10}\mathcal{Z}>2$ is considered strong evidence, equivalent to $>5\sigma$ significance.
Table 2 of \cite{benneke_how_2013} present a similar, albeit slightly more conservative interpretation of the Bayes factor, with a similar threshold of $\log_{10}\mathcal{Z}=2.1$ for `strong' evidence in favour of one hypothesis over another, equivalent to 3.6$\sigma$ significance.

\subsection{Bayesian model averaging}
We used the techniques of Bayesian Model Averaging (BMA) in order to combine estimates of a single parameter over a range of models, following the review of \cite{fragoso_bayesian_2018}.
This has recently been applied to exoplanet spectroscopy in \cite{nixon_methods_2023}, demonstrating that these methods provide more realistic posterior uncertainties.
They highlight that to naively use BMA, the use of multiple duplicate models must be avoided to avoid the repeated contribution of that model to the average. 
As we do not have any identical models in our retrieval suite, BMA remains a valid approach.

Consider Bayes theorem for the $i^{\rm th}$ model $M_{i}$ for data $\vec{D}$, with parameters $\vec{\theta}_{i}$:
\begin{equation}\label{eqn:bayestheorem}
    P(\vec{\theta}_{i}|\vec{D},M_{i}) = \frac{P(\vec{D}|\vec{\theta}_{i},M_{i})P(\vec{\theta}_{i}|M_{i})}{P(\vec{D}|M_{i})}.
\end{equation}
We are interested in obtain a joint posterior probability distribution $P(\vec{\theta}|\vec{D},M)$ for the subset of parameters $\vec{\theta}$ that are shared between the set of models.
From each model we require posterior probability distribution $P(\vec{\theta}_{i}|\vec{D},M_{i})$, the likelihood $P(\vec{D}|\vec{\theta}_{i},M_{i})$, and the evidence $P(\vec{D}|M_{i}) \equiv \mathcal{Z} = \int P(\vec{D}|\vec{\theta}_{i},M_{i})P(\vec{\theta}_{i}|M_{i})d\vec{\theta}_{i}$.
We then assume a prior probability distribution over the full set of models under consideration, and therefore each model has an associated prior probability $P(M_{i})$.
The choice of this prior probability should reflect the prior knowledge of the system under consideration. 
For example, the prior probability of a clear atmosphere model should be lower than that of a cloudy atmosphere model for the HR 8799 planets.
However, quantifying this degree of certainty is highly subjective.
We choose instead to use an uninformative prior distribution across $N$ models for each planet:
\begin{equation}
    P(M_{i}) = \frac{1}{N}
.\end{equation}
This allows the data to determine which models should be favoured based on the evidence.

Considering all models in the range 1 to $N$, the posterior model probabilities given the data are
\begin{equation}
    P(M_{i}|\vec{D}) = \frac{P(\vec{D}|M_{i})P(\vec{M_{i})}}{\sum_{j=1}^{N}P(\vec{D}|M_{j})P(M_{j})}.
\end{equation}
The marginal posterior distribution for a single parameter $\theta$ present in all of the models is thus 
\begin{equation}\label{eqn:combinedposteriors}
    P(\theta|\vec{D}) = \sum_{j=1}^{N}P(\theta|\vec{D},M_{j})P(M_{j}|\vec{D}).
\end{equation}
This combined posterior distribution folds in both the uncertainty from the data and prior distributions, but from the model uncertainty as well, providing a more robust estimate of the overall uncertainty on the inferred parameter.


\subsection{Self-consistent forward modelling}\label{sec:selfconsistent}

In order to ensure that the retrieval results are robust and insensitive to the details of {\tt pRT}, we fit each of the planet's spectra using several grids of 1D self-consistent models: ATMO \citep{phillips_new_2020,petrus_x-shyne_2023}, Sonora Bobcat, Cholla ,and Diamondback \citep{marley_sonora_2021,karalidi_sonora_2021}, Morley et al. (in prep), Exo-REM \citep{charnay_self-consistent_2018}, and petitCODE \citep{molliere_model_2015,molliere_observing_2017}.
These models represent the current state-of-the-art in both cloudy and cloud-free self-consistent 1D models. 
The boundaries and intervals of each of these grids is presented in Table \ref{tab:grid_boundaries}. 

\subsubsection{ATMO}
We used an up-to-date grid of ATMO of models from \cite{petrus_x-shyne_2023}, which in turn is based on prior versions from \cite{tremblin_fingering_2015} and \cite{phillips_new_2020}.
ATMO is a clear atmosphere model, based on the hypothesis that diabatic convection \citep{tremblin_cloudless_2016,tremblin_cloudless_2017}, not clouds, are responsible for the reddening of the near-infrared spectra of directly imaged exoplanets and brown dwarfs. 
This convection is instigated by disequilibrium chemical processes that reduce the temperature gradient, thus reddening the atmosphere.
In ATMO, this is parameterised through an effective adiabatic index $\gamma_{\rm ad}$, which modifies the temperature gradient.
The inclusion of this parameter meant that ATMO is the only clear atmosphere grid that produced a reasonable fit to the spectra of the HR~8799 companions.

\subsubsection{Exo-REM}
The Exoplanet Radiative-convective Equilibrium Model (Exo-REM, \cite{baudino_interpreting_2015, baudino_toward_2017}) is a self-consistent model used to study directly imaged exoplanets and brown dwarfs 
\citep{charnay_self-consistent_2018}, but has also been extended to lower mass transiting planets \citep{blain_exorem_2021}.
It implements a cloud microphysics model by combining \citetalias{ackerman_precipitating_2001} with the timescale approach of \citet{rossow1978}, which allows it to reproduce the L-T brown dwarf spectral sequence as a function of effective temperature.
\cite{bonnefoy_first_2016} used this grid to explore the atmospheres of the HR~8799 companions, finding atmospheres mildly enriched in metals ([M/H]=0.5) and well constrained effective temperatures.
However, they developed a set of custom grids that implement detailed cloud properties to model the atmospheres, which are likely more suited to the HR~8799 planets than the more general publicly available grid.

\subsubsection{Sonora}
The newly developed suite of Sonora models are designed to model the spectra and evolution of substellar atmospheres, covering the L-T-Y spectral sequence
\citep{marley_sonora_2021}. Sonora comes in several flavours, implementing equilibrium chemistry in Sonora Bobcat \citep{marley_sonora_2021}, disequilibrium chemistry in Sonora Cholla  \citep{karalidi_sonora_2021}, and cloudy atmospheres in Sonora Diamondback \citep{morley_diamondback_2024}.
Like Exo-REM and ATMO, the Sonora models are a 1D, radiative-convective equilibrium model that couples hydrostatic and thermochemical equilibrium temperature structure with a radiative transfer scheme to compute the atmospheric emission spectrum.

\begin{figure*}[t]
\includegraphics[width=\linewidth]{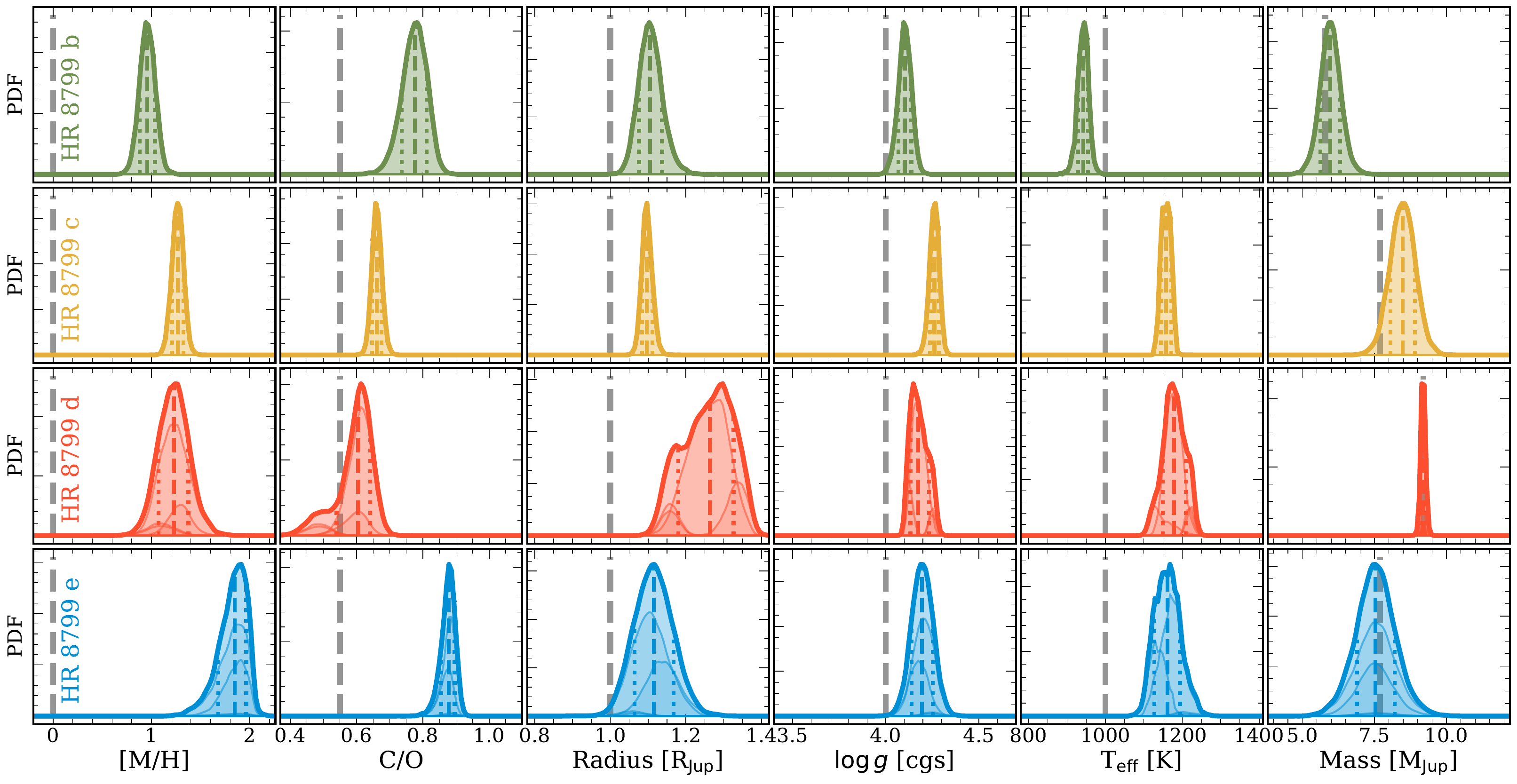}
    \caption{Bayesian-averaged posterior parameter distributions for each of the HR 8799 planets based on the group $A\cap B$ set of retrievals. In faint lines beneath the total posterior distribution are the individual contributions from different retrievals. The coloured dashed and dotted lines indicate the median and $\pm34.1$\% confidence regions respectively. The vertical grey lines indicate typical parameter values (e.g. solar metallicity and C/O) and serve as a visual reference for each parameter. For the planet mass, they indicate the dynamical mass estimate from \cite{zurlo_orbital_2022}.}
    \label{fig:retrieval_parameter_summary}
\end{figure*}

Sonora Bobcat and Cholla did not fit the HR~8799 spectra at all, validating the necessity of cloudy (or similar) atmospheric models.
Thus we continued only with the cloudy Sonora Diamondback models in order to interpret the HR~8799 atmospheres.
While similar to Exo-REM in implementing clouds, Diamondback currently fixes the C/O ratio to a solar value of 0.458 \citep{lodders_abundances_2009}, preventing the measurement of this parameter, and potentially leading to biases in the remaining parameters.

\subsubsection{petitCODE}
{\tt petitCODE} is a radiative-convective and chemical equilibrium code used to compute the structures and spectra of exoplanet atmospheres \citep{molliere_model_2015,molliere_observing_2017}. We used the {\tt cool-cloudy} and {\tt hot-cloudy} grids computed for \cite{stolker_miracles_2020_species}, spanning temperatures from 500-850 ({\tt cool-cloudy}) and 1000-2000~K ({\tt hot-cloudy}). The code setups are based on the work presented in \citet{samland_spectral_2017,lindermordasini2019}. Both grids implement the cloud model described in \citet{ackerman_precipitating_2001}. While the cool grid only assumes Na$_2$S and KCl clouds, and the hot grid adds Mg$_2$SiO$_4$ and Fe clouds.

\subsubsection{Grid fits}
We performed Bayesian fits using {\tt species} to interpolate the grids \citep{stolker_miracles_2020_species}, and {\tt MultiNest} to sample the parameter space.
400 live points were used for these fits, with uniform priors on all parameters covering the grid ranges as described in Table \ref{tab:grid_boundaries}, and an additional Gaussian prior on the planet mass.
We fit for covariance width and strength for all IFS datasets following the method of \cite{wang_accreting_2020}, as the empirical covariance matrices cannot be incorporated in {\tt species}, other than for GRAVITY data.
Fitting for the covariance parameters was universally favoured by the Bayes factor, and thus we only present the full fits.
Consistent with the expectations of \cite{greco_measurement_2016} and \cite{nasedkin_impacts_2023}, including these parameters also tended to broaden the posterior distributions, though posterior widths remain far narrower than the variation between the models.

In addition to the Bayesian fits, we performed a simple $\chi^{2}$ minimisation over each grid to avoid potential issues with interpolating the spectra along the different parameter axes.
Using this framework, we identify the single best-fit spectra, as presented in Table \ref{tab:gridchi2}. 

\section{Results}\label{sec:retrievals_all}
\begin{figure*}[ht]
    \centering
    \includegraphics[width=0.95\linewidth]{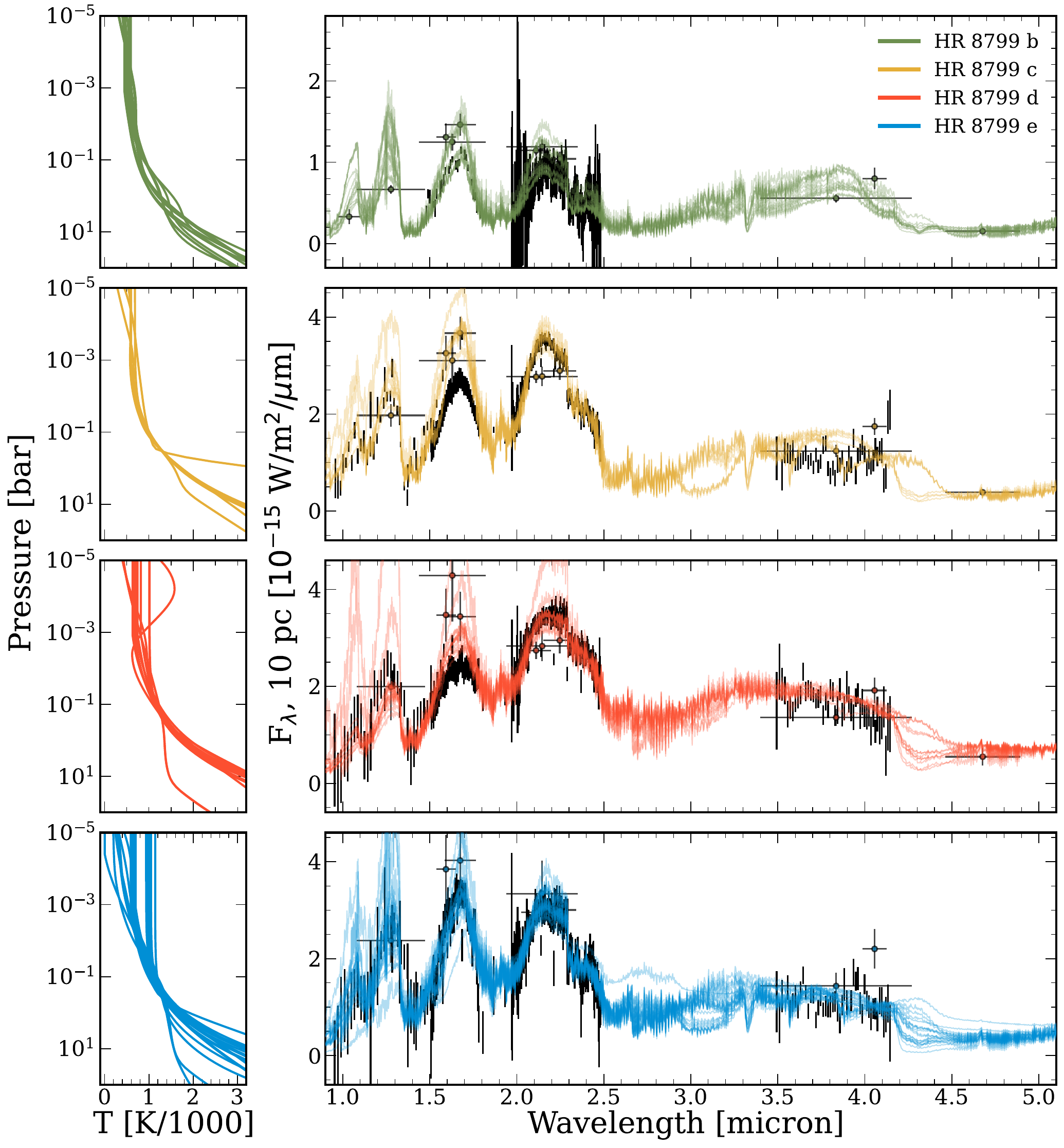}07
    \caption{Best-fit temperature profiles and spectra for the group $A\cap B$ retrievals. From top to bottom are HR 8799 b, c, d, and e.}
    \label{fig:best_fit_retrieval_spectra}
\end{figure*}

\begin{figure*}[t]
    \centering
    \includegraphics[width=0.21\linewidth]{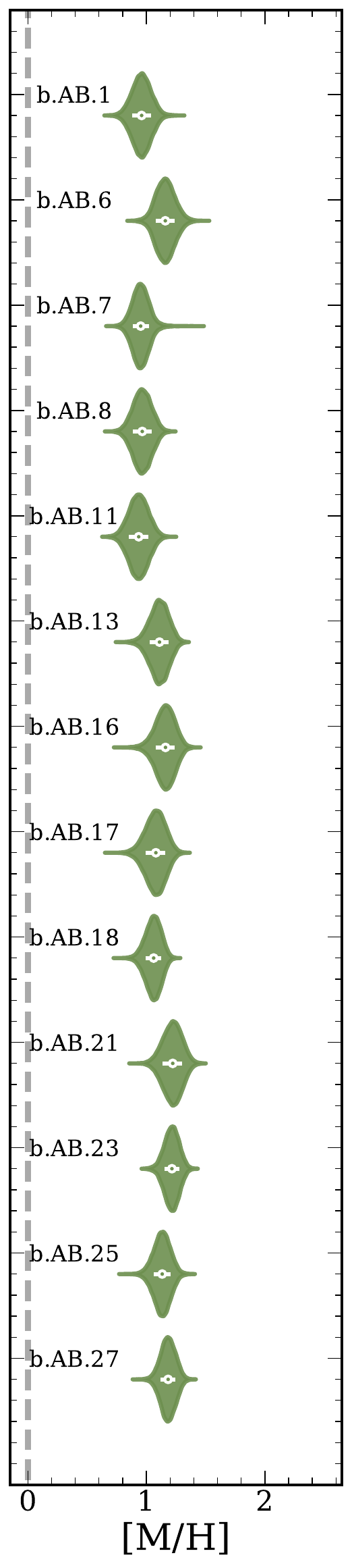}
    \includegraphics[width=0.21\linewidth]{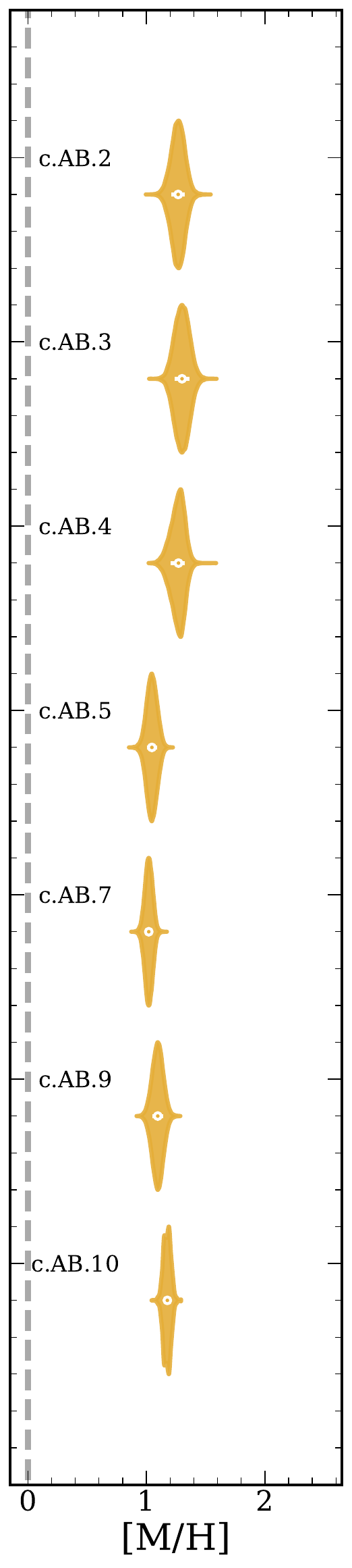}
    \includegraphics[width=0.21\linewidth]{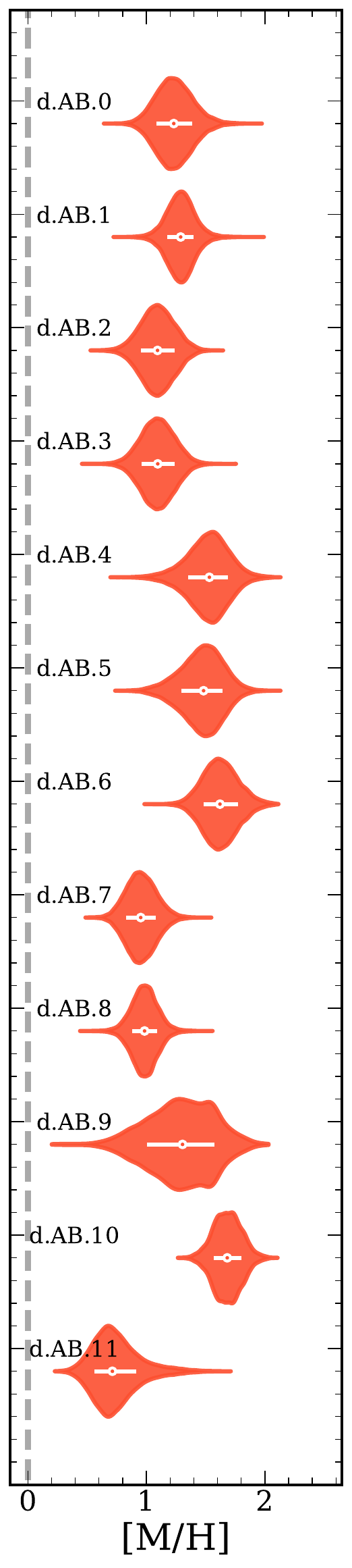}
    \includegraphics[width=0.21\linewidth]{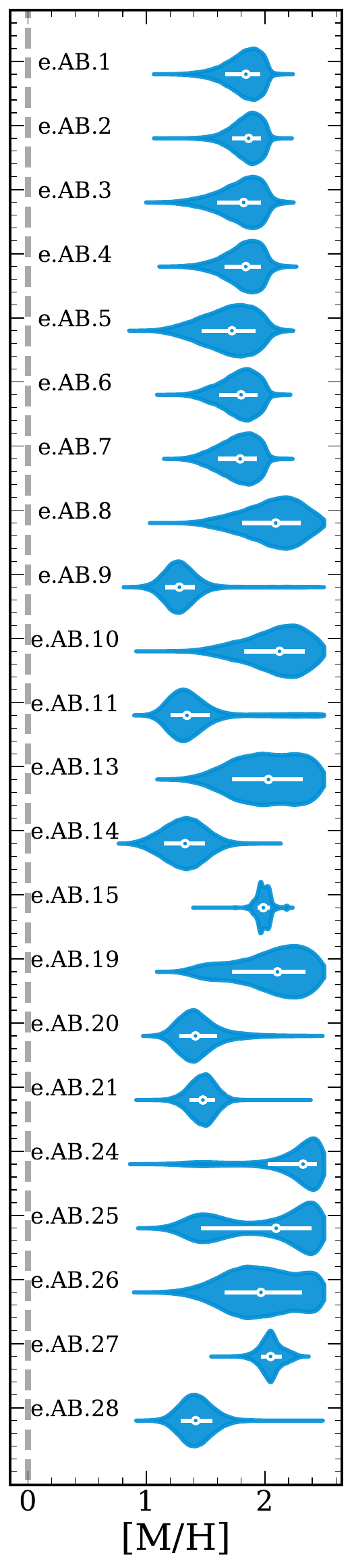}
    \caption{[M/H] posterior distributions for all retrievals in group $A\cap B$. From left to right are the distributions for b, c, d, and e. The vertical line indicates solar metallicity. Model keys are as in tables \ref{tab:full_results_b}--\ref{tab:full_results_e}, and are sorted by the Bayes factor from top to bottom.}
    \label{fig:retrieval_violins_feh}
\end{figure*}

\begin{figure*}[t]
    \centering
    \includegraphics[width=0.21\linewidth]{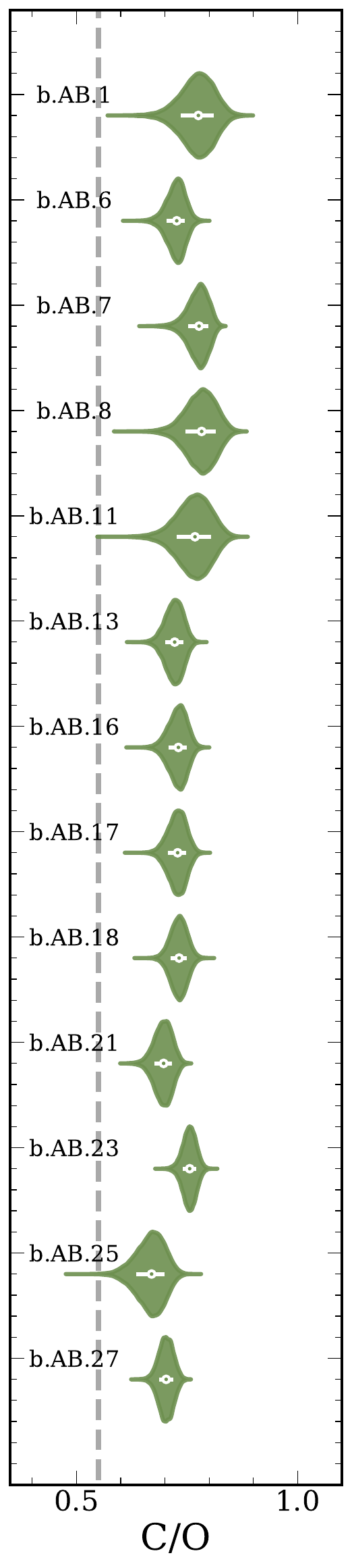}
    \includegraphics[width=0.21\linewidth]{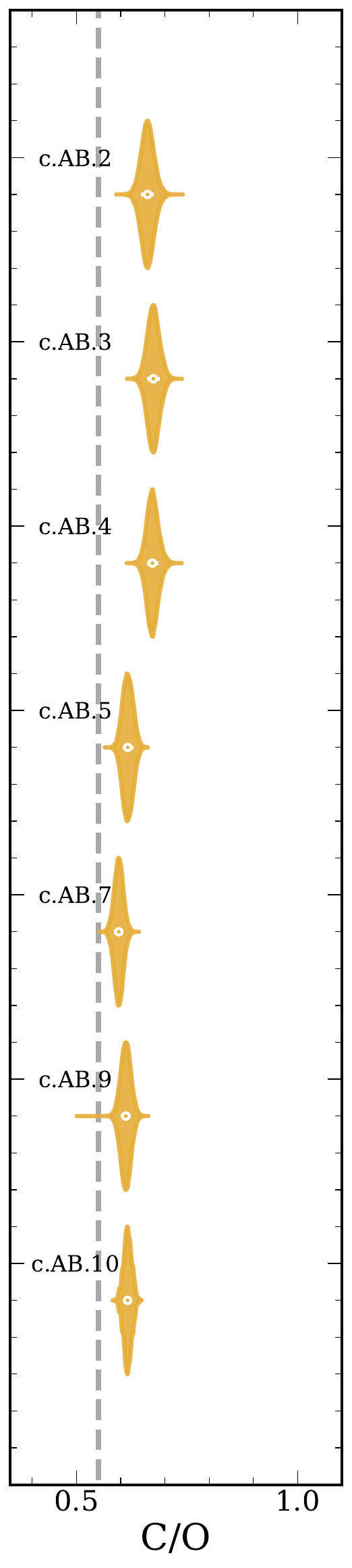}
    \includegraphics[width=0.21\linewidth]{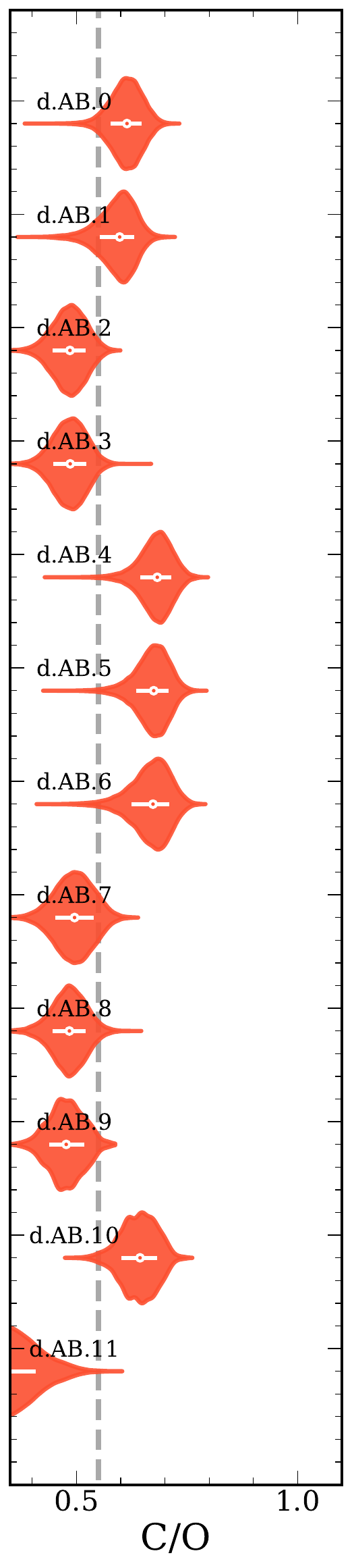}
    \includegraphics[width=0.21\linewidth]{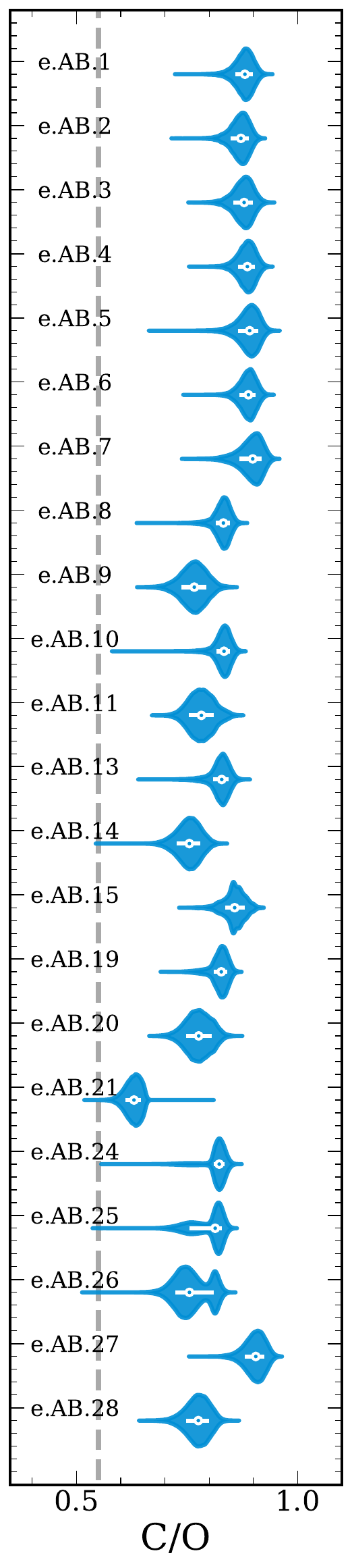}
    \caption{C/O posterior distributions for all retrievals in group $A\cap B$. From left to right are the distributions for b, c, d, and e.  The vertical line indicates  stellar C/O. Model keys are as in tables \ref{tab:full_results_b}--\ref{tab:full_results_e}, and are sorted by the Bayes factor.}
    \label{fig:retrieval_violins_c-o}
\end{figure*}

\begin{figure*}[t]
    \centering
    \includegraphics[width=0.21\linewidth]{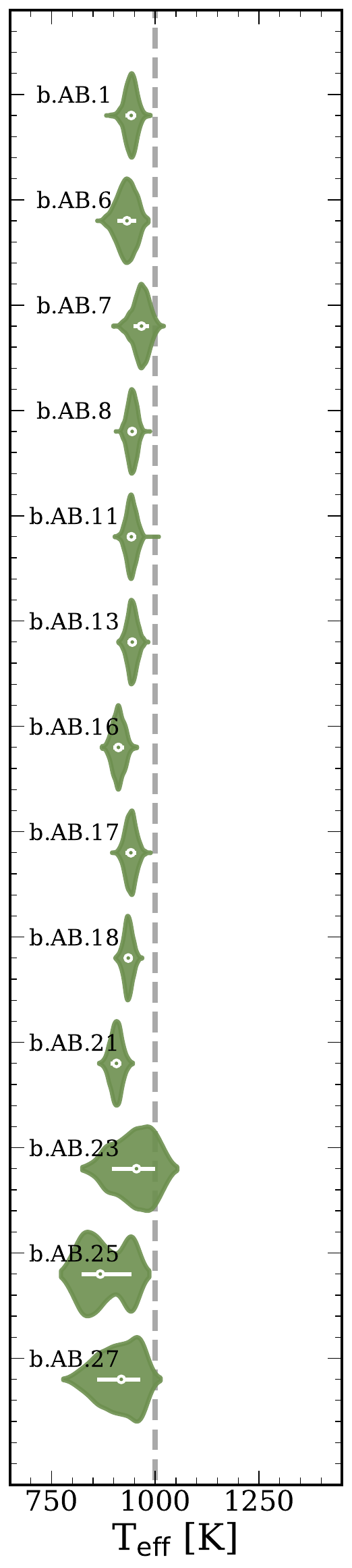}
    \includegraphics[width=0.21\linewidth]{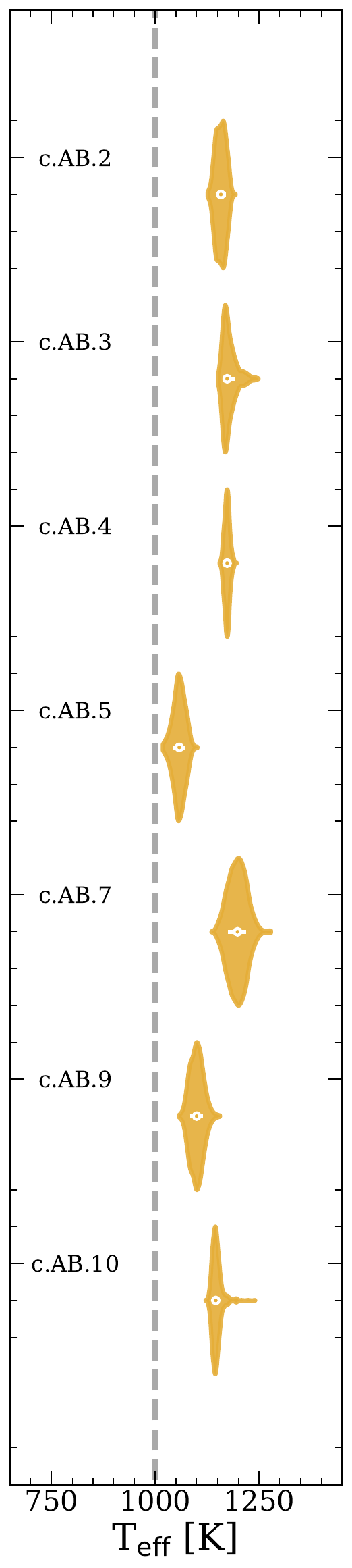}
    \includegraphics[width=0.21\linewidth]{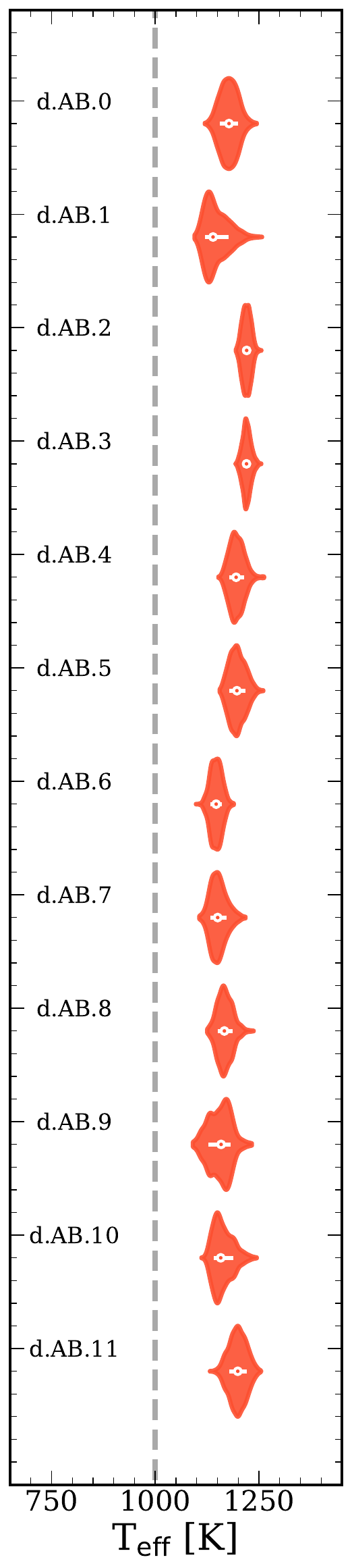}
    \includegraphics[width=0.21\linewidth]{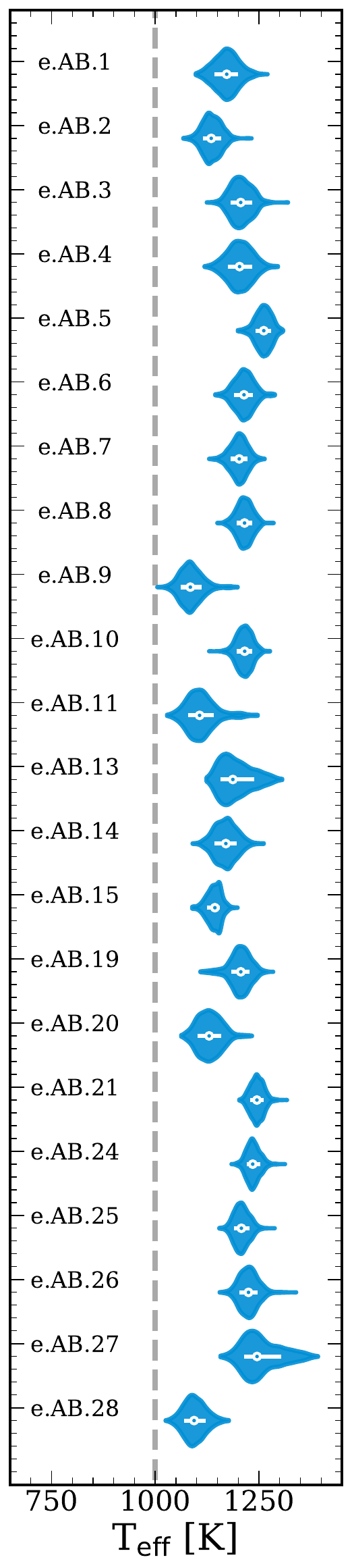}
    \caption{T$_{\rm eff}$ posterior distributions for all retrievals in groups $A\cap B$. From left to right are the distributions for b, c, d, and e.  The vertical line indicates 1000~K. Model keys are as in tables \ref{tab:full_results_b}--\ref{tab:full_results_e}, and are sorted by the Bayes factor.}
    \label{fig:retrieval_violins_teff}
\end{figure*}

\begin{figure*}[t]
    \includegraphics[width=0.95\linewidth]{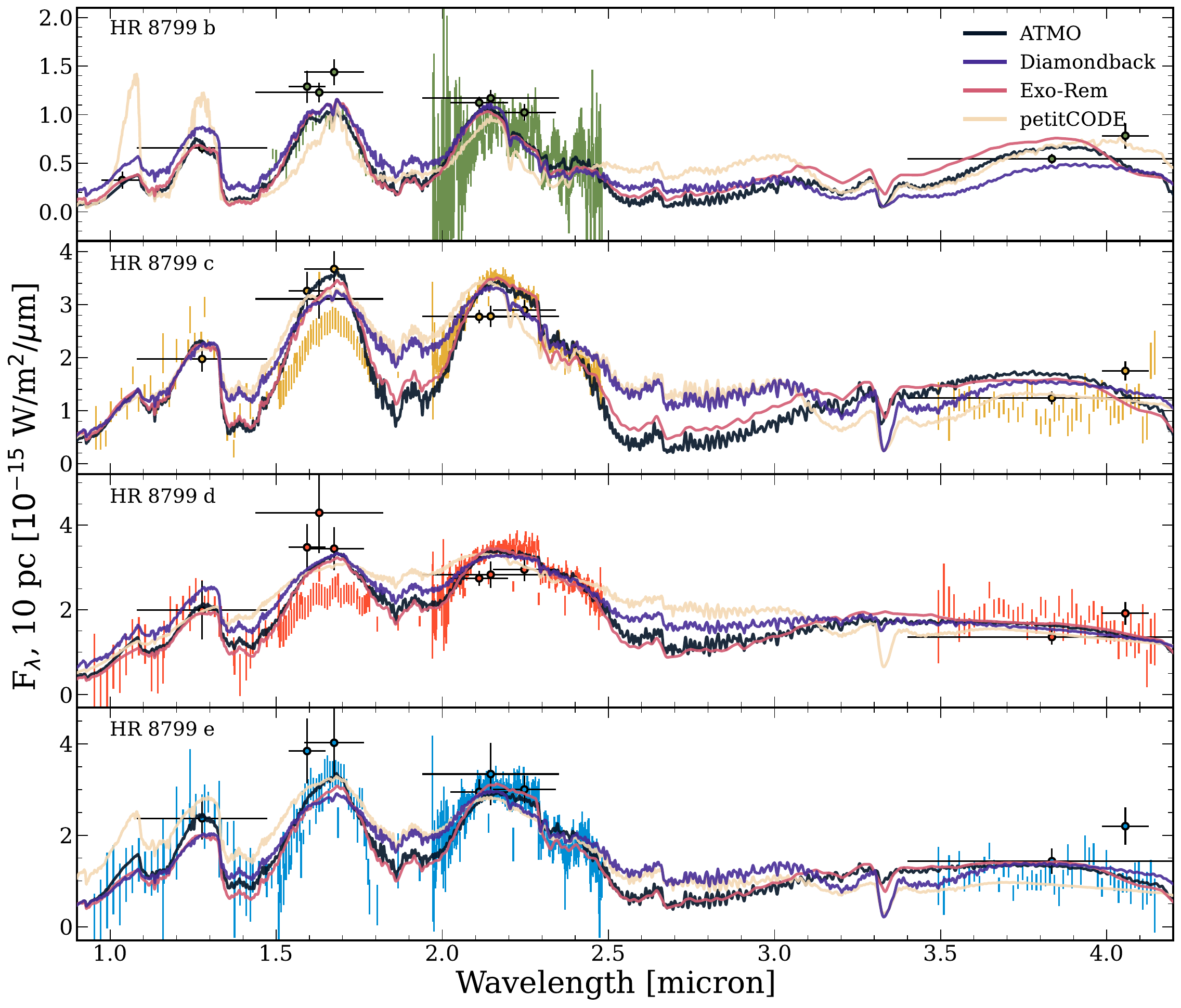}
    \caption{Grid fits from Exo-REM, Sonora Diamondback, ATMO, and petitCODE. ALES data for HR~8799 d are scaled by the overall best-fit scaling parameter (Exo-REM, 1.18). Fits are the single best-fit $\chi^{2}$ model from the grid.}
    \label{fig:gridfits}
\end{figure*}

\begin{table*}[t]
\centering
\begin{threeparttable}
    \centering
    \begin{small}
    \caption{Grid-retrieval results}
    \label{tab:grid-retrievals}
    \begin{tabular}{l|llllllllll}
    \toprule
    \textbf{Planet} & \textbf{Model} & $\bm{\Delta}\log_{\mathbf{10}} \bm{\mathcal{Z}}$ &$\bm{T_{\rm eff}}$  & $\bm{\log g}$ & \textbf{[M/H]} & $\bm{f_{\rm sed}}$ & \textbf{C/O} & \textbf{Radius} & $\bm{\log L/L_{\odot}}$ & \textbf{Mass}\\
     & & & [K] & [cgs] & & & & [R$_{\rm Jup}$] & & [M$_{\rm Jup}$] \\
    \midrule
    b & ATMO & 
        -63 &
        ${1020}_{-7}^{+9}$ & 
        ${4.42}_{-0.04}^{+0.04}$ & 
        ${>0.6}$ & 
        \ldots&
        ${<0.3}$ & 
        ${0.73}_{-0.02}^{+0.02}$ & 
        ${-5.278}_{-0.009}^{+0.008}$ & 
        ${5.6}_{-0.4}^{+0.4}$ 
    \\[3pt] 
    & Diamondback & 
        -98 &
        ${959}_{-8}^{+10}$ & 
        ${4.18}_{-0.03}^{+0.03}$ & 
        ${>0.5}$ & 
        ${<1.0}$ & 
        0.458&
        ${0.95}_{-0.02}^{+0.02}$ & 
        ${-5.155}_{-0.007}^{+0.007}$ & 
        ${5.6}_{-0.4}^{+0.4}$ 
    \\[3pt] 
    & Exo-Rem & 
        0 &
        ${862}_{-6}^{+21}$ & 
        ${4.0}_{-0.03}^{+0.02}$ & 
        ${0.69}_{-0.03}^{+0.11}$ & 
        \ldots &
        ${0.50}_{-0.02}^{+0.10}$ & 
        ${1.19}_{-0.03}^{+0.03}$ & 
        ${-5.14}_{-0.02}^{+0.02}$ & 
        ${5.5}_{-0.4}^{+0.4}$ 
    \\[3pt] 
    & petitCODE$^{a}$ & 
        -105 &
        ${>850}$ & 
        ${3.48}_{-0.02}^{+0.01}$ & 
        ${>1.4}$ & 
        ${2.53}_{-0.02}^{+0.01}$ & 
        0.55&
        ${2.00}_{-0.04}^{+0.02}$ & 
        ${-4.73}_{-0.02}^{+0.01}$ & 
        ${4.8}_{-0.2}^{+0.8}$ 
    \\[3pt] 

    \midrule
    c& ATMO & 0 &
    ${1195}_{-6}^{+5}$ & 
    ${4.34}_{-0.03}^{+0.03}$ & 
    $>0.6$ & 
    \ldots&
    ${0.425}_{-0.014}^{+0.012}$ &
    ${1.01}_{-0.01}^{+0.01}$ & 
    ${-4.723}_{-0.003}^{+0.003}$ & 
    ${9.1}_{-0.5}^{+0.5}$ 
    \\[3pt]
    & Diamondback & -245 &
    ${1237}_{-2}^{+2}$ & 
    ${<3.5}$ & 
    ${>0.5}$ & 
    ${2.32}_{-0.02}^{+0.02}$ & 
    0.458 &
    ${0.99}_{-0.01}^{+0.01}$ & 
    ${-4.683}_{-0.002}^{+0.002}$ & 
    ${1.25}_{-0.01}^{+0.01}$ 
    \\[3pt]
    & Exo-Rem & -30 &
    ${1065}_{-1}^{+1}$ & 
    ${4.15}_{-0.02}^{+0.02}$ & 
    $>1.0$ & 
    \ldots&
    ${0.501}_{-0.002}^{+0.003}$ & 
    ${1.314}_{-0.005}^{+0.005}$ & 
    ${-4.694}_{-0.002}^{+0.002}$ & 
    ${9.8}_{-0.5}^{+0.5}$ 
    \\[3pt]
    & petitCODE & -622 &
    ${1400}_{-0.5}^{+0.2}$ & 
    ${4.00}_{-0.01}^{+0.01}$ & 
    ${>0.3}$ & 
    ${>4.5}$ & 
    ${0.55}_{-0.00}^{+0.00}$ &
    ${0.80}_{-0.00}^{+0.00}$ & 
    ${-4.652}_{-0.002}^{+0.002}$ & 
    ${2.57}_{-0.02}^{+0.02}$ 
    \\[3pt]

    \midrule

d & ATMO & -38 &${1300}_{-2}^{+2}$ & 
            ${4.39}_{-0.01}^{+0.01}$ & 
            ${0.07}_{-0.02}^{+0.02}$ & 
            \ldots &
            ${<0.3}$ & 
            ${0.926}_{-0.004}^{+0.004}$ & 
            ${-4.652}_{-0.003}^{+0.003}$ & 
            ${8.4}_{-0.2}^{+0.2}$ 
            \\[3pt]
    & Diamondback & -76 &${1234}_{-7}^{+8}$ & 
            ${4.27}_{-0.02}^{+0.02}$ & 
            ${>0.5}$ & 
            ${1.95}_{-0.06}^{+0.04}$ & 
            0.458 &
            ${1.05}_{-0.02}^{+0.01}$ & 
            ${-4.632}_{-0.002}^{+0.002}$ & 
            ${8.4}_{-0.2}^{+0.2}$ 
            \\[3pt]
    & Exo-Rem & 0 &${1155}_{-4}^{+5}$ & 
            ${4.23}_{-0.01}^{+0.01}$ & 
            ${0.78}_{-0.03}^{+0.03}$ & 
            \ldots &
            ${0.200}_{-0.002}^{+0.002}$ & 
            ${1.168}_{-0.005}^{+0.006}$ & 
            ${-4.656}_{-0.005}^{+0.007}$ & 
            ${9.3}_{-0.2}^{+0.6}$ 
            \\[3pt]
    & petitCODE & -185 &${1400}_{-1}^{+1}$ & 
            ${4.47}_{-0.01}^{+0.01}$ & 
            $>0.3$ & 
            ${2.99}_{-0.07}^{+0.06}$ & 
            ${0.55}_{-0.000}^{+0.000}$ & 
            ${0.859}_{-0.003}^{+0.003}$ & 
            ${-4.587}_{-0.002}^{+0.003}$ & 
            ${8.9}_{-0.2}^{+0.2}$ 
            \\[3pt]

    \midrule
    e& ATMO & 0 &
    ${1251}_{-20}^{+22}$ & 
    ${4.38}_{-0.07}^{+0.06}$ & 
    ${>0.6}$ & 
    \ldots &
    ${0.52}_{-0.03}^{+0.03}$ & 
    ${0.87}_{-0.03}^{+0.03}$ & 
    ${-4.778}_{-0.006}^{+0.006}$ & 
    ${7.2}_{-0.8}^{+0.8}$
    \\[3pt]
    & Diamondback & -16 &
    ${1244}_{-6}^{+6}$ & 
    ${<3.5}$ & 
    ${>0.5}$ & 
    ${2.31}_{-0.05}^{+0.05}$ & 
    0.458 &
    ${0.92}_{-0.01}^{+0.01}$ & 
    ${-4.738}_{-0.005}^{+0.005}$ & 
    ${1.12}_{-0.05}^{+0.07}$
    \\[3pt]
    & Exo-Rem & -14 &
    ${1064}_{-3}^{+3}$ & 
    ${4.13}_{-0.04}^{+0.04}$ & 
    ${>1.0}$ & 
    \ldots &
    ${0.50}_{-0.01}^{+0.01}$ & 
    ${1.24}_{-0.01}^{+0.01}$ & 
    ${-4.747}_{-0.005}^{+0.006}$ & 
    ${8.0}_{-0.8}^{+0.8}$
    \\[3pt]
    & petitCODE & -56 &
    ${1399}_{-4}^{+2}$ & 
    ${4.00}_{-0.03}^{+0.02}$ & 
    $>0.3$ & 
    ${4.48}_{-0.03}^{+0.02}$ &
    ${0.55}_{-0.00}^{+0.00}$ & 
    ${0.755}_{-0.004}^{+0.005}$ & 
    ${-4.703}_{-0.004}^{+0.004}$ & 
    ${2.3}_{-0.1}^{+0.1}$ 
    \\[3pt]
    \bottomrule
    \end{tabular}
    \end{small}
    \begin{tablenotes}
    \small
    \item\textbf{Notes}
    \item All values presented are the median values from the fits, with uncertainties given as the $\pm 34.1$\% percentiles. Values without uncertainties were fixed during the fit.
    \item $a$ the {\tt petitcode-cool-cloudy} grid was used to fit HR 8799 b.
    
    \end{tablenotes}
\end{threeparttable}
\end{table*}

\begin{table*}[t]
\centering
\begin{threeparttable}
    \centering
    \begin{small}
    \caption{Grid-fit $\chi^{2}$ results}
    \label{tab:gridchi2}
    \begin{tabular}{l|lllllllll}
    \toprule
    \textbf{Planet} & \textbf{Model} & $\bm{\chi^{2}}$ &$\bm{T_{\rm eff}}$  & $\bm{\log g}$ & \textbf{[M/H]} & $\bm{f_{\rm sed}}$ & \textbf{C/O} & $\bm{\gamma_{\rm ad}}$ & \textbf{Radius}\\
     & & & [K] & [cgs] & & & & & [R$_{\rm Jup}$] \\
    \midrule
    b & ATMO & 744 & 1000 & 4.5 & 0.6 & \ldots & 0.3 & 1.05 &  1.00\\
    & Diamondback & 879 & 1100 & 3.5 & 0.5 & 2.0 & 0.458 & \ldots&1.07\\
    & Exo-REM & 510 & 850 & 3.5 & 0.5 & \ldots & 0.55 & \ldots&1.05\\
    & petitCODE$^{a}$ & 968 & 850 & 3.5 & 1.4 & 2.0 & \ldots &\ldots& 0.79 \\
    \midrule
    c & ATMO & 2162 & 1200 & 4.5 & 0.6 & \ldots & 0.3 & 1.01 & 0.97 \\
    & Diamondback & 3900 & 1200 & 3.5 & 0.5 & 2.0 & 0.458 & \ldots &1.08\\
    & Exo-REM  & 1868 & 1100 & 3.5 & 1.0 & \ldots & 0.8 & \ldots &1.22\\
    & petitCODE & 7374 & 1200 & 3.5 & 0.3 & 3.0 & 0.55 & \ldots &1.09\\ 
    \midrule
    d$^{b}$ & ATMO & 943 & 1300 & 3.0 & 0.6 & \ldots & 0.55 & 1.01 & 0.94\\
    & Diamondback & 1336 & 1200 & 3.5 & 0.5 & 1.0 & 0.458 & \ldots &1.12\\
    & Exo-REM & 936 & 1200 & 3.0 & 1.0 & \ldots & 0.55 & \ldots &1.12 \\
    & petitCODE & 2254 & 1200 & 4.0 & 0.0 & 1.5 & 0.55 & \ldots &1.21 \\
    \midrule
    e & ATMO & 669 & 1300 & 4.0 & 0.6 & \ldots & 0.55 & 1.03 & 0.81\\
    & Diamondback & 819 & 1200 & 3.5 & 0.5 & 2.0 & 0.458 & \ldots &1.02\\
    & Exo-REM & 623 & 1100 & 3.5 & 1.0 & \ldots & 0.8 & \ldots &1.15\\
    & petitCODE & 1136 & 1400 & 4.0 & 0.3 & 4.5 & 0.55 & \ldots &0.74\\
    \bottomrule
    \end{tabular}
    \end{small}
    \begin{tablenotes}
    \small
    \item\textbf{Notes}
    \item All values presented are the single best-fit value according to the $\chi^{2}$. All of the models share a similar number of parameters, which are listed in Table \ref{tab:grid_boundaries}. 
    Additional parameters for data scaling and covariance fitting are shared between all models.
    \item $^{a}$ for HR 8799 b, the {\tt petitcode-cool-cloudy} grid was used rather than the {\tt petitcode-hot-cloudy} grid.
    \item $^{b}$ the fits for HR 8799 d included a scaling parameter for the ALES data.
    
    \end{tablenotes}
\end{threeparttable}
\end{table*}

Based on both the atmospheric retrievals and the self consistent grid fits, the HR 8799 planets share enriched atmospheres, with stellar-to-superstellar C/O ratios.
The properties of each atmosphere for subsets of the ensemble of retrievals are summarised in tables \ref{tab:literature_values_b}--\ref{tab:literature_values_e} for planets b--e, respectively.
These tables also contain ranges of plausible parameter values for each planet based on the aggregate of the self-consistent models, while Fig.~\ref{fig:retrieval_parameter_summary} shows the distributions for a subset of the key parameters.
These estimates are synthesised from the results of the Bayesian fits and the $\chi^{2}$ minimisation, rejecting solutions with unphysical masses and radii.  
The full results of the grid fits are found in Table~\ref{tab:grid-retrievals} for the Bayesian fits and Table~\ref{tab:gridchi2} for the single-best $\chi^{2}$ fits.
In these tables, an index is assigned to each retrieval, with the format \textit{planet.group.index}, which serves as the retrieval identifier throughout the text.

Nearly 100 retrievals were performed for this analysis across the four companions in order to derive robust constraints on primary planetary properties.
The aggregate results of the Bayesian model average of group $A\cap B$ retrievals are presented in Fig.~\ref{fig:retrieval_parameter_summary}, with the best fit models for the same sample of retrievals compared to the data in Fig.~\ref{fig:best_fit_retrieval_spectra}.
Overall we find consistently good fits for all four planets, with best-fit $\chi^{2}/\nu<2$ for each planet.
The self-consistent grid-fits incorporate additional physical processes and require fewer free parameters than the retrievals, making them less flexible.
The additional processes, such as radiative-convective equilibrium, chemistry, and cloud physics act as narrow priors on the bulk atmospheric parameters.
While the lack of flexibility leads to worse fits when compared to retrievals, the fits of multiple self-consistent grids still result in mutually consistent results. 
Notably, {\tt petitCODE} gives consistent parameter estimates even though it is strongly disfavoured by the goodness-of-fit metrics.

Both the {\tt pRT} retrievals and the self-consistent grid fits show that all of the atmospheres are strongly enriched in metals, finding median [M/H]$\gtrsim$1.0 for each planet.
This is driven by the carbon and oxygen abundances as measured through \h2o and CO, and particularly through the CO absorption feature at 2.3~$\upmu$m. 
The C/O ratios decrease with decreasing separation from $0.78\pm0.04$ in HR 8799 b to $0.60\pm0.05$ for HR 8799 d, with HR 8799 e breaking the trend with carbon rich C/O ratio of $0.87\pm0.02$. 
In addition to water and CO, HCN is confidently detected in HR 8799 c and e, while \ch4 is detected in HR 8799 c.

We find that cloudy atmospheres are universally favoured over clear atmospheres.
Extended silicate clouds, together with an iron cloud deck are the preferred models for the three inner planets, with the cooler HR 8799 b showing evidence for Na$_{2}$S clouds.

The retrievals found masses, radii, and surface gravities consistent with evolutionary models. 
By construction, these form Group A of our retrievals.
In general, retrieving the planet radius is challenging: the radius of directly imaged exoplanets is commonly underestimated by both retrievals and grid fits \citep[e.g.][]{bonnefoy_first_2016}.
These unphysical solutions are found in several of the grid fits, while nearly every retrieval finds plausible values of the mass and radius.
The small radius is often compensated for by adjusting the temperature, metallicity, or cloud properties.
Nevertheless, in all cases the most favoured retrieval produced masses consistent with dynamical mass measurements and radii consistent with evolutionary models.
In the group $A\cap B$ retrievals that used dynamical mass priors, the retrieved mass estimate is within $1\sigma$ of the dynamical mass for b, d, and e, with the posterior width consistent with the prior width which. 
For HR 8799 c the retrieved mass is moderately higher than the dynamical mass estimate at $8.5\pm0.5$ \mj compared to the dynamical mass of $7.7\pm0.7$ \mj.
However, if the dynamical mass is not used as a constraint and $\log g$ is freely retrieved, the resulting mass estimate is found to be much larger than the dynamical mass estimate, highlighting the importance of including additional constraints in a retrieval framework. 
The estimate of the effective temperature of each planet is consistent with previous findings, with e, c, and d sharing temperatures around 1100 K, and b being cooler at around 950 K.
While we do not perform a comparison to spectral templates due to the inferred high metallicity, low surface gravity, and potential complications from the viewing angle of the planets, we find that the spectral types found in \cite{bonnefoy_first_2016} remain a good description of the four planets.
HR 8799 d and e are similar to late-L type brown dwarfs, consistent with their inferred effective temperature.
HR~8799~c is more likely fit by an early-T spectral type as it is a few kelvin cooler than e and d and is beginning to show spectral features from \ch4.
At 950~K HR 8799 b is solidly in the T dwarf regime.

In this section, we present the measured properties for each of the four planets individually.
Following that, in Section \ref{sec:modelsarehard}  we present a detailed discussion of the results and challenges of comparing different thermal structures, chemistry and cloud parameterisations, and data inclusion.

\subsection{HR~8799~b}\label{sec:b_retrievals}
For HR~8799~b we included the GRAVITY and OSIRIS spectra, together with the full set of photometry, allowing the OSIRIS data to float as the published data are not flux calibrated. 
For the grid fits we included additional parameters to describe the covariance of the OSIRIS data.

HR 8799 b is the coldest and lowest mass planet in the HR 8799 system. 
The best estimate of these parameters via Bayesian averaging of group $A\cap B$ retrievals finds an effective temperature of $942^{+12}_{-16}$~K and a mass of $6.0^{+0.4}_{-0.3}$~M$_{\rm J}$, driven by the use of the dynamical mass estimate of \cite{zurlo_orbital_2022} as a prior.
The radius is slightly inflated compared to jupiter, with ($R_{pl} = 1.10^{+0.03}_{-0.03}$); combining the mass and radius estimates leads to $\log g = 4.10^{+0.03}_{-0.04}$.
This is consistent with the Bayesian grid fits, though the single-best $\chi^{2}$ fits found both lower ($\log g=3.5$) and higher ($\log g$=4.5) solutions.
The Bayesian averaged results (for group $A\cap B$) are driven by a single retrieval, with a $\Delta\log_{10}\mathcal{Z}$ of 4 relative to the next best retrieval. 
This single best retrieval uses the \citetalias{zhang_elemental_2023} temperature profile and free chemistry, finding a metallicity of $0.96^{+0.08}_{-0.08}$ and a C/O of $0.78^{+0.03}_{-0.04}$. 

The grid fits find temperatures between 850~K$-$1100~K, with the single best fit, found using Exo-REM, finding $T_{\rm eff}=850$ K.
Using the Bayesian fit, Exo-REM is again the most favoured model by the Bayes factor, and finds $T_{\rm eff}=862^{+21}_{-6}$ K and a radius of ${1.19}_{-0.03}^{+0.03} R_{\rm Jup}$, consistent with expectations from evolutionary models \citep[e.g.][]{marley_masses_2012}. 
The ATMO and Sonora Diamondback models finds a somewhat higher temperature and a smaller radius, but are disfavoured by the Bayes factor.
petitCODE is divided into {\tt cool-cloudy} and {\tt hot-cloudy} grids, with the cool grid extending up to 850~K, and the hot grid beginning at 1000~K. 
As the effective temperature of HR~8799~b likely falls between these grids, it poorly fit the data, although the remaining parameters of $\log g$ and \fsed are compatible with the other self-consistent models.

Figure \ref{fig:retrieval_violins_feh} shows the variation of the metallicity across the different models considered.
We see that strong enrichment solution is almost always found, particularly by models preferred by the Bayes factor. 
The degree of enrichment does not systematically vary between disequilibrium and free chemistry retrievals.
All of the self-consistent models favour high metallicity solutions, reaching the upper bounds of the grid in all cases.

\begin{figure*}[t]
    \includegraphics[width=\linewidth]{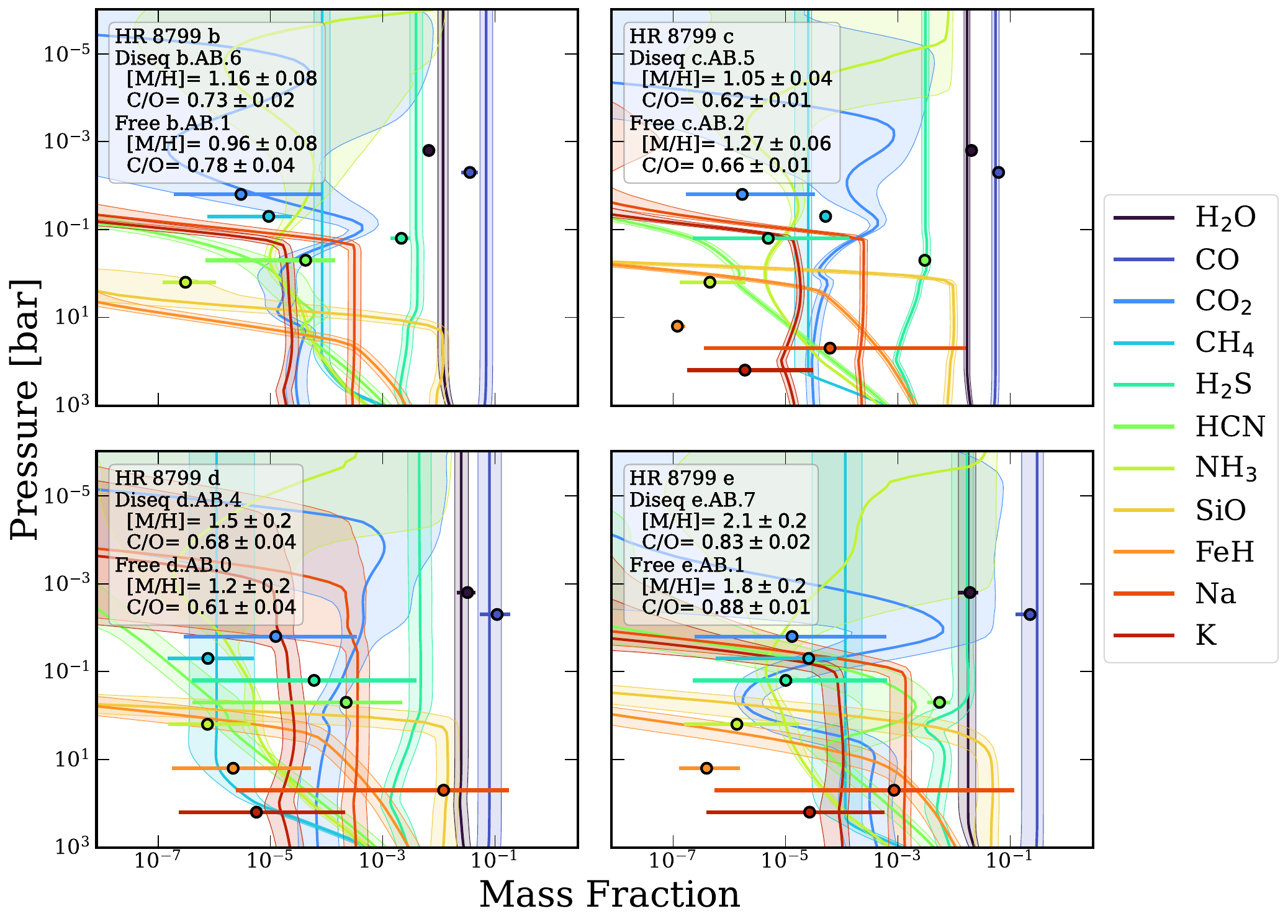}
    \caption{Mass fraction abundance profiles as a function of pressure in the atmospheres of HR~8799 b, c, d, and e. The solid lines indicate the median values of the most favoured disequilibrium retrieval for each planet, with the shaded region indicating the 90\% confidence interval. The circular markers show the median values for each species from the most favoured free chemistry retrieval, with the error bars indicating the 90\% confidence interval. The position along the pressure axis is arbitrary. The minimum mass fraction allowed in the free retrievals was $10^{-7}$.}
    \label{fig:chemical_profiles}
\end{figure*}

Similarly to the metallicity, the C/O ratio is shown for the different retrievals in Fig. \ref{fig:retrieval_violins_c-o} and $T_{\rm eff}$ in \ref{fig:retrieval_violins_teff}. The C/O ratio is generally constrained to between 0.6 and 0.8. 
However, for free chemistry retrievals, the inferred C/O ratio only indicates the gas-phase composition. 
Additional oxygen is sequestered in the silicate clouds: accounting for this sequestration would result in a lower C/O ratio.
Among the grid fits, shown in Fig. \ref{fig:gridfits}, the C/O ratio shows more variation, ranging from the lower bound of the ATMO grid at 0.3 to 0.55 from Exo-REM. 
The best fit models from Exo-REM are consistent with the stellar value of 0.54, and additional data covering the near infrared water features is likely necessary to improve these constraints. 
From Fig. \ref{fig:chemical_profiles} we find that both the free retrieval and disequilibrium retrievals display similar trends in the retrieved chemical abundances, finding that nearly 10\% of the atmosphere is CO by mass, while water has a lower abundance of around 1\% by mass. 
The best-fit free retrieval finds systematically lower abundances for both of these species compared to the best-fit disequilibrium retrieval.
H$_{2}$S and CH$_{4}$ are found to be the next most abundant species in both the disequilibrium and free chemistry retrievals. 
However, even though their abundances are constrained by the posterior distribution, there is no evidence for their detection when comparing between the full free chemistry retrieval (b.AB.1) and retrievals without these species (b.A.0 and b.A.2). 
The b.AB.1 retrieval uses the Z23 temperature profile, free chemistry, and clouds condensing at their equilibrium position.
This same setup is used for b.A.0 and b.A.2, apart from the exclusion of CH$_{4}$ 
 and H$_{2}$S respectively.
While the inferred CH$_{4}$ abundance of $\log X_{{\rm CH}_4}=-5.0\pm0.4$ is compatible with \cite{barman_simultaneous_2015}, there is no statistical evidence to support the detection.

In order to compare the cloud composition for HR 8799 b, we considered a set of retrievals using the same temperature structure (\citetalias{molliere_retrieving_2020}) and cloud parameterisation, and vary the cloud optical properties and condensation curve.
We found that Na$_{2}$S clouds are preferred over silicate, iron, and KCl clouds ($\Delta\log_{10}\mathcal{Z}\geq2)$.
Patchy clouds were also explored, but no evidence was found for patchiness, regardless of cloud composition.
At the temperature of HR 8799 b silicate clouds are expected to occur below the photosphere, with Na$_{2}$S or KCl clouds becoming the primary aerosol opacity source.
This result is likely driven by the condensation temperature of Na$_{2}$S rather than the optical properties; unlike silicate clouds which have strong absorption features in the mid-infrared, crystalline Na$_{2}$S is featureless out to 15~$\upmu$m, apart from a characteristic scattering slope.


\subsection{HR~8799~c}\label{sec:c_retrievals}
In our standard retrieval setup, we included data from SPHERE, GPI, CHARIS, GRAVITY, and ALES for HR~8799~c. 
We omitted the OSIRIS data, as it overlaps nearly completely with the GRAVITY data, is not flux calibrated, and requires either binning to lower resolution to be fit with the {\tt c-k} opacity tables, or the use of the higher-resolution line-by-line opacities to fit the full resolution data, dramatically increasing computation time.
Nevertheless, we performed several retrievals incorporating the OSIRIS data rather than the GRAVITY data to determine how this choice impacts the retrieved chemistry and clouds, and to determine if we can reproduce the findings of \cite{konopacky_detection_2013}.
Overall, HR~8799~c proved challenging to fit: many retrieval setups either required far more model computations before convergence than the other three planets, or failed to converge entirely. 
The grid-fits for HR~8799~c also displayed the greatest variation between models.

Like HR~8799~b, the group $A\cap B$ retrievals of HR~8799~c are dominated by a single retrieval, with $\Delta\log_{10}\mathcal{Z}=2$ compared to the next best retrieval.
This retrieval used the \citetalias{zhang_elemental_2023} temperature profile and free chemistry, and requires high-altitude silicate clouds ($\log P_{\rm base} = -3.4\pm1.8$~bar). 
The inferred effective temperature of $1159^{+11}_{-12}$~K is consistent with the range of temperatures found by the self-consistent grids, which spans from 1100~K to 1200~K, with the Bayesian fits averaging around $\sim 1200$~K.
The retrieved mass ($8.5^{+0.4}_{-0.5}~M_{\rm Jup}$) is slightly higher than the dynamical mass estimate, though radius ($1.10\pm0.01~R_{\rm Jup}$) is compatible with evolutionary models, and from these we derive a $\log g$ of $4.26^{+0.02}_{-0.03}$.
The grid-fits also found plausible radii, favouring values slightly larger than 1 \rj.
All of the self-consistent models found temperatures of 1100$-$1200~K, and $\log g$ between 3.5 and 4.5. 
From the single-best $\chi^{2}$ fits, the ATMO model found a $\log g$ of 4.5, while the remaining models find a lower solution of 3.5. 

As with the other HR~8799 planets the retrievals favour highly enriched solutions, finding [M/H] = $1.27^{+0.05}_{-0.05}$. 
The disequilibrium chemistry retrievals find slightly lower metallicities than the free chemistry retrievals, with the most-favoured disequilibrium retrieval finding a metallicity of $1.05\pm0.04$.
The data for HR~8799~c are highly discrepant in the H band (Figure \ref{fig:allspectra}), with the CHARIS data and photometry being around 50\% brighter than the SPHERE and GPI data.
As the metallicity is highly sensitive to the amplitude of the J, H, and K band peaks, such discrepancies need to be resolved in the data to ensure reliable measurement of this parameter.
All of the grid-fit solutions require high metallicity and are limited by the grid boundaries.
The C/O ratios between the Bayesian fits and the $\chi^{2}$ minimisation are consistent, typically favouring stellar to slightly substellar C/O.
The retrievals present a more consistent picture, with most retrievals favouring a mildly super-stellar C/O ratio, with the average group $A\cap B$ C/O of $0.60^{+0.02}_{-0.01}$.
The most favoured free chemistry retrieval for HR~8799~c find water and CO abundances consistent with most favoured disequilibrium retrieval.
The free retrieval also finds an extremely high HCN mass fraction of $\log X_{\rm HCN}=-2.54\pm0.05$, orders of magnitude higher than the predictions from equilibrium chemistry.
This finding is strongly favoured by the Bayes factor, with $\log_{10}\mathcal{Z}=30$ in favour of including HCN (c.AB.3 over c.A.8), with both retrievals using the Z23 temperature profile and clouds condensing at their equilibrium locations.
The detection of HCN was largely driven by the ALES data; the wavelength dependence of the detection is discussed further in Section \ref{sec:chemistry}.
If HCN is excluded, the overall metallicity is also increased, mostly due to a 3 dex increase in the H$_{2}$S abundance to $\log X_{\rm H_2S}=-2.38\pm0.06$.
While this solution is disfavoured, the higher H$_{2}$S abundance is more compatible with the equilibrium chemistry predictions.
High resolution spectroscopy is likely required to precisely characterise the sulphur and nitrogen elemental abundances, and determine reliable abundances for these trace species.
Although the \ch4 abundance is relatively low, with $\log X_{\rm CH_{4}}=-4.3\pm0.06$, it is precisely constrained and detected with high confidence, $\Delta\log_{10}\mathcal{Z}=11.5$ (c.AB.3 over c.A.6). 

HR 8799 c is host to a highly cloudy atmosphere. 
The most favoured retrieval finds high altitude ($\log P_{\rm MgSiO_{3}}=-3.4\pm1.8$~bar) MgSiO$_{3}$ cloud with a mass fraction of $\log X_{\rm MgSiO_{3}}=-4.7\pm1.2$, together with a deep iron deck.
The vertical mixing strength for the clouds is $\log \kzz=8.0\pm0.9$, while the \fsed for both the silicate and iron clouds are compatible, with values between 5-6.
Cloud composition could not be robustly determined for HR 8799 c, due to difficulties in retrieval convergence. 
However, we find that crystalline MgSiO$_{3}$ clouds (c.AB.5) provide a better fit by the $\chi^{2}$, and are favoured by the Bayes factor over patchy amorphous MgSiO$_{3}$ (c.AB.8), both using the M20 profile and disequilibrium chemistry.
The use of patchy cloud layers may improve this fit, allowing individual layers to impact the spectrum independently, but this would come at the cost of substantially increasing the number of parameters to fit the continuum shape.
The crystalline morphology provide a marginally better fit to the MIRI photometric data, but spectroscopic characterisation of the silicate feature is likely necessary to robustly distinguish these cases.

We ran a disequilibrium and free chemistry retrieval using the OSIRIS data in place of the GRAVITY data to check for consistency and to determine if we could reproduce the findings of \cite{konopacky_detection_2013}.
In order to use the correlated-k method of {\tt pRT}, we binned the OSIRIS data by a factor of 4 to a spectral resolving power of $R\sim1000$.
We find that a high metallicity solution is still found using these data, with effective temperatures, and surface gravities consistent with the GRAVITY retrievals. 
For the disequilibrium retrieval, the C/O ratio is found to be significantly higher than any of the GRAVITY retrievals, as well as the free chemistry OSIRIS retrieval.
While the free retrievals using the GRAVITY data find a slightly higher metallicity overall, the OSIRIS data finds a slightly higher \ch4 abundance of $-3.81\pm0.09$, as well as a much higher H$_{2}$S abundance.
Overall, we find that the main findings of metal-rich atmospheres are reproducible regardless of whether we use the GRAVITY or OSIRIS data, and confirm the detection of water and CO in the atmosphere of HR 8799 c.

\subsection{HR~8799~d}\label{sec:d_retrievals}
For HR~8799~d we included all available spectroscopic data as described in Section \ref{sec:obs}, but include a scaling factor for the ALES data set as otherwise it is incompatible with NACO photometric observations.
This shifts the mean L-band flux of HR~8799~d to a similar magnitude as e and c.

Unlike HR~8799~b or c, d is well fit by a broad selection of models, and no single retrieval dominates the Bayesian average of figure \ref{fig:retrieval_parameter_summary}. 
All of the preferred models $(\Delta\log_{10}\mathcal{Z}<2)$ used disequilibrium chemistry and the \citetalias{molliere_retrieving_2020} or \citetalias{zhang_elemental_2023} temperature profiles. 
Using the Bayesian average, the retrieved effective temperature is  $1179^{+31}_{-28}$~K, compatible with the ranges found by the grid fits, which find $T_{\rm eff}$ from 1150~K -- 1300~K.
The planet mass ($9.2\pm0.1~M_{\rm Jup}$) is tightly constrained by the dynamical mass prior, which in turn allowed for precise measurement of the planet radius ($1.26\pm0.07~R_{\rm Jup}$) and $\log g$ ($4.18\pm0.05$). 
This surface gravity is consistent with the estimates from the Bayesian grid fits, but is significantly higher than the 3.0-4.0 range found by the $\chi^{2}$ fits.
The self-consistent fits find marginally smaller radii than the retrievals, generally between 1.1 and 1.2 \rj.

The metallicity of HR 8799 d is consistent with HR 8799 b and c, with [M/H] = $1.2\pm0.2$.
While most grid-fit solutions also favoured high-metallicity atmospheres, the Bayesian fit with the ATMO model find a solution consistent with solar metallicity, though the radius was inconsistent with evolutionary models ($0.926\pm0.004$~\rj).
The best fit Exo-Rem model find a metallicity of 0.78$\pm0.03$, a radius of $1.168\pm0.006$~\rj and an effective temperature of 1155$\pm5$~K.
The C/O ratio is always found to be consistent with the stellar value, with retrievals finding C/O = $0.60^{+0.04}_{-0.06}$.
The grid-fits are also typically consistent with stellar, though the best fit Exo-Rem model found a substellar C/O ratio of 0.2.

Patchy clouds are marginally disfavoured by the Bayes factor, and the patchiness is poorly constrained, finding $f_{\rm cloud}=0.45\pm0.3$. The most favoured solutions require either amorphous or crystalline MgSiO$_{3}$ or crystalline MgSiO$_{3}$, with a marginal preference for the amorphous structure.
Each of these compositions displays slightly different near infrared slopes, shown in Figure \ref{fig:cloud_opacities}. 
However, such a slope can be induced by various sources of continuum opacity that may not be fully accounted for in the retrieval. 
Thus mid-infrared observations of the silicate absorption features are necessary to robustly constrain the composition and particle geometry.
There is no preference for a free cloud base pressure compared to the equilibrium position, suggesting that the \citetalias{ackerman_precipitating_2001} model is sufficient to describe the clouds in this atmosphere.

While the disequilibrium retrievals are favoured over the free chemistry retrievals, there is excellent agreement between the freely retrieved abundances and the disequilibrium chemical profiles.
Water and CO are both highly abundant, and the freely retrieved abundances agree with the disequilibrium profiles to within $1\sigma$.
No other species are both highly abundant and well constrained, so we do not perform leave-one-out retrievals to test for their presence. However, even at low abundances the freely retrieved CH$_{4}$ abundance is compatible with the disequilibrium profile.

\subsection{HR~8799~e}\label{sec:e_retrievals}
The measurements of HR~8799~e largely reinforce existing literature values.
The single most favoured retrieval, which also dominates the group $A\cap B$, used free chemistry and the \citetalias{zhang_elemental_2023} temperature profile, together with a cloud base calculated using equilibrium condensation.
This lead to similar results as for HR~8799~c and d, and is consistent with the results of the grid fits.
An effective temperature of $1138^{+30}_{-22}$~K is retrieved, compatible to within the uncertainties of the grid-fits, which found a temperature range of 1100~K to 1200~K.
The mass posterior was determined by the dynamical mass prior, as was the planet radius, finding $M_{pl} = 7.5^{+0.6}_{-0.6}$ and $R_{pl}=1.13^{+0.05}_{-0.05}$ respectively. 
This leads to a $\log g$ of $4.18^{+0.06}_{-0.05}$, slightly higher than the self-consistent estimates of 3.5--4.0.
The overall best self-consistent model by the $\chi^{2}$ was Exo-Rem, which finds an effective temperature of 1100K, a radius of 1.15 \rj, and a somewhat low log $g$ of 3.5.
ATMO is the most favoured self-consistent model when using the Bayesian framework, though it found an unphysically small radius and higher temperature than other models.
HR 8799 e is the only companion for which the MIRI photometry is not convincingly fit, as seen below in figure \ref{fig:mrs_predictions}. 
Every model underestimated the flux beyond 10~$\upmu$m relative to the measurements, though this may be due to contamination from the host star or inner disc \citep{boccaletti_imaginghr8799_2023}.

Compared to the other three planets, HR~8799~e is found to have an even more metal rich atmosphere, with [M/H]=$1.9^{+0.1}_{-0.2}$.
Metallicities $>$1 were a universal feature of the retrievals for e.
This was reinforced by the grid-fits, which uniformly find strong enrichment, running into the upper grid boundaries.
Free chemistry retrievals are always preferred over the disequilibrium retrievals; from these we found the group $A\cap B$ C/O ratio is $0.87^{+0.02}_{-0.02}$.
As the free chemistry retrieval C/O ratio only accounts for the gas-phase abundances, there is additional oxygen sequestered in the silicate clouds that could reduce the C/O ratio.
However, the most favoured disequilibrium retrieval finds a similar value of 0.83$\pm0.02$, suggesting that HR 8799 e is somewhat of an outlier compared to the other three planets.
Using a similar setup to \cite{molliere_retrieving_2020} (e.AB.11), we find C/O=$0.78^{+0.03}_{-0.03}$.
The grid fits tend to find C/O ratios compatible with the stellar value, though the overall best fit Exo-Rem model also finds a higher value of 0.8. 

In addition to the water and CO rich atmosphere, HCN is found to be highly abundant, with $\log X_{\rm HCN}=-2.26\pm0.11$. 
This detection is strongly favoured by the Bayes factor, with $\log_{10}\mathcal{Z}=7.5$.
As with HR 8799 c and shown in Fig. \ref{fig:HCNDetection}, this detection was driven by using the HCN opacity to fit the ALES spectrum, though changes to the shape in the H and K-bands also provide a slightly better fit as well.
This is slightly enriched compared to equilibrium chemistry predictions, but is expected for a metal rich planet with a relatively high C/O ratio \citep{giacobbe_five_2021}, and can be produced through photochemical reactions \citep{moses_chemical_2013}.
In contrast, the presence of \ch4 is poorly constrained.

As with HR~8799~c and d, the most favoured retrieval for e favours silicate clouds with a deeper iron deck, both condensing at their equilibrium locations, with no preference for patchy clouds.
A broad range of disequilibrium retrievals (e.AB.7--11) found consistent cloud properties, with condensation at the equilibrium location preferred, with a silicate abundance lower than predicted by equilibrium chemistry (between 10$\times$ to 100$\times$ less than equilibrium), and iron abundances consistent with equilibrium.
Individual \fsed parameters for the silicate and iron clouds were not required, with variations in the Bayes factor driven more by the choice of patchiness and temperature profile.

\subsection{Impacts of modelling choices}\label{sec:modelsarehard}
While we examined the primary atmospheric characteristics of each of the four HR~8799 planets, it is crucial to understand how the choice of model impacts these measurements.
As described in Section \ref{sec:modelling}, we performed retrievals using a broad selection of thermal structures, chemical parameterisations, and cloud models, each of which we is examined in detail below.

\subsubsection{Thermal structure}

\begin{figure}[t]
    \centering
    \includegraphics[width=0.95\linewidth]{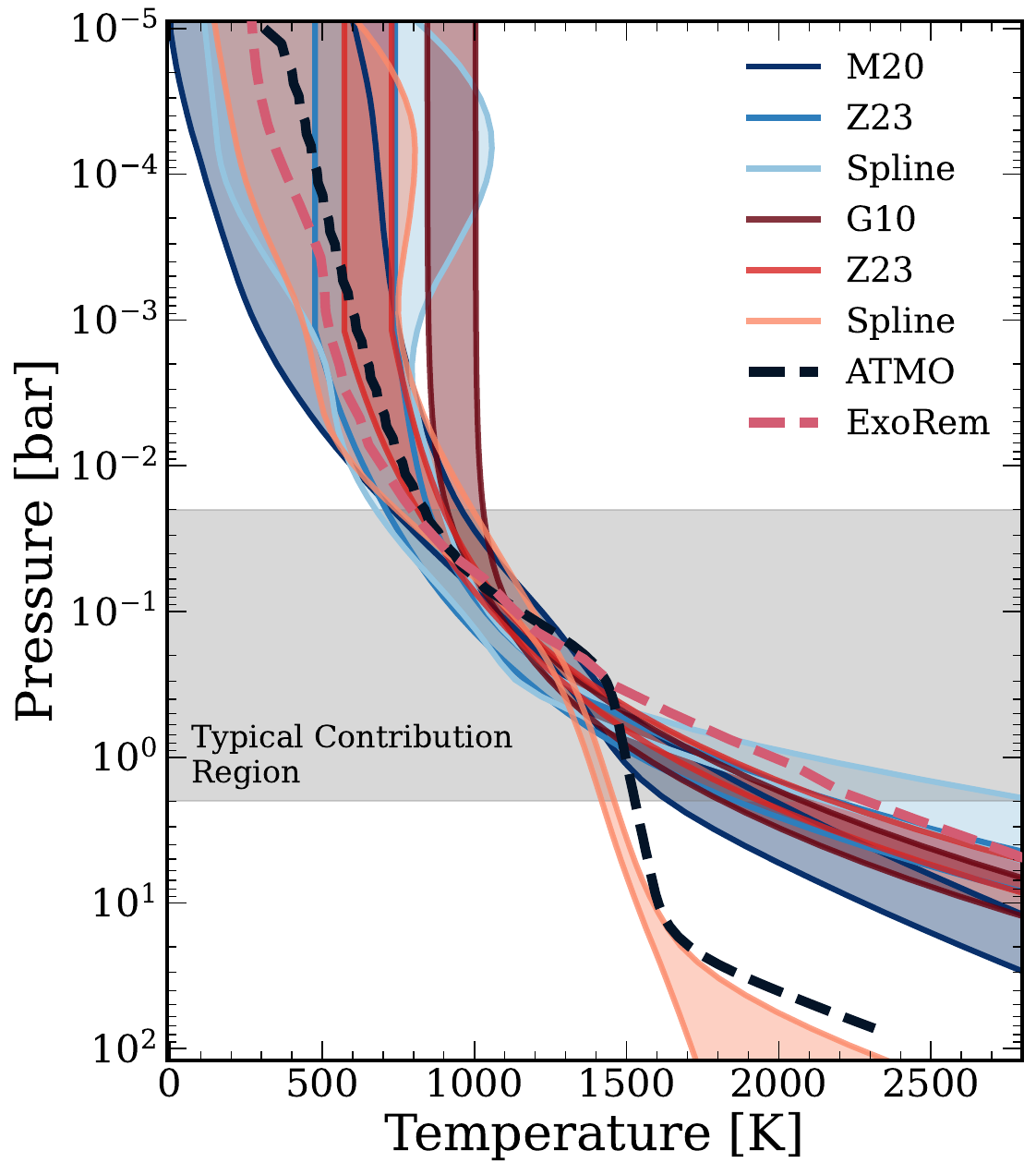}
    \caption{Temperature profiles for HR 8799 e. In blue are temperature profiles from disequilibrium retrievals, while in red are free chemistry retrievals. The shaded regions indicate 90\% confidence intervals. Also included are the temperature profiles from the best fit self-consistent models. 
    }
    \label{fig:HR8799e_PT_Profile}
\end{figure}


\begin{figure*}[t]
    \includegraphics[width=\linewidth]{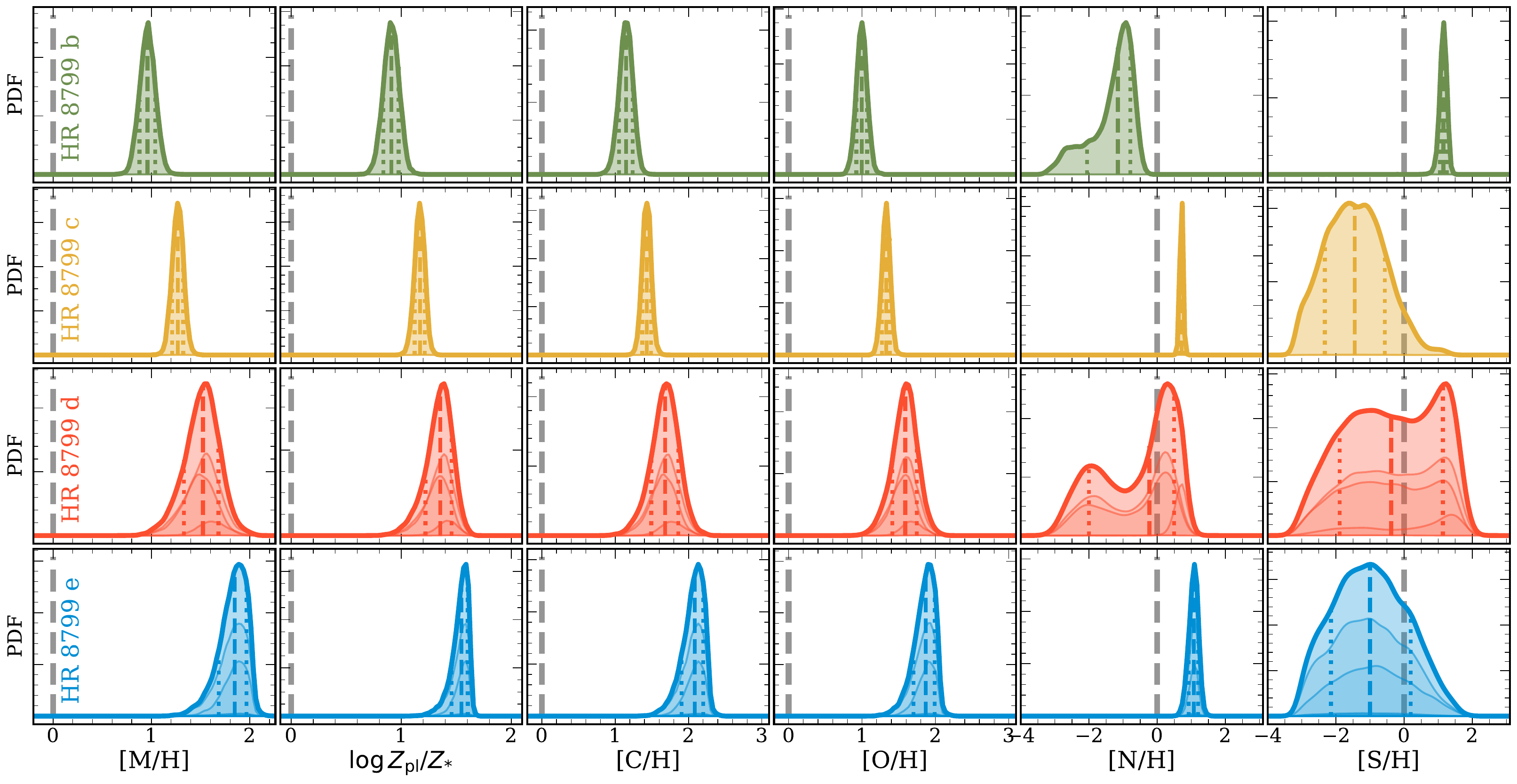}
    \caption{Metallicities and elemental number ratios for the HR 8799 planets derived from Baysian averaged group $A\cap B$ free chemistry retrievals.}
    \label{fig:elemental_ratios}
\end{figure*}

For each planet we performed retrievals using four different temperature profile parameterisations. 
The \citetalias{zhang_elemental_2023} profiles is used in the most favoured retrieval for three of the four planets, while for HR 8799 d there is equal evidence for the \citetalias{zhang_elemental_2023} and \citetalias{molliere_retrieving_2020} profiles.
The Guillot profile is found to be the second most preferred profile for HR~8799~c and e, while the spline profile using six nodes is strongly disfavoured by the Bayes factor.
In Figure \ref{fig:HR8799e_PT_Profile} we compare the retrieved and self-consistent temperature profiles for HR~8799~e.
We find that there is excellent agreement between nearly all of the retrievals in the photosphere region, as well as with the best-fit Exo-Rem temperature profile.
There is little variation in the photosphere region between the disequilibrium and free chemistry models.
Only one model - the spline profile with free chemistry - found a profile more similar to that of the clear ATMO profile, though it is strongly disfavoured by the Bayes factor compared to the other free chemistry retrievals.
The spline profile is the only profile that does not explicitly assume an adiabat deep in the atmosphere or rely on assumptions from self-consistent models, and so we cannot fully rule out the diabatic profiles of \cite{tremblin_fingering_2015}.
The bulk atmospheric properties are also reasonably consistent across the different temperature parameterisations.
While there are statistically significant variations in the C/O ratio between the different parameterisations, they remain broadly consistent between 0.7 and 0.9.

Although the spline profile is disfavoured by the retrievals, it is a useful parameterisation to determine the amount of flexibility required by the model, and to explore the known degeneracies between the atmospheric thermal structure and clouds \cite{tremblin_fingering_2015,tremblin_cloudless_2016}.
We performed a series of retrievals on HR 8799 b, varying the number of nodes in the spline profile and observing how the retrieved profile changes with the increased flexibility.
We repeated this test for both a clear atmosphere model and an model with clouds condensing at the equilibrium base pressure.
We find no significant differences in the temperature profiles between the clear and cloudy atmospheres.
For HR 8799 b, the Bayes factor favours retrievals with three or four nodes in the spline profile.
Fewer nodes mean the profile cannot be accurately modelled, while more nodes add additional parameters without improving the fit to the spectra.


\subsubsection{Chemistry}

For all four planets we performed retrievals using a grid derived from an equilibrium chemistry solver with disequilibrium \h2o-CO-\ch4 quenching, as well as free chemistry retrievals where we directly retrieved the mass-fraction abundance of various species.
Fig. \ref{fig:chemical_profiles} shows the abundance profiles from the best fit disequilibrium and free chemistry retrievals for each planet.
Both types of retrievals produce consistent metallicities and C/O ratios for each planet: overall there is excellent agreement in the water and CO abundances, which are the primary opacity sources in these atmospheres.
Only HR 8799 b shows statistically significant discrepancy between the two methods for these species, with the free retrieval finding a slightly lower metallicity than the disequilibrium retrieval.
Even for trace species, the free retrievals and disequilibrium retrievals are largely compatible, though only a few species have statistically significant detections in the free retrievals.

The strongest trace species detections are HCN in HR 8799 c and e, at abundances far higher than predicted by the equilibrium model.
HR 8799 b also has a well constrained H$_{2}$S measurement, though it is not statistically significant.
\ch4 is significantly detected in the atmosphere of HR 8799 c, demonstrating that with sufficient $S/N$ and wavelength coverage, it is possible to constrain abundances at below $10^{-4}$ by mass.
The free chemistry detection of \ch4 is at a moderately higher abundance than in the best fit disequilibrium model.
While it is likely also present in the cooler atmosphere of b, additional wavelength coverage or higher $S/N$ observations are required for a significant detection.
For HR 8799 c, we include in Figures \ref{fig:CH4Detection} and \ref{fig:HCNDetection} in the appendix comparisons between the HR 8799 c data and models both with and without the contribution of \ch4 and HCN opacity, demonstrating the impact of these species on the spectral shape. 
While the HCN detection is driven primarily by the low flux of the ALES data, there is a significant change in the H-band shape, as well as a slight change in the peak amplitude of the K-band.
As the ALES data are relatively low $S/N$, additional H and L band data should be obtained to confirm this detection.
However, the \ch4 detection is driven by modest improvements in the fit throughout the K-band.
Several abundant species predicted by the equilibrium network are not confidently detected by the free chemistry retrievals, such as \co2, NH$_{3}$, and  H$_{2}$S.
Additional wavelength coverage or higher spectral resolution may allow for the characterisation of such species.

If we take the averaged free retrieval results at face value, we can derive elemental abundance ratios for each of the four planets, using a similar method to calculating the metallicity.
Taking the volume mixing ratios of each molecular species, we can count the total number of C, N, O, and S atoms, and calculate the ratio relative to the planetary hydrogen abundance.
Thus for example
\begin{equation}
    \mathrm{C/H} = \frac{\rm X_{CO} + X_{\rm CO_{2}} + X_{\rm CH_{4}}+ X_{\rm HCN}}{X_{\rm H_{2}}+ 2X_{\rm H_{2}O} + 2X_{\rm H_{2}S} +X_{\rm HCN} +4X_{\rm CH_{4}}+3X_{\rm NH_{3}}},
\end{equation}
where all abundances are measured in number fraction. 
These ratios are then normalised to the solar values from \cite{asplund_chemical_2009}.

In Fig. \ref{fig:elemental_ratios}, we present the elemental abundance ratios for each of the four planets.
We find that most elements are enhanced relative to solar for all four planets.
HR 8799 b appears depleted in nitrogen relative to the other planets, likely due to the nondetection of NH$_{3}$, which will require observations of the 10~$\upmu$m feature to characterise.
The HCN detections in HR 8799 c and e tightly constrain the nitrogen enhancement, though these planets still appear less enriched in nitrogen than in carbon or oxygen, though this is again likely due to additional nitrogen stored in N$_{2}$ and NH$_{3}$, whose opacities are inaccessible at these wavelengths.
The sulphur elemental ratio is poorly constrained for all of the planets apart from HR 8799 b, which has a precise - though not statistically significant - constraint on the H$_{2}$S abundance. 
HR 8799 b appears sulphur rich, while the remaining planets appear consistent with the solar value, or slightly depleted in sulphur, though this is largely due to a lack of measured chemical species.

The C/O ratio is a key consideration for planetary atmospheres.
However, measuring the atmospheric C/O ratio and linking it to the bulk planet composition is far from trivial.
\cite{lodders2002} and \cite{lodders_solarsystem_2003} explore the chemistry and condensation of substellar atmospheres, identifying the condensation sequence of refractory species throughout these atmospheres, finding that at typical L-dwarf temperatures there will be silicate clouds in the photosphere region.
\cite{fonte_oxygen_2023} demonstrate how oxygen is sequestered in silicate clouds and other refractory species.
This was followed by the recent work from \cite{calamari_predicting_2024}, who calculate the bulk planet C/O ratio from the atmospheric ratio, finding that the median sequestration of oxygen due to this condensation is $17.8^{+1.7}_{-2.3}$\%.
They also identify a relation between the bulk and observed C/O ratio:
\begin{equation}
    \left(\mathrm{C/O}\right)_{\rm obs} \approx \frac{\left(\mathrm{C/O}\right)_{\rm bulk}}{1-0.371\left(\mathrm{C/O}\right)_{\rm bulk}}.
\end{equation}
Solving for $\left(\mathrm{C/O}\right)_{\rm bulk}$, we find that for HR 8799 e, with an observed C/O ratio of $0.88^{+0.02}_{-0.02}$, should have a bulk C/O ratio of 0.66, much closer to the stellar value of 0.54. 
Likewise, HR 8799 d has the lowest observed C/O ratio of $0.61^{+0.03}_{-0.04}$, which translates to a modestly substellar bulk C/O ratio of 0.50.
In general, this relation reduces the variation between the four planets, and brings the planetary C/O ratio more in line with the known stellar value. 

While both chemistry models are compatible, they also share similar biases.
The free chemistry model measures the gas phase abundance in the photosphere, and is primarily impacted by the atmosphere above the silicate clouds.
Conversely, the underlying equilibrium model does remove oxygen from the gas phase due to condensation, though the additional flexibility in the cloud parameterisation means that it is only exactly correct for \fsed$=1$. 
By parameterising disequilibrium via fixing the chemical abundances above a quench point, the model may lose this sensitivity, and therefore measures the abundances of CO, \h2o, and \ch4 in a similar fashion to the free chemistry model.
Thus the C/O ratio as inferred by both models will be strongly impacted by the oxygen-depleted region above the silicate clouds, leading to over-estimates of the C/O ratio.
Throughout this work we present these measurements, but we note that the adjustment introduced by \cite{calamari_predicting_2024} is likely a more accurate estimate of the bulk planet composition.

In addition to the elemental ratios, we also computed $\Zpl/\Zstar$, which allows us to directly compare our metallicities to literature values, such as those of \cite{thorngren_mass-metallicity_2016}.
We converted the metallicity [M/H] of each atmosphere to $\Zpl$ using the methods of \cite{thorngren_connecting_2019}, adapting for our own notation:
\begin{equation}\label{eqn:thorngren_zpl}
   10^{[\mathrm{M/H}]} = \frac{1 + Y/X}{\left(\Zpl^{-1}-1\right)\left(\frac{\mu_{Z}}{\mu_{H}}\right)},
\end{equation}
where $X$, $Y$, and $Z$ are the solar hydrogen, helium and metal mass fractions and $\mu$ is the mean molecular weight of the metal content of the atmosphere. 
Rearranging and substituting in the measured atmospheric metallicity [M/H], we find:
\begin{equation}\label{eqn:thorngren_rearr}
    \Zpl = \left(1+\frac{1 + Y/X}{10^{[\mathrm{M/H}]}\left(\frac{\mu_{Z}}{\mu_{H}}\right) (Z/H_{\odot})}\right)^{-1}.
\end{equation}
We take the same assumptions as \cite{thorngren_connecting_2019}, taking $\mu_{Z}$ to be 18, assuming most of the metal content is in water, $\mu_{H}$ to be 1 for atomic hydrogen, and Y/X to be 0.3383 as in \cite{asplund_chemical_2009}.
As the metallicity of HR 8799 A is near solar, $\Zstar/H$ is taken to be the solar value of $Z/H_{\odot}=1.04\times10^{-3}$.
To normalise to the stellar metallicity we follow \cite{thorngren_mass-metallicity_2016} and calculate the $\Zstar$ as 
\begin{equation}\label{eqn:thorngren_zplstar}
    \Zstar = 0.014\times 10^{\mathrm{Fe/H}}.
\end{equation}
For HR 8799, we used solar metallicity to calculate Fe/H, but refer to the discussion in Section \ref{sec:hoststar}.

Disequilibrium chemistry has long been thought to play a key role in shaping the composition of the HR 8799 atmospheres \citep[e.g.][]{marois_direct_2008}.
With well-constrained chemical abundances, we can start to place limits on the strength of vertical mixing that drives this disequilibrium.
The quench pressure we retrieve is defined as the level below which (in pressure) the abundances of \h2o, CO, and \ch4 become vertically constant.
This parameterises dynamical mixing that homogenises the upper layers of the atmosphere.
More rigorously, the quench point is defined as the point at which the chemical timescale $t_{\rm chem}$ and the mixing $t_{\rm mix}$ are equal.
Following \cite{zahnle_methane_2014}, the mixing timescale is defined through the ratio of the local atmospheric scale height $H$ to the vertical eddy diffusion coefficient, $\kzz$:
\begin{equation}
    t_{\rm mix} = \frac{H^{2}}{\kzz}.
\end{equation}
The chemical timescale depends on the reaction rates involved. Considering the CO--\ch4 reaction chain, \cite{zahnle_methane_2014} derive a timescale at the quench point ($t_{q}$) for CO.
For strong mixing, pulling from material at depths below the point where the atmosphere is 1000 K, they find the timescale well described by an Arrhenius relation for quench pressure $p$ in bar, metallicity $m$, where $m=10^{\rm [M/H]}$, and temperature $T$in kelvin:
\begin{equation}
    t_{\rm q1} = 1.5\times10^{-6}p^{-1}m^{-0.7}\exp{\left(42000/T\right)} \ {\rm s}.
\end{equation}
For weak mixing, and therefore drawing from low temperatures with little CO, the timescale is found to be
\begin{equation}
    t_{\rm q2} = 40p^{-2}\exp{\left(25000/T\right)}\ {\rm s},
\end{equation}
Combining the two, the total chemical timescale is defined as:
\begin{equation}
    t_{\rm CO} = \left(\frac{1}{t_{\rm q1}} + \frac{1}{t_{\rm q2}}\right)^{-1},
\end{equation}
which will favour the lower of the two values $t_{\rm q1}$ and $t_{\rm q2}$.
Equating the mixing and CO reaction timescales, we can infer the strength of vertical mixing in the atmospheres of the HR 8799 planets:
\begin{equation}
    \kzz = \frac{H^{2}}{t_{\rm CO}},
\end{equation}
where the scale height is defined as
\begin{equation}
    H = \frac{k_{\rm B}T}{\mu g},
\end{equation} 
for temperature $T$, surface gravity $g$, mean molecular mass $\mu$, and the Boltzmann constant $k_{\rm B}$.
In order to calculate these quantities for the HR~8799 planets, we take $T_{\rm eff}$ as a representative temperature to calculate the timescales and scale height, $g$ as the measured surface gravity, and $\mu$ as the average mean molecular weight of the atmosphere.
As \kzz is exponential in temperature, the choice of what temperature to use strongly influences the measured value.
Using the temperature at the quench pressure, typically deep in the atmosphere, results in unphysically strong vertical mixing, with $\log \kzz\approx20$.
A more thorough analysis could try to measure the vertical mixing as a function of temperature throughout the atmosphere, but the current data quality is of insufficient resolution or $S/N$ for such measurements.
Thus we treat $T_{\rm eff}$ as a representative temperature with which to determine the vertical mixing strength.
We include the results of these calculations, together with the retrieved \kzz used to parameterise the \citetalias{ackerman_precipitating_2001} clouds in Table \ref{tab:Kzz}.

\begin{table}[t]
\centering
\begin{threeparttable}
    \caption{Quench pressures, vertical mixing parameters, and sedimentation fractions for the HR 8799 planets.}
    \label{tab:Kzz}

    \begin{tabular}{r|lllll}
    \toprule
       \textbf{Planet} & $\log \mathbf{P_{\rm \textbf{q}}}$ &  $\mathbf{K_{\rm \textbf{zz\,q}}}$  & $\mathbf{K_{\rm \textbf{zz,\,\citetalias{ackerman_precipitating_2001}}}}$ & \multicolumn{2}{c}{$\bm{f_{\rm sed}}$}\\
       & [bar]& [cm$^{2}$/s] & [cm$^{2}$/s] & MgSiO$_{3}$ & Fe\\
    \midrule
        b& $1.7^{+0.2}_{-0.2}$ & $2.9^{+0.6}_{-0.7}$ & $8.6^{+0.8}_{-1.0}$ & $1.1^{+0.3}_{-0.3}$ & $4^{+2}_{-1}$\\[2pt]
        c      & $2.3^{+0.1}_{-0.1}$ & $6.3^{+0.3}_{-0.5}$ & $9.2^{+0.4}_{-0.4}$ & $3.3^{+0.3}_{-0.3}$&$6^{+2}_{-2}$\\[2pt]
        d      & $1.0^{+0.6}_{-0.7}$ & $4.8^{+0.9}_{-1.3}$ & $9.2^{+0.7}_{-0.9}$ & $2^{+4}_{-1}$&$6^{+2}_{-2}$\\[2pt]
        e      & $2.1^{+0.4}_{-1.2}$ & $5.9^{+0.9}_{-0.6}$ & $8.6^{+0.7}_{-0.8}$ & $1.4^{+0.4}_{-0.3}$&$ 6^{+2}_{-3}$\\
    \bottomrule
    \end{tabular}
    \begin{tablenotes}
    \item\textbf{Notes}
    \item Measured from the group $A\cap B$ retrievals, providing median and $\pm34$\% confidence regions.
    
    \end{tablenotes}
\end{threeparttable}
\end{table}

\begin{figure*}[t]
    \centering
    \includegraphics[width=0.49\linewidth]{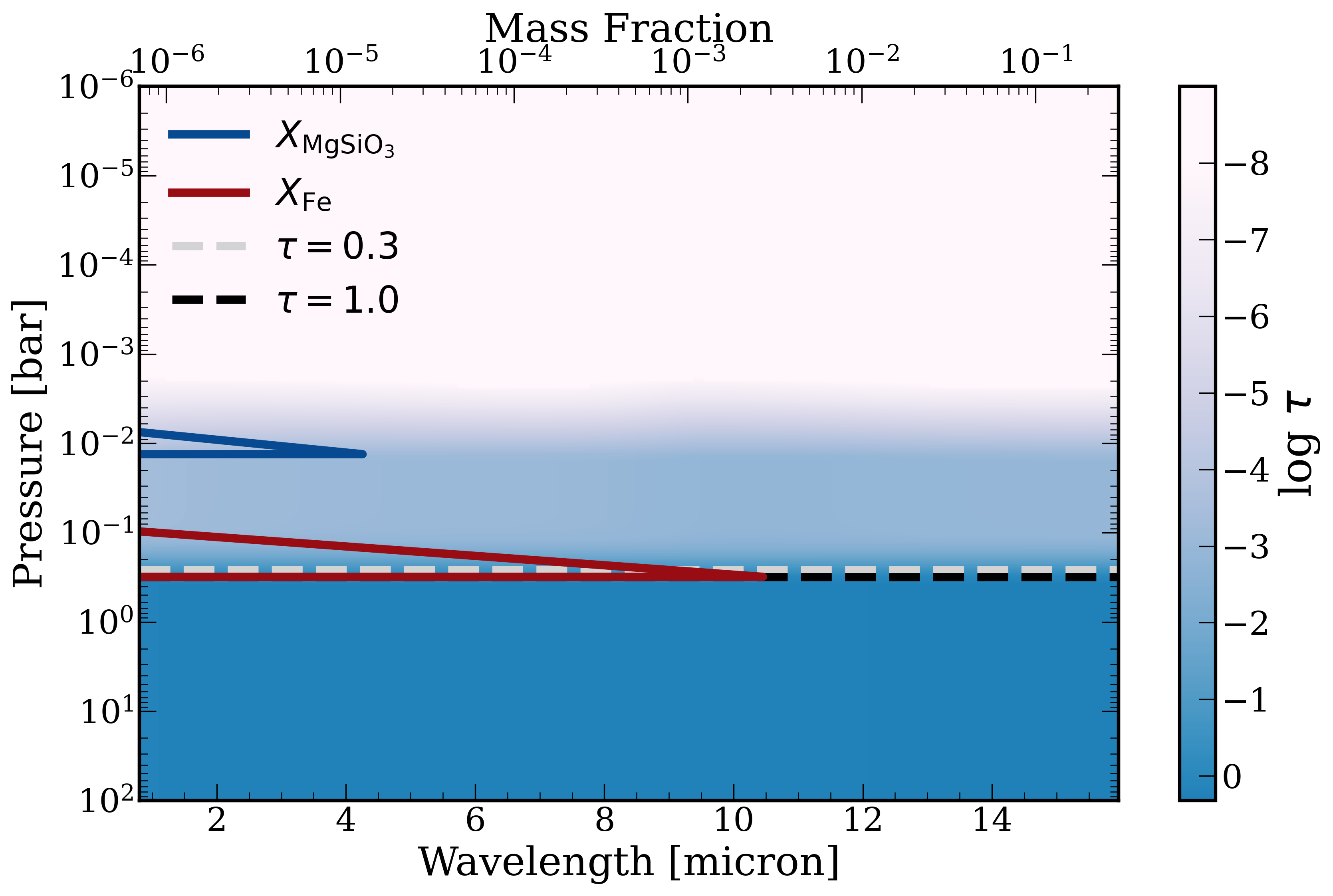}
    \includegraphics[width=0.49\linewidth]{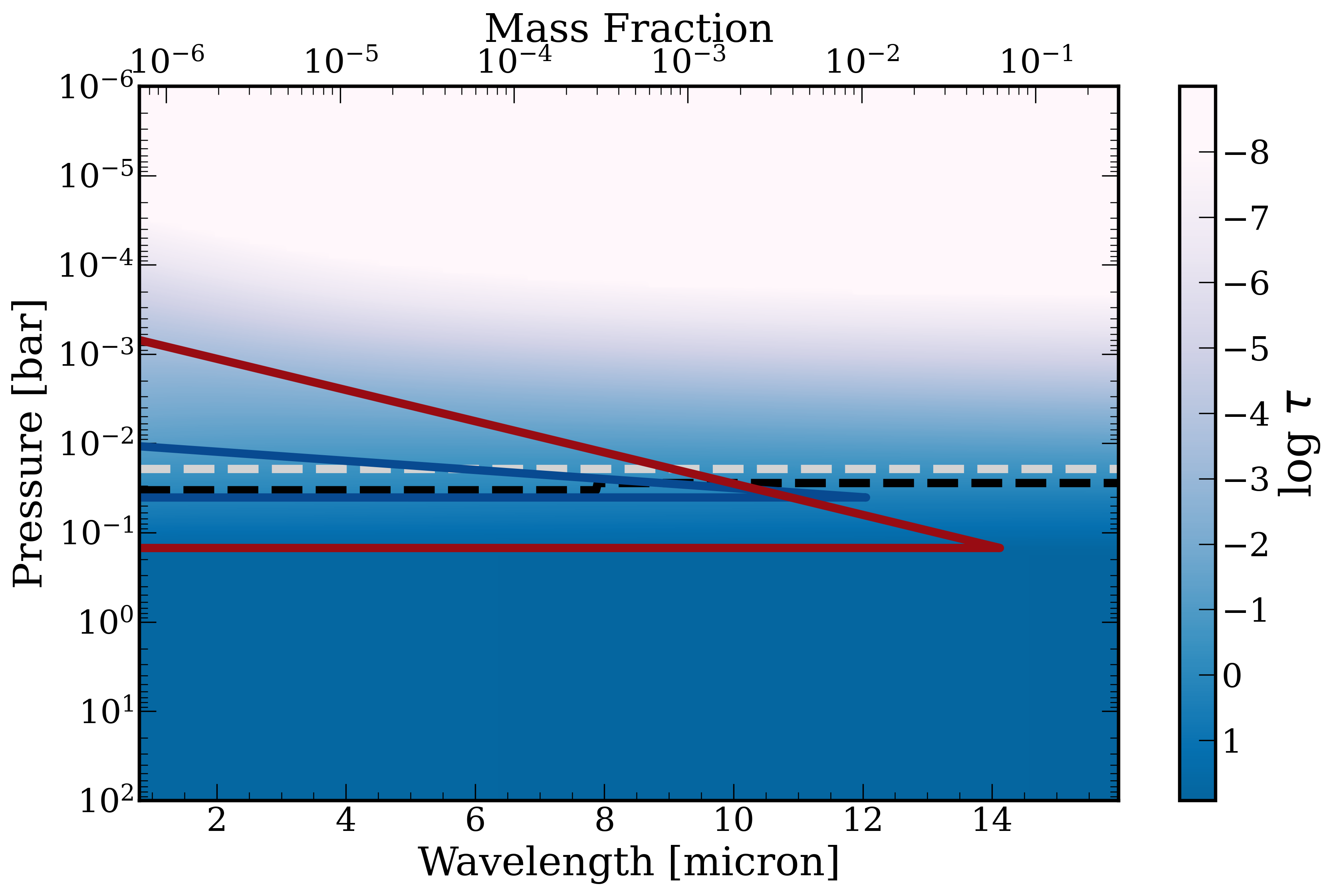}
    \caption{Cloud properties for HR 8799 e, showing the optical depth due to clouds as a function of pressure (colour map), with the $\tau=0.3$ and $\tau=1.0$ contours highlighted by the dashed lines. The solid lines indicate the mass fraction abundance of the MgSiO$_{3}$ and Fe clouds in blue and red respectively. \textbf{Left:} Cloud properties e.AB.13, using a freely retrieved cloud base pressure and abundances. \textbf{Right:} The same, but for Case e.AB.2, which uses the equilibrium condensation to determine the location of the cloud base. The abundances are determined using equilibrium chemistry, and retrieving a scaling factor, ($\log S_{\rm Fe}=0.0\pm1.1$, $\log S_{\rm MgSiO_{3}}=-0.8\pm1.0$).}
    \label{fig:cloud_optical_depth}
\end{figure*}

The quench pressure is well constrained in the Bayesian average of group $A\cap B$ retrievals for all four planets; the values of which are listed in Table \ref{tab:Kzz}.
All of the planets quench below the photosphere, with d quenching at the highest altitude, around 10 bar.
Including only the disequilibrium retrievals in the Bayesian average of group $A\cap B$, we derived \kzz from the quench pressure.
We then turn to \cite{soni_effect_2023} for a comparison, who provide predictions for \ch4 and CO abundances for varying \kzz, $T_{\rm eff}$, and log $g$ across a range of metallicities.
For our measured  $T_{\rm eff}$ and \ch4 abundances, we should expect $\log\kzz$ of around 6 for the warmer three companions, regardless of whether we use the measured CO or \ch4 abundance. 
For HR 8799 b a much lower value (less than $\sim2$) is expected, assuming a 10$\times$ solar metallicity.
Our inferred \kzz values for c, d, and e are compatible with this prediction, finding $\log$\kzz between 5 and 6.
For HR 8799 b we also measure weak mixing, with  $\log\kzz = 2.9^{+0.6}_{-0.7}$, which is again compatible with the predictions of \cite{soni_effect_2023}.
These measurements from the quench pressure are also inconsistent with the parameter used in the \citetalias{ackerman_precipitating_2001} clouds, which require stronger vertical mixing of $\log\kzz\sim9$.
This discrepancy is perhaps not surprising: 3D modelling predicts that \kzz should vary with altitude throughout the atmosphere, and the larger cloud particles likely respond to the atmospheric motion differently than the gas phase constituents.
Ultimately, more precise constraints on the thermal structure and chemical abundances, as well as trace species detections are necessary to derive a more precise vertical mixing strength.
Further modelling work is also necessary to provide a more physically motivated transport model than a vertically constant eddy diffusion coefficient.

\subsubsection{Clouds}\label{sec:retrievedclouds}

For the inner three planets, we find the most favoured solution is an optically thin silicate cloud lying above a deeper, optically thick iron cloud deck, as seen in figure \ref{fig:cloud_optical_depth}.
In our standard setup, the clouds were parameterised as in \citetalias{ackerman_precipitating_2001}, with the clouds condensing at the intersection of their condensation curve and the temperature profile, with their extent determined by \fsed.
The cloud mass fraction was allowed to scale from equilibrium.
In totally free retrievals, the cloud abundances, locations, and vertical extents were all free parameters of the model.
This decouples the clouds from both the chemistry and the atmospheric thermal structure, allowing them to fit the spectral shape, but in potentially nonphysical configurations.
In this framework the cloud extent was then parameterised as in \citetalias{ackerman_precipitating_2001}, determined by \fsed and \kzz.
Using this setup we find an optically thin silicate cloud lying above a compact iron cloud, while the \citetalias{ackerman_precipitating_2001} setup finds an iron cloud that extends high above the silicate cloud.
Depending on the choice of other parameters, either the free cloud base or the equilibrium base can be preferred by the Bayes factor. 
HR 8799 e free chemistry retrievals strongly favour the equilibrium condensed clouds, while the disequilibrium retrievals favour the free cloud base setup.
However, in general the clouds condensing at equilibrium are the most favoured setup for each planet.
Without broad wavelength coverage and high spectral resolution to probe a high dynamic range in pressure, it is difficult to robustly distinguish between the different potential cloud structures.

There is a marginal preference for clouds parameterised using a \cite{hansen_multiple_1971} particle size distribution over a log-normal distribution (e.AB.20 over e.AB.25, $\Delta\log_{10}\mathcal{Z}=0.7$), though this was only compared for HR 8799 e using the Z23 profile, and assuming the clouds condense at their equilibrium saturation location.
In this case $a_{h}$ is calculated from  \fsed and \kzz, and a lower \fsed for the MgSiO$_{3}$ cloud was retrieved than with the log normal distribution ($f_{\rm sed,\,Hansen} = 1.18\pm0.20$).
The effective distribution width parameter, $b_{h}$ was found to be 0.016, which is narrower than the distributions found by \cite{burningham_cloud_2021}.

In general our clouds are comparable to those of \cite{burningham_cloud_2021} and other similar studies \citep[e.g.][]{molliere_retrieving_2020,vos_patchy_2023,balmer_vlti_2023}. \cite{burningham_cloud_2021} find a combination of MgSiO$_{3}$, SiO$_{2}$, and Fe clouds provides the best-fit model to an ultracool field dwarf,  2MASSW J2224438--015852. 
Figure \ref{fig:cloud_optical_depth} shows that the MgSiO$_{3}$ clouds in their model are located at $10^{-3}$~bar with a maximum optical depth of $\tau=0.3$ at 1~$\upmu$m  and an effective particle radius of $\sim0.04~\upmu$m.
The SiO$_{2}$ clouds are slightly deeper, at $10^{-2}$~bar, and are optically thick at 1~$\upmu$m. 
This is the same location where we find MgSiO$_{3}$ clouds condensing in the atmospheres of HR 8799 c, d, and e.
The iron cloud is found to be deep in the atmosphere, though with an extended structure, with some contribution at the same altitude as the silicate clouds.
While we did not fit for a three cloud model, the similar locations and optical depths of these clouds suggests similar structures between the objects, even though they differ in effective temperature by hundreds of kelvin.

\begin{figure}[t]
    \centering
    \includegraphics[width=\linewidth]{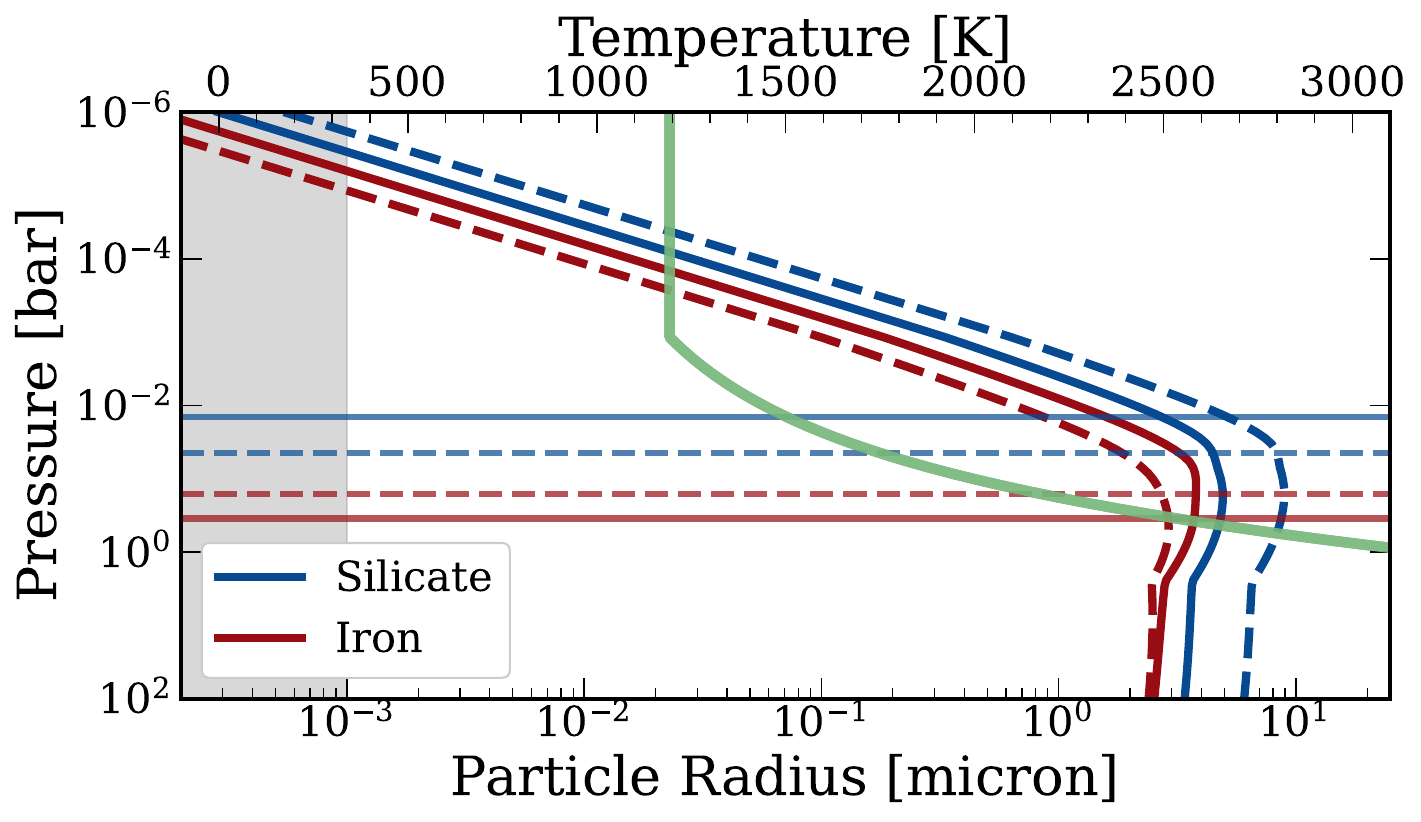}
    \caption{Effective particle radii as a function of altitude for silicate (blue) and iron (red) clouds. The solid lines indicate the radii for Case e.AB.13, which used a free cloud base pressure and abundance, while the dashed lines are for Case e.AB.2, which used equilibrium condensation and scaled equilibrium abundances. The horizontal lines indicate the cloud base pressure. The green line indicates the temperature profile. The shaded regions indicate unphysical particle sizes where the opacity contribution is set to 0.}
    \label{fig:cloud_particle_radius}
\end{figure}

Figure \ref{fig:cloud_particle_radius} highlights how the difference in particle radius contribute to the difference in the wavelength dependence of the cloud optical depth.
The cloud particle radius in the \citetalias{ackerman_precipitating_2001} model is a function of many atmospheric factors, including the temperature, mixing strength, \fsed, and particle number density.
We see that changes in the particle radius are correlated with changes in both the temperature and the particle density. 
\cite{luna_empirically_2021} explore the impact of particle size and composition on the spectral signatures of clouds in young brown dwarfs in the mid-infrared, finding that small particle sizes will lead to visible features in the planetary spectrum.
However, we see in Figure \ref{fig:cloud_particle_radius} that the particle sizes in regions of the atmospheres with significant cloud mass fraction tend to be larger ($>1~\upmu$m), and that there are no indications of deep cloud absorption features in the mid-infrared, even though silicate clouds are present in the atmosphere.
Small silicate particles should produce deeper absorption features, which are not observed in the mid-infrared, suggesting that the impact of the small mean particle size in the upper atmosphere does not contribute strongly to the cloud opacity.

\begin{figure}[t]
    \centering
    \vspace{2em}
    \includegraphics[width=\linewidth]{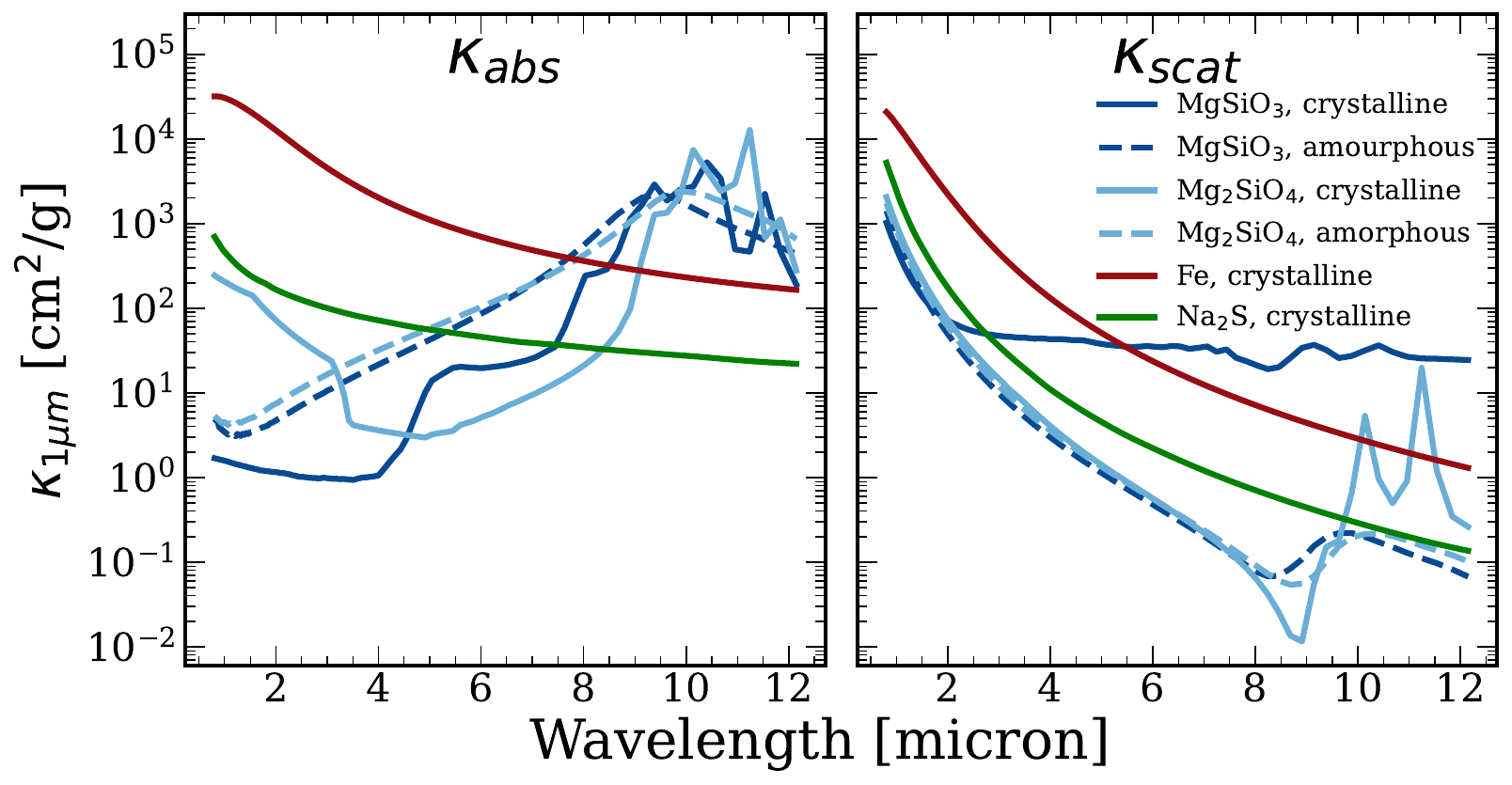}
    \caption{Cloud absorption (left) and scattering (right) opacities for different condensate compositions and structure, for 1-$\upmu$m particles. Dark blue indicates MgSiO$_{3}$, light blue is Mg$_{2}$SiO$_{4}$. Solid (dashed) lines are for crystalline (amorphous) substances. The solid red (green) line is for crystalline iron (Na$_2$S).}
    \label{fig:cloud_opacities}
\end{figure}

Due to the different slopes in the near-infrared opacity as a function of wavelength, as shown in Figure \ref{fig:cloud_opacities}, there is some sensitivity to different compositions and particle geometries.
This can explain the mild preference for amorphous MgSiO$_{3}$ clouds in certain like-for-like retrieval comparisons.
In reverse, the lack of features leads to a preference for Na$_{2}$S clouds in HR 8799 b. 
Performing additional tests on HR~8799~e, we find that Mg$_{2}$SiO$_{4}$ (e.AB.14) are mildly disfavoured by the retrievals, and Al$_{2}$O$_{3}$ (e.AB.24) clouds are strongly disfavoured, though this is again more likely from their condensation location rather than from the impact of aerosol spectral features.
Observations of the silicate absorption features at 10~$\upmu$m would allow more precise measurement of this wavelength dependence, and in turn place better constraints on the cloud structure and composition.



\section{Discussion}\label{sec:discussion}
\subsection{Highly enriched atmospheres}

\begin{figure*}[t]
    \centering
    \includegraphics[width = 0.95\linewidth]{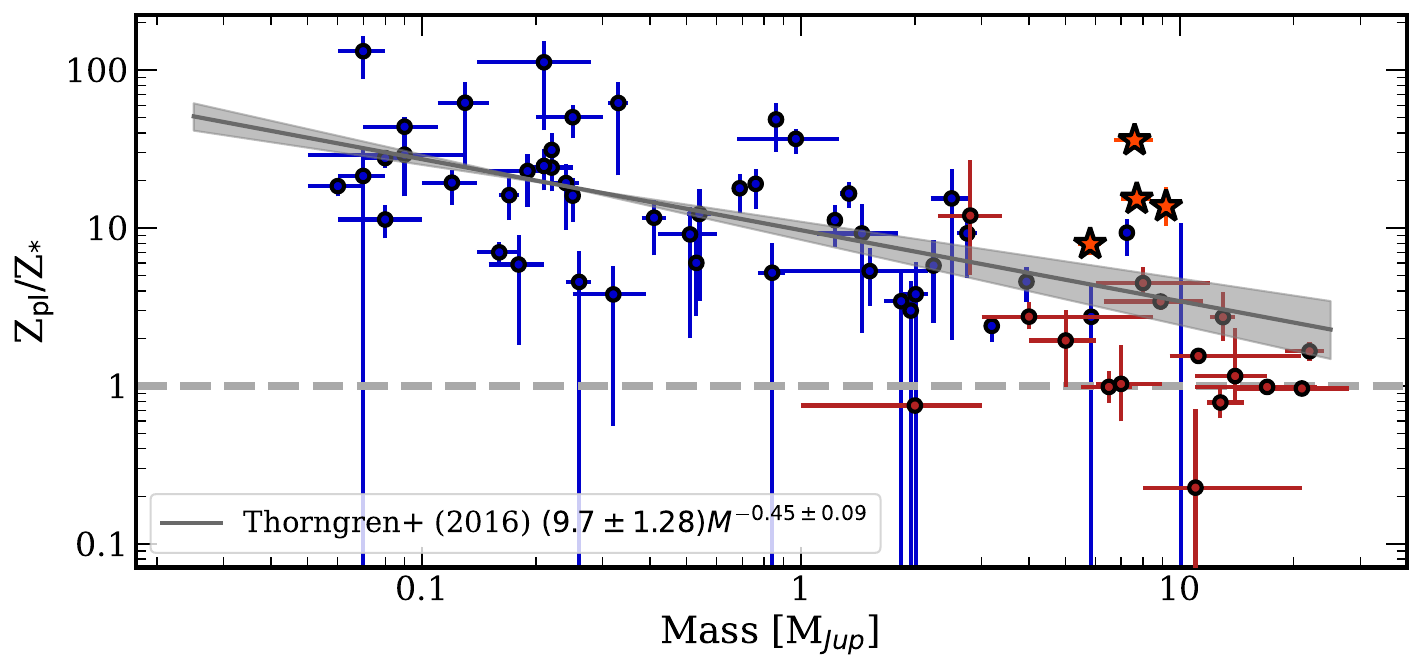}
    \caption{Metallicities of the exoplanet planet population. In blue are transiting planets, adapted from \cite{thorngren_mass-metallicity_2016} (dark blue). In red are directly imaged planets, with references listed below. Indicated with stars are the HR 8799 planets as measured in this work. The grey line indicates the fit by \citet{thorngren_mass-metallicity_2016}.\protect\\
    \textbf{Notes:}
    $\beta$ Pic b \citep{gravity_collaboration_peering_2020};
    PDS 70 b \citep{wang_constrainingPDS70_2021};
    51 Eri b \citep{whiteford_retrieval_2023};
    VHS 1256 b \citep{hoch_moderate-resolution_2022};
    HIP 65426 b \citep{petrus_medium-resolution_2021};
    $\kappa$ And b \citep{bonnefoy_characterization_2014,wilcomb_moderate-resolution_2020};
    YSES 1 b \citep{zhang_13co-rich_2021}
    AB Pic b \citep{palma-bifani_peering_2023};
    AF Lep b \citep{zhang_elemental_2023};
    GJ 504 b \citep{bonnefoy_gj504_2018};
    HD 95086 b \citep{desgrange_in-depth_2022};
    Ross 458 c \citep{burgasser_clouds_2010};
    DH Tau b \citep{patience_spectroscopy_2012};
    HN Peg b \citep{legget_HNPegB_2008,suarez_ultracool_2021_HNPeg};
    CT Cha b \citep{schmidt_direct_2008};
    GQ Lup b \citep{demars_emission_2023}.}
    \label{fig:DI_metallicities}
\end{figure*}

While some enrichment is expected in giant planets formed through core accretion, metallicities of nearly 100$\times$ the stellar value are far beyond the expectations for planets with masses larger than that of Jupiter.
Having transformed the atmospheric metallicities from the retrievals to $\Zpl/\Zstar$ using equations \ref{eqn:thorngren_rearr} and \ref{eqn:thorngren_zplstar}, we can compare the HR 8799 planets to the broader population.
Figure \ref{fig:DI_metallicities} shows how the inferred metallicities of the HR 8799 system compare to those of other directly imaged planets, and to a fit from \cite{thorngren_mass-metallicity_2016} derived from a sample of transiting exoplanets.
The HR~8799 planets are clear outliers amongst the directly imaged planets, whose metallicities were taken from the literature. 
We again used equations \ref{eqn:thorngren_rearr} and \ref{eqn:thorngren_zplstar} to convert from a measurement of [M/H] to $\Zpl/\Zstar$. 
Where stellar metallicities are not available, we assumed a metallicity [Fe/H]=0. 
Only the 2 \mj planet Af Lep b has a comparable degree of enrichment to the HR 8799 planets \citep{zhang_elemental_2023}.
However, they are comparable to hot jupiters observed in transmission, such as WASP 39 b \citep{rustamkulov_wasp39prism_2023}.
\cite{looper_discovery_2008} and \cite{stephans_spectra_2009} demonstrate how high metallicity atmospheres facilitate condensation, in turn leading to the strong reddening seen in a subset of L dwarfs. 
The retrieval of highly metal rich and very cloudy atmospheres is consistent with this picture from brown dwarfs.
Further supporting the high [M/H] retrievals are their consistency with the self-consistent grids, which always favour their upper limits. 

\begin{figure*}[ht]
    \centering
    \includegraphics[width = 0.95\linewidth]{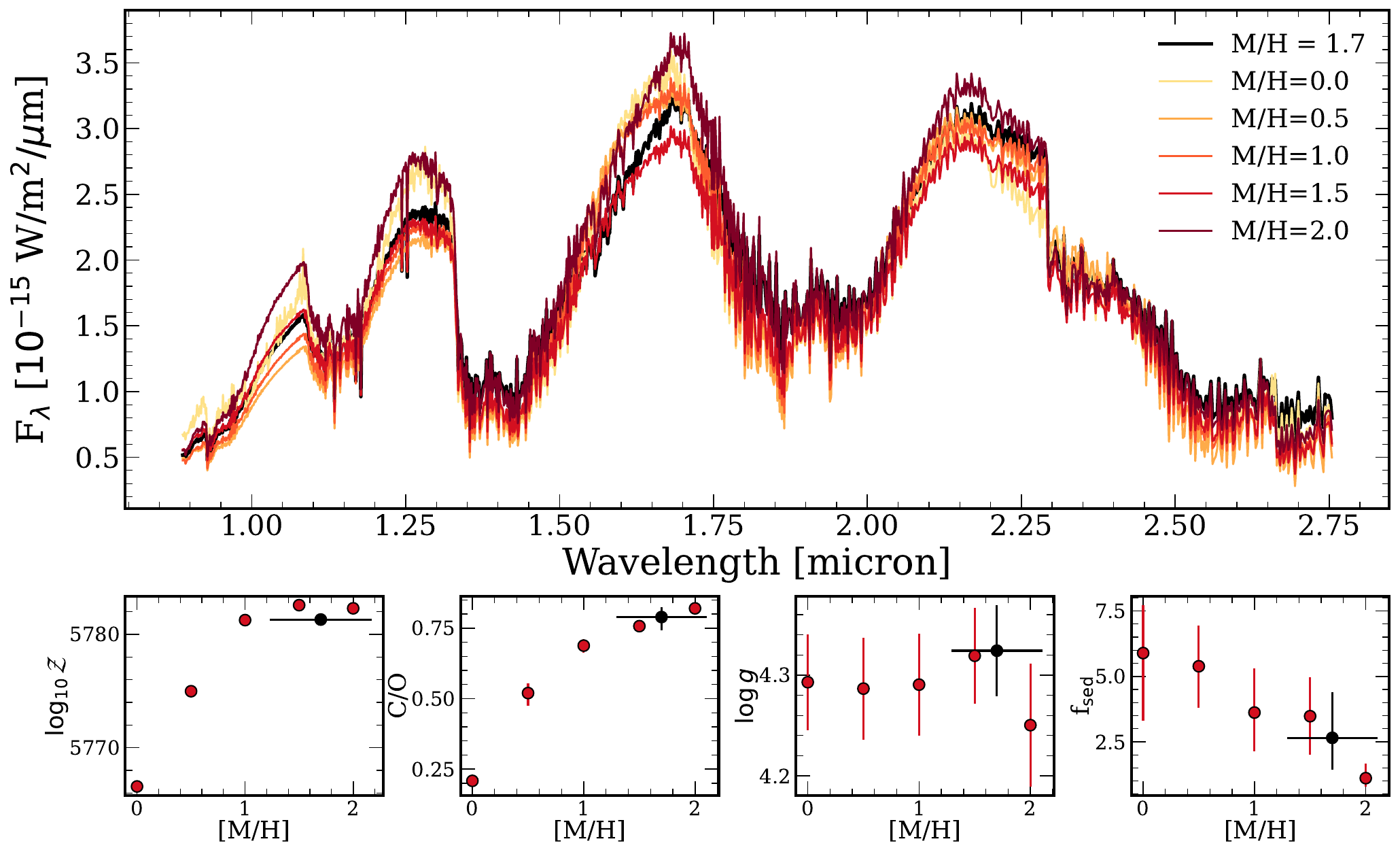}
    \caption{Comparison retrievals of HR 8799 e  where the metallicity is fixed to a constant value. Black points denote the equivalent retrieval where the metallicity is a free parameter. The top panel shows how the best fit spectra differ across the set of retrievals, while the bottom row shows how different atmospheric parameters vary as a function of metallicity.}
    \label{fig:M/H_ParameterSweep}
\end{figure*}

To validate these findings, we performed a series of test retrievals for HR~8799~e, fixing the metallicity to solar composition, and compared cases where [M/H] is fixed to values between 0.0 and 2.0 in steps of 0.5 dex.
Figure \ref{fig:M/H_ParameterSweep} shows the best fit spectra from each of these retrievals, showing the clear differences in the J, H, and K bands between the different metallicity cases that are unable to be compensated for by varying other atmospheric parameters.
We find that the cases of [M/H] = 1.5 and 2.0 are strongly favoured over the cases between 0.0 and 1.0, with solar composition disfavoured at $\Delta\log\mathcal{Z}>10$.
The remaining atmospheric parameters also significantly varied between the different retrievals: the C/O ratio increases with increasing [M/H], maintaining a relatively constant abundance of H$_{2}$O in the atmosphere while allowing the CO abundance to increase.
This combines with the decreasing \fsed, increasing the cloudiness of the planet to dampen the stronger molecular features at high metallicity.
A more thorough treatment of the condensation and chemistry in the retrieval framework are likely necessary to accurately infer both of these parameters.

The spectra shown in Figure~\ref{fig:M/H_ParameterSweep} show that the retrieved metallicity is strongly dependant on both the height of the J, H, and K band peaks, as well as the shape of these features.
However, the J bands is only covered by the low-resolution SPHERE data, with relatively poor $S/N$, and different instruments measure significantly different flux in the H band.
Without compatible measurements in this spectral regime, robust conclusions about the metallicity are hard to draw.
Additional constraints can be obtained from other wavelength regions; \cite{lodders2002} find that the CO$_{2}$ abundance scales proportionally to [M/H]$^{2}$.
With strong features between 3 and 4~$\upmu$m, as well as in the mid-infrared, future observations should be able to place robust constraints on this parameter.
Section \ref{sec:jwst} discusses the potential of JWST to make such observations.

Finally, we performed independent comparisons to the moderate resolution OSIRIS spectra presented in \cite{ruffio_deep_2021}.
Following their methods, we used the parameters of the single best-fit disequilibrium models for HR 8799 b, c, and d to compute a high spectral resolution using line-by-line opacity lists.
We convolved this model to the OSIRIS instrumental spectral resolution, and binned the model to the OSIRIS wavelength grid.
The resulting spectra was multiplied by the atmospheric transmission function provided by \cite{ruffio_deep_2021}.  
The continuum was measured by high-pass filtering the spectrum, and was subsequently subtracted from the model.
We fit a scaling factor between the model and the OSIRIS data and computed the resulting $\chi^{2}$.
This exercise was repeated, setting the metallicity of the model to solar.
For HR 8799 c and b we find that the high metallicity model provides a better fit to the OSIRIS data than the solar model, with the caveat that fixing the metallicity during the retrieval may result in a better fit than setting it a posteriori, without changing other atmospheric parameters.
For HR 8799 d the high metallicity model is only a marginally better fit than the solar metallicity model.
In general, the fits from the GRAVITY and remaining archival data provide reasonable fits to the R\,$\approx4000$ OSIRIS data, of similar quality to the fits displayed in \cite{ruffio_deep_2021}, and are included in the appendix in Figure \ref{fig:OSIRIS_Comp}.

\subsection{Impacts of data selection}
Given the inhomogeneity in the data in terms of spectral resolution, $S/N$, observing strategy and more, we performed a series of retrievals to examine the impact of different datasets on the retrieval results.
We first performed retrievals using only the GRAVITY data to determine the constraining power of this new dataset.
We found that using only the GRAVITY data we could rule out clear atmosphere solutions at $\Delta\log\mathcal{Z}>7$ (e.g. e.A.31 over e.A.32), using the Z23 profile and disequilibrium chemistry.
Using only the GRAVITY observations for HR~8799~e, we could obtain estimates of the metallicity ($2\pm0.3$), effective temperature ($1143^{+38}_{-32}$~K), and C/O ratio ($0.71^{+0.08}_{-0.2}$), which are broadly consistent with the results from the combined dataset.
We obtain similarly reliable estimates for b (b.A.32), but using only the GRAVITY data for d (d.A.12) and c (c.A.11), resulted in significantly lower metallicities.
All of these retrievals use the same setup of the Z23 profile, disequilibrium chemistry, and silicate and iron clouds condensing at their equilibrium saturation point.
Given the higher spectral resolution and $S/N$ of the GRAVITY data, this leads to the conclusion that solutions to the full retrievals are largely driven by the fits to the GRAVITY spectra. 
The C/O ratios for d and b are also incompatible when using only the GRAVITY data, finding substellar values for both planets.

Conversely, we also performed retrievals that exclude the GRAVITY spectra. For HR~8799~e (e.A.33) and d (d.A.13), we find that the retrieved parameters again broadly agree with the retrieval including the GRAVITY data.
Thus even if the retrievals are dominated by the GRAVITY data, the conclusions we draw are robust even when excluding the GRAVITY spectra.

\subsection{Formation}
The formation mechanism of the HR~8799 planets has seen much debate since their discovery: simply put, how can one form four such massive planets in the same system?
The C/O ratio and metallicity are the best formation tracers observed to date in these planets. Nevertheless, neither gravitational instability nor core accretion scenarios have been ruled out.
The high degree of enrichment in these objects would seem to suggest a core accretion formation scenario.
However, \cite{wang_accretion_2023} finds that even for [M/H]=0.5, approximately 100 earth masses of solids are required to enrich the atmosphere of HR~8799~e. 
With the even higher metallicities measured in this work, this number would increase, and it is unclear if it is possible to have nearly 1000 earth masses of metals available in a protoplanetary disc and accreted with high efficiency.
\cite{wang_accretion_2023} additionally find that late accretion of planetesimals would require less material to result in a similar degree of enrichment, potentially only requiring a few hundred earth masses of solids to achieve the high metallicities of all four planets, though this material should quickly settle out of the atmosphere.

\begin{figure}[t]
    \includegraphics[width=\linewidth]{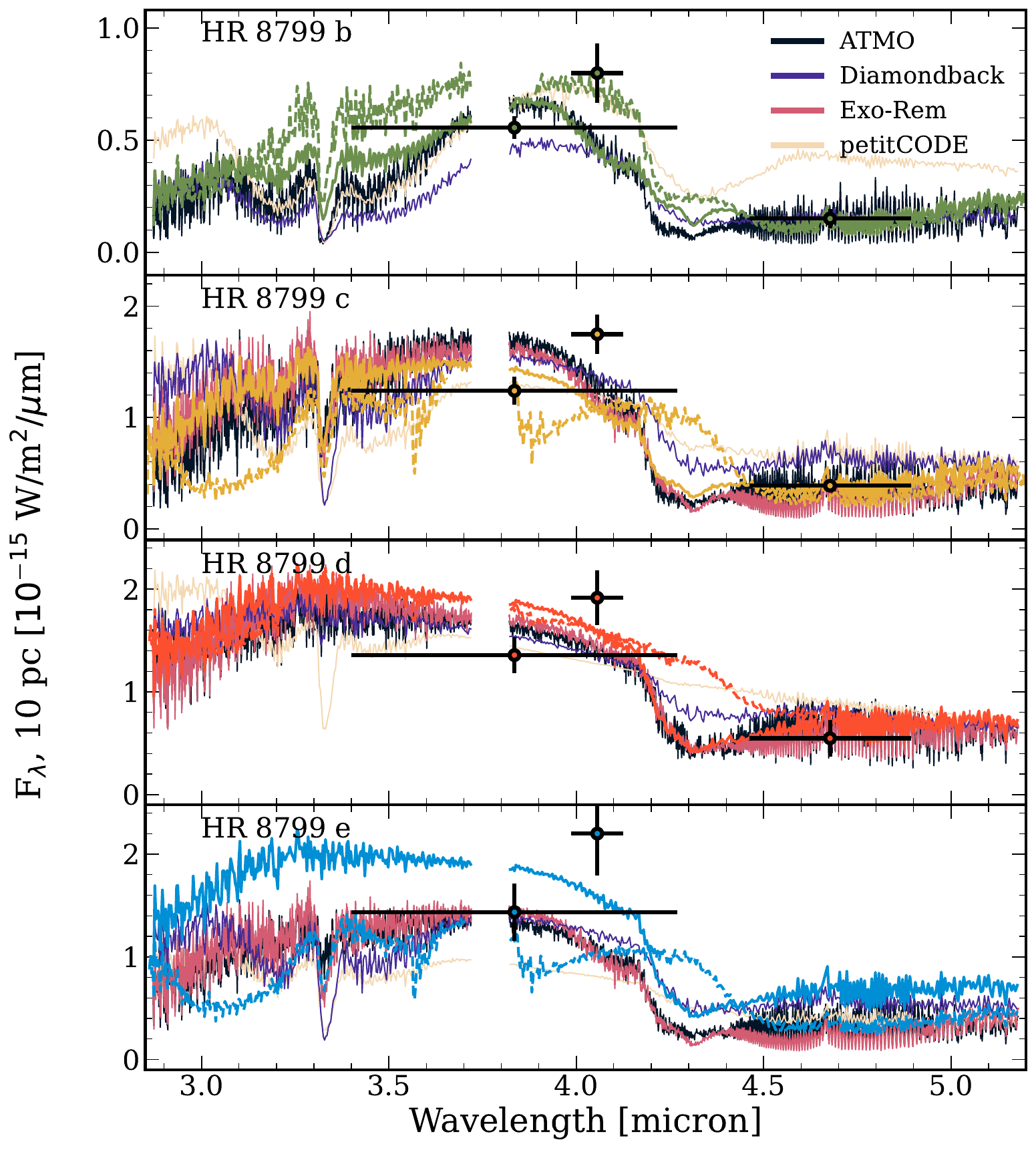}
    \caption{Predictions for NIRSpec/G395H based on most favoured disequilibrium (solid) and free chemistry (dashed) retrievals, together with the best-fit self-consistent models from each grid.}
    \label{fig:g395h_predictions}
\end{figure}

\begin{figure}[t]
    \includegraphics[width=0.99\linewidth]{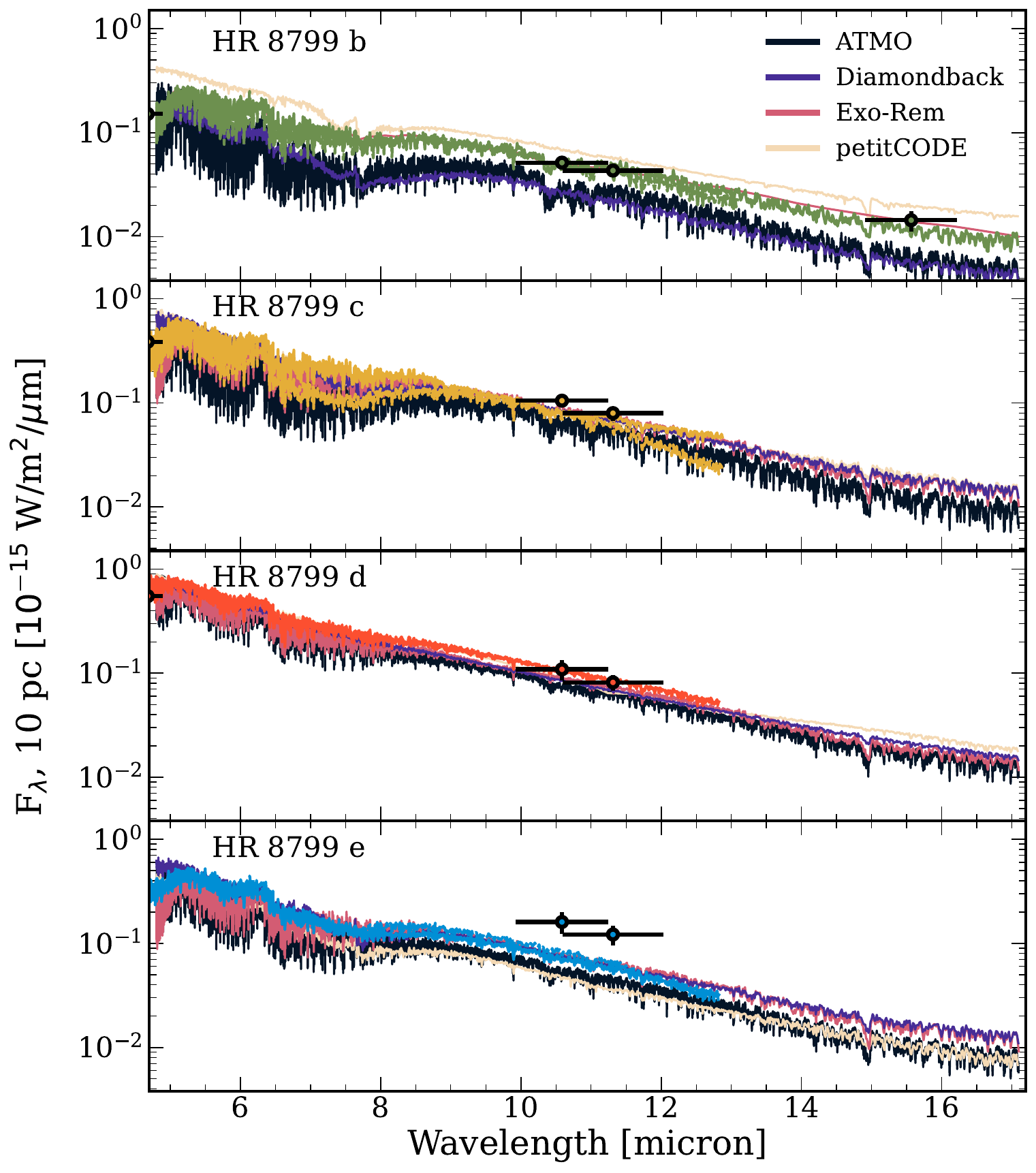}
    \caption{Predictions for MIRI/MRS based on the most favoured disequilibrium (solid) and free chemistry (dashed) retrievals, together with the best-fit self-consistent models from each grid.}
    \label{fig:mrs_predictions}
\end{figure}

All four companions have stellar C/O ratios or higher, and the C/O ratio in this system varies with separation, decreasing from b to d, before a sharp increase in the C/O ratio for the innermost planet.
Super-stellar C/O ratios have been tied to core-accretion formation together with pebble drift and evaporation \citep{schneider_driftingI_2021, schneider_driftingII_2021, molliere_interpreting_2022}. 
A pathway to significant metal enrichment was found by \cite{bitsch_enriching_2023}, though it predicts that the CO rich pebbles should evaporate near the CO iceline, which is outside the radius of even HR~8799~b.
Planetesimal accretion cannot be ruled out either: if large amounts of solids, with near-stellar composition, are accreted, more metal-rich planets will have C/O ratios that approach the stellar value. This is consistent with the trends in metallicity and C/O between the b, c, and d planets, though e remains an exception. 
Such a transition could be explained by the outer three planets, particularly d, trapping water ice and preventing these solids from reaching the innermost planet.
Some combination of these mechanisms could explain both the atmospheric enrichment and the trends in the C/O ratio: for example, early enrichment from evaporating pebbles could lead to the high planetary metallicities, while late accretion of planetesimals could then drive the C/O ratio down towards the solar value.
Alternatively, \cite{chen_gap_2023} demonstrate that the opening of gaps in a protoplanetary disc can significantly alter the composition of the gas and ices available to accrete onto forming planets, and it seems likely that substructure induced by the four HR 8799 planets would strongly impact their eventual composition.

With effective temperatures well over 1000 K and radii significantly larger than that of jupiter, the coldest initial condition scenarios of \cite{marley_luminosity_2007} can be excluded.
Beyond this constraint, the masses, luminosities, and radii seem largely consistent with a broad range of potential evolutionary tracks \citep[e.g.][]{baraffe_evolutionary_2003,saumon_evolution_2008,mordasini_characterization_2017}.
Further work to measure more formation tracers is clearly necessary to unravel this system.
Mid-infrared spectroscopic observations could characterise the NH$_{3}$ abundance, at least in HR~8799~b, as well as the debris disc observed in \citet{boccaletti_imaginghr8799_2023}.
Higher spectral resolution could enable the measurement of carbon isotopes (e.g. as well as place better constraints on the metallicity of each of these atmospheres).
Such measurements, combined with dedicated formation models of giant planets outside the water iceline, are necessary to determine whether these four planets share a formation pathway, or whether there were different mechanisms impacting different regions of the protoplanetary disc.

\subsection{Predictions for JWST}\label{sec:jwst}
The high spectral resolution modes of JWST may allow us to verify the measurements made in this work, particularly through observation of CO$_{2}$, CO, and CH$_{4}$ features in the near infrared and the silicate absorption features near 10~$\upmu$m.
To this end, we present the range of model predictions in these wavelength regions at the spectral resolution of the JWST instruments, for NIRSpec in figure \ref{fig:g395h_predictions} and for the MIRI/MRS in figure \ref{fig:mrs_predictions}.
The comparisons to the MIRI photometry shown in figure \ref{fig:mrs_predictions} demonstrate the typical degree of compatibility between the models and the data in these mid-infrared wavelengths.
In the NIRSpec/G395H wavelength range we see significant discrepancy between models for the same planet in the amplitude of the \co2 feature at 3.8~$\upmu$m, as well as the \ch4 feature at 3.3~$\upmu$m and the CO lines between 4.5~and 5~$\upmu$m. 
These observations will also be able to confirm the presence of HCN in the atmospheres of HR 8799 c and e.
Precise measurement of these features should provide robust constraints on the metallicity of these objects, verifying the degree of enrichment found via the ground-based observations.
While silicate clouds are preferred in the retrieval comparison, none of the models show signs of deep silicate absorption features near 10~$\upmu$m, but spectroscopic observations are required to validate these models.
Mid-infrared observations will be particularly valuable for HR 8799 b, and will allow the clear detection of ammonia. 
If combined with a chemical model to determine the ratios of NH$_{3}$:HCN:N$_{2}$, this will allow for the measurement of the N/O ratio, which can also be used as a formation diagnostic \citep{turrini_tracing_2021, pacetti_chemical_2022}.
Recently \cite{ruffio_highcontrast_2023} demonstrated the potential for high-contrast imaging with NIRSpec; even without the use of a coronagraph it should be possible to obtain flux calibrated, moderate resolution spectroscopy of HR~8799 b, c, and d using the NIRSpec IFU through a combination of forward modelling and reference differential imaging.
In the case where the planet signal is unable to be separated from that of the host star, \cite{patapis_direct_2022} demonstrated that it will be possible to at least identify trace species through molecular mapping in the mid-infrared, though this will be unable to characterise the broad wavelength features of the silicate clouds.

\section{Conclusions}
After more than 15 years of study, the HR~8799 planets remain mysterious, though increasing data quality is allowing us to peer deeper into these atmospheres than ever before.
We present new K-band spectra from the VLTI/GRAVITY, which together with a large set of archival data form the basis of the our atmospheric analysis.
Using {\tt petitRADTRANS} retrievals and fits to self-consistent grids, we inferred the atmospheric properties of all four companions, with reasonable agreement between the two methods.
Our results are broadly consistent with the literature in terms of effective temperature, mass, surface gravity, and radius for all four planets.
The use the dynamical mass as a prior in the retrievals when determining $\log g$ allows us to  reliably retrieve physically reasonable planet radii.

We find that all four planets are strongly enriched in metals, though there is still discrepancy between different models in constraining the precise value.
This was validated by running retrievals using different temperature profiles and chemical models, and comparing to self-consistent grids.
Further self-consistent modelling is necessary, particularly to extend model grids out to high metallicities.
The C/O ratio is stellar to superstellar for all four planets.
It decreases from the outermost planet to HR 8799 d, while HR 8799 e has a higher C/O ratio than the other companions.
We confidently detect HCN in HR 8799 c and e, at abundances far higher than predicted by equilibrium chemistry; though this detection is largely driven by low-$S/N$ data from LBT/ALES.
\ch4 is also confidently detected in HR 8799 c for the first time.
From the disequilibrium chemistry retrievals, H$_{2}$S appears to be a highly abundant species in all of the planets, but higher S/N and spectral resolution are required for a confident detection in a free retrieval framework.
Using our retrieved quench pressure and chemical abundances, we are able to derive a vertical mixing strength, finding \kzz values compatible with high-metallicity predictions from \cite{soni_effect_2023}.
The mixing strength is stronger for the warmer planets, at $\log\kzz\approx6$, and is lower for HR 8799 b with $\log\kzz\approx2$.

All of the planets are highly cloudy.
For the inner three planets, these clouds are composed of silicate clouds lying above the photosphere, and deep, dense iron clouds forming the base of the photosphere.
Cooler than the other three planets, the most favoured model for HR 8799 b requires Na$_{2}$S clouds.
All of the planets have effective temperatures consistent with literature values, with HR~8799~b still unique in its lower temperature and mass compared to its siblings.

We emphasise the use of robust model comparison in this work: while it may be difficult to present precise measurements of certain properties, the use of multiple methods and models allows us to draw a robust portrait of each of these atmospheres.
We also note that our conclusions rely on data with significant incompatibilities, particularly in the H-band flux. 
While we performed extensive analysis to mitigate the influence of any individual dataset, further observations are required to obtain reliable spectroscopic measurements in the near-infrared.
While the HR 8799 planets share many similarities, much like our own Solar System there are differences in their atmospheric properties, which require further study.

\begin{acknowledgements}
We would like to thank Quinn Konopacky, Alice Zurlo, Alex Greenbaum, Beth Biller, Olivier Flasseur, David Doelman, and Pengyu Lui for providing the archival datasets used in this work.
We are also grateful to Gilles Loupe and Malavika Vasist for providing helpful suggestions on how to implement the spline temperature profile.
Thanks as well to Daniel Thorngren for providing the model for calculating the planet metallicity $Z$.
Finally, we are grateful to our anonymous reviewer for their thorough and insightful report.

Software used: {\tt petitRADTRANS, pyKLIP, species, VIP-HCI, pyMultiNest, phot\_utils, Python, numpy, matplotlib, astropy, sympy, Aspro}.

SL acknowledges the support of the French Agence Nationale de la Recherche (ANR), under grant ANR-21-CE31-0017 (project ExoVLTI).
J.J.W., A.C., and S.B.\ acknowledge the support of NASA XRP award 80NSSC23K0280.
G.-D.M.\ acknowledges the support of the DFG priority program SPP 1992 ``Exploring the Diversity of Extrasolar Planets'' (MA~9185/1) amd from the Swiss National Science Foundation under grant 200021\_204847 ``PlanetsInTime''.
Parts of this work have been carried out within the framework of the NCCR PlanetS supported by the Swiss National Science Foundation.
This work is based on observations collected at the European Southern Observatory under ESO programme 1104.C-0651.
This publication makes use of VOSA, developed under the Spanish Virtual Observatory project supported by the Spanish MINECO through grant AyA2017-84089.
VOSA has been partially updated by using funding from the European Union's Horizon 2020 Research and Innovation Programme, under Grant Agreement nº 776403 (EXOPLANETS-A)
\end{acknowledgements}
%
%
\bibliographystyle{aa}  
\bibliography{hr8799_gravity}

\begin{appendix}
\section{Data logs}\label{ap:photometry}
\begin{table}[h]
\centering
\begin{threeparttable}
    \begin{small}
    \caption{Near infrared stellar Photometry of HR8799, using apparent flux normalised to 10~pc, retrieved from the Spanish Virtual Observatory \citep{bayo_vosa_2008}.}
    \label{tab:stellarphotometry}
    \begin{tabular}{rlll}
        \toprule
        \textbf{Filter}  & $\bm{\lambda}$  & \textbf{Flux} & \textbf{Ref.}\\
         & [$\upmu$m] & [erg/s/cm$^{2}$/\r{A}] & \\
        \midrule
        2MASS J & 1.235 & $2.198\pm 0.055\times 10^{-12}$ &S06\\
        2MASS H & 1.662 & $8.754\pm 0.145\times 10^{-13}$&S06\\
        2MASS Ks & 2.159 & $3.433\pm 0.057\times 10^{-13}$ &S06\\
        WISE W1 & 3.353 & $6.840\pm 1.367\times 10^{-14}$ & W10\\
        WISE W2 & 4.603 & $2.316\pm 0.166\times 10^{-14}$ & W10\\
        \bottomrule
    \end{tabular}
    \begin{tablenotes}
    \small
    \item\textbf{Notes}
    \item References: 
    S06 \cite{skrutskie_2mass_2006}; 
    W10 \cite{wright_wise_2010}; 
    \end{tablenotes}
    \end{small}
\end{threeparttable}
\end{table} 
\begin{table}
\centering
\begin{threeparttable}
    \begin{small}
    \caption{Photometric data for HR 8799 bcde.}
    \label{tab:phot_data}
    \begin{tabular}{lll}
        \toprule
        \textbf{Instrument/Filter} & \textbf{m} & \textbf{Ref.}\\
        \midrule
        \multicolumn{3}{c}{HR 8799 b}\\
        \midrule
            Keck/NIRC2.H & $18.05\pm 0.09$ & C12\\
            Keck/NIRC2.Ks & $17.03\pm 0.08$ & M10\\
            Keck/NIRC2.Ms & $16.05\pm 0.3$ & G11\\
            Paranal/NACO.Lp & $15.52\pm 0.1$ & C14\\
            Paranal/NACO.NB405 & $14.82\pm 0.18$ & C14\\
            Paranal/SPHERE/IRDIS.B\_J & $19.78\pm 0.09$ & Z16\\
            Paranal/SPHERE/IRDIS.D\_H23\_2 & $18.08\pm 0.14$ & Z16\\
            Paranal/SPHERE/IRDIS.D\_H23\_3 & $17.78\pm 0.1$ & Z16\\
            Paranal/SPHERE/IRDIS.D\_K12\_1 & $17.15\pm 0.06$ & Z16\\
            Paranal/SPHERE/IRDIS.D\_K12\_2 & $16.97\pm 0.09$ & Z16\\
            Subaru/CIAO.z & $21.22\pm 0.29$ & C11\\
            JWST/MIRI.F1065C & $13.54\pm0.04$ & B23\\
            JWST/MIRI.F1140C & $13.64\pm0.07$ & B23\\
            JWST/MIRI.F1550C & $13.49\pm0.25$& B23\\
        \midrule
        \multicolumn{3}{c}{HR 8799 c}\\
        \midrule
            Keck/NIRC2.H & $17.06\pm 0.13$ & C12\\
            Keck/NIRC2.Ks & $16.11\pm 0.08$ & M10\\
            Keck/NIRC2.Ms & $15.03\pm 0.14$ & G11\\
            Paranal/NACO.Lp & $14.65\pm 0.11$ & C14\\
            Paranal/NACO.NB405 & $13.97\pm 0.11$ & C14\\
            Paranal/SPHERE/IRDIS.B\_J & $18.6\pm 0.13$ & Z16\\
            Paranal/SPHERE/IRDIS.D\_H23\_2 & $17.09\pm 0.12$ & Z16\\
            Paranal/SPHERE/IRDIS.D\_H23\_3 & $16.78\pm 0.1$ & Z16\\
            Paranal/SPHERE/IRDIS.D\_K12\_1 & $16.19\pm 0.05$ & Z16\\
            Paranal/SPHERE/IRDIS.D\_K12\_2 & $15.86\pm 0.07$ & Z16\\
            JWST/MIRI.F1065C & $12.97\pm0.18$ & B23\\
            JWST/MIRI.F1140C & $13.59\pm0.26$ & B23\\
            JWST/MIRI.F1550C & $11.88\pm0.23$ & B23\\
        \midrule
        \multicolumn{3}{c}{HR 8799 d}\\
        \midrule
            Keck/NIRC2.H & $16.71\pm 0.24$ & C12\\
            Keck/NIRC2.Ks & $16.09\pm 0.12$ & M10\\
            Keck/NIRC2.Ms & $14.65\pm 0.35$ & G11\\
            Paranal/NACO.Lp & $14.55\pm 0.14$ & C14\\
            Paranal/NACO.NB405 & $13.87\pm 0.15$ & C14\\
            Paranal/SPHERE/IRDIS.B\_J & $18.59\pm 0.37$ & Z16\\
            Paranal/SPHERE/IRDIS.D\_H23\_2 & $17.02\pm 0.17$ & Z16\\
            Paranal/SPHERE/IRDIS.D\_H23\_3 & $16.85\pm 0.16$ & Z16\\
            Paranal/SPHERE/IRDIS.D\_K12\_1 & $16.2\pm 0.07$ & Z16\\
            Paranal/SPHERE/IRDIS.D\_K12\_2 & $15.84\pm 0.1$ & Z16\\
            JWST/MIRI.F1065C & $12.98\pm0.14$& B23\\
            JWST/MIRI.F1140C & $12.98\pm0.17$& B23\\
            JWST/MIRI.F1550C & $11.88\pm0.23$& B23\\
        \midrule
        \multicolumn{3}{c}{HR 8799 e}\\
        \midrule
            Keck/NIRC2.Ks & $15.91\pm 0.22$ & C12\\
            Keck/NACO.Lp & $14.49\pm 0.21$ & M10\\
            Keck/NACO.NB405 & $13.72\pm 0.2$ & C14\\
            Paranal/SPHERE/IRDIS.B\_J & $18.4\pm 0.21$ & Z16\\
            Paranal/SPHERE/IRDIS.D\_H23\_2 & $16.91\pm 0.2$ & Z16\\
            Paranal/SPHERE/IRDIS.D\_H23\_3 & $16.68\pm 0.21$ & Z16\\
            Paranal/SPHERE/IRDIS.D\_K12\_1 & $16.12\pm 0.1$ & Z16\\
            Paranal/SPHERE/IRDIS.D\_K12\_2 & $15.82\pm 0.11$ & Z16\\
            JWST/MIRI.F1065C & $12.52\pm0.26$& B23\\
            JWST/MIRI.F1140C & $12.52\pm0.23$& B23\\
            JWST/MIRI.F1550C & $11.01\pm0.42$& B23\\
        \bottomrule
    \end{tabular}
    \begin{tablenotes}
    \item\textbf{Notes:}
    \item References: C12 \cite{currie_direct_2012}; M10 \cite{marois_images_2010}; G11 \cite{galicher_m-band_2011}; C14 \cite{currie_deep_2014}; Z16 \cite{zurlo_first_2016}; C11 \cite{currie_combined_2011}; 
    B23 \cite{boccaletti_imaginghr8799_2023}.
    \end{tablenotes}
    \end{small}
\end{threeparttable}
\end{table}

\begin{table*}[t]
\centering
\begin{threeparttable}
    \centering
    \begin{small}
    \caption{Spectroscopic Observation Log.}
    \begin{tabular}{l|llllllllll}
        \toprule
        \textbf{Planet} & \textbf{Instrument} & \textbf{Date} & $\boldsymbol{\lambda}$  & $\boldsymbol{\lambda/\Delta\lambda}$ & $\mathbf{\Delta}$PA &  \textbf{Seeing} & \textbf{Airmass} & \textbf{DIT} & \textbf{NEXP} & \textbf{Ref.}\\
        & & & & [$\mu$m] & [\textdegree] & [as], Med. & Min. & [s] & & \\
        \midrule
        b & GRAVITY & 2019-11-11 & 2.0$-$2.4 & 500 & \ldots& 0.98 & 1.54 & 8x100 & 3 & \ldots\\      
          & GRAVITY & 2021-08-26 & 2.0$-$2.4 & 500 & \ldots& 0.88 & 1.48 & 8x100 & 3 & \ldots\\
          & GRAVITY & 2021-08-27 & 2.0$-$2.4 & 500 & \ldots& 0.86 & 1.45 & 8x100 & 2 & \ldots\\      
          & OSIRIS  & 2009-07-22 & 2.0$-$2.4 & 60 &\ldots & \ldots& 1.0 & 900 & 30 & B11 \\
          & OSIRIS  & 2009-07-23 & 1.5$-$1.8 & 60 & \ldots&\ldots & 1.0 & 900 & 30 & B11\\
          & OSIRIS  & 2009-07-30 & 1.5$-$1.8 & 60 &\ldots & \ldots& 1.0 & 900 & 30 & B11\\
          & OSIRIS  & 2010-07-11 & 2.0$-$2.4 & 60 &\ldots &\ldots & 1.0 & 900 & 30 & B11 \\
          & OSIRIS  & 2010-07-13 & 1.5$-$1.8 & 60 &\ldots &\ldots & 1.0 & 900 & 30 & B11\\ 
        \midrule
        c & GRAVITY & 2019-11-11 & 2.0$-$2.4 & 500 &\ldots & 1.01 & 1.62 & 8x100 & 3 & \ldots\\       
          & GRAVITY & 2021-08-26 & 2.0$-$2.4 & 500 & \ldots& 1.04 & 1.50 & 8x100 & 3 & \ldots\\          
          & GRAVITY & 2021-08-27 & 2.0$-$2.4 & 500 &\ldots & 1.00 & 1.45 & 8x100 & 2 & \ldots\\         
          & GRAVITY & 2022-08-19 & 2.0$-$2.4 & 500 &\ldots & 0.64 & 1.54 & 4x100 & 10 & \ldots\\         
          & GRAVITY & 2023-07-02 & 2.0$-$2.4 & 500 &\ldots & 0.75 & 1.45 & 4x100 & 6 & \ldots\\        
          & SPHERE  & 2015-07-04 & 0.9$-$1.6  & 30  & 16.4 & 1.43 & 1.44 & \ldots& 46 & F20\\
          & CHARIS & 2018-09-01 & 1.2$-$2.4 & 19 & 202.24 & 0.47 & 1.0 & 20 & 1201 & W22\\
          & CHARIS & 2018-09-02 & 1.2$-$2.4 & 19 & 206.55 & 0.42 & 1.0 & 20 & 1253 & W22\\
          & GPI & 2016-09-19 & 1.5$-$1.8 & 45 & 20.93 & 0.97 & 1.61 & 60 & 60 & G18 \\
          & ALES & 2019-09-18 & 2.8$-$4.2 & 35 & 85.64 & 0.8-1.1 & 1.02 & 3.934 & 1300 & D22\\
          & OSIRIS & 2010-2011 & 1.97$-$2.38 & 4000 &\ldots &\ldots &\ldots & 600 & 33 & K13\\
        \midrule
        d & GRAVITY & 2019-11-09 & 2.0$-$2.4 & 500 &\ldots & 0.85 & 1.63 & 8x60  & 4 & \ldots\\
          & GRAVITY & 2019-11-11 & 2.0$-$2.4 & 500 &\ldots & 1.14 & 1.70 & 8x100 & 3 & \ldots\\
          & GRAVITY & 2021-08-26 & 2.0$-$2.4 & 500 &\ldots & 0.96 & 1.54 & 8x100 & 3 & \ldots\\
          & GRAVITY & 2021-08-27 & 2.0$-$2.4 & 500 &\ldots & 1.18 & 1.48 & 8x100 & 2 & \ldots\\
          & GRAVITY & 2022-09-15 & 2.0$-$2.4 & 500 &\ldots & 0.67 & 1.54 & 4x100 & 12 & \ldots\\        
          & SPHERE & 2014-08-12 & 0.9$-$1.6 & 30 & 29.65 & 0.87 & 1.43 & 100 & 32  & Z16\\
          & SPHERE & 2014-08-12 & 0.9$-$1.6 & 30 & 15.37 & 0.87 & 1.43 & 60  & 48  & Z16\\
          & CHARIS & 2018-09-01 & 1.2$-$2.4 & 19 & 202.24 & 0.47 & 1.0 & 20 & 1201 & W22\\
          & CHARIS & 2018-09-02 & 1.2$-$2.4 & 19 & 206.55 & 0.42 & 1.0 & 20 & 1253 & W22\\
          & GPI & 2016-09-19 & 1.5$-$1.8 & 45 & 20.93 & 0.97 &1.61 & 60 & 60 & G18 \\
          & ALES & 2019-09-18 & 2.8$-$4.2 & 35 & 85.64 & 0.8-1.1 & 1.02 & 3.934 & 1300 & D22\\
        \midrule
        e & GRAVITY & 2018-08-28 & 2.0$-$2.4 & 500 &\ldots & 0.67 & 1.44 & 10x100 & 7 & G19\\ 
          & GRAVITY & 2019-11-09 & 2.0$-$2.4 & 500 &\ldots & 0.84 & 1.55 & 8x60   & 3 & M20\\ 
          & GRAVITY & 2019-11-11 & 2.0$-$2.4 & 500 &\ldots & 1.15 & 1.47 & 8x100  & 3 & M20\\ 
          & GRAVITY & 2021-08-26 & 2.0$-$2.4 & 500 &\ldots & 0.66 & 1.50 & 8x100  & 2 & \ldots\\
          & GRAVITY & 2021-08-27 & 2.0$-$2.4 & 500 &\ldots & 0.67 & 1.47 & 8x100  & 2 & \ldots\\
          & GRAVITY & 2021-09-27 & 2.0$-$2.4 & 500 &\ldots & 0.82 & 1.55 & 4x100  & 6 & \ldots\\ 
          & GRAVITY & 2023-07-02 & 2.0$-$2.4 & 500 &\ldots & 0.75 & 1.46 & 4x100  & 6 & \ldots\\     
          & SPHERE & 2014-08-12 & 0.9$-$1.6 & 30 & 29.65 & 0.87 & 1.43 & 100 & 32  & Z16\\
          & SPHERE & 2014-08-12 & 0.9$-$1.6 & 45  & 15.37 & 0.87 & 1.43 & 60 & 48  & Z16\\
          & CHARIS & 2018-09-01 & 1.2$-$2.4 & 19 & 202.24 & 0.47 & 1.0 & 20 & 1201 & W22\\
          & CHARIS & 2018-09-02 & 1.2$-$2.4 & 19 & 206.55 & 0.42 & 1.0 & 20 & 1253 & W22\\
          & GPI & 2016-09-19 & 1.5$-$1.8 & 45 & 20.93 & 0.97 & 1.61 & 60 & 60 & G18 \\
          & ALES & 2019-09-18 & 2.8$-$4.2 & 35 & 85.64 & 0.8-1.1 & 1.02 & 3.934 & 1300 & D22\\
        \bottomrule        
    \end{tabular}
    \begin{tablenotes}
    \small
    \item\textbf{Notes}
    \item References: 
    B11: \cite{barman_clouds_2011}; 
    K13: \cite{konopacky_detection_2013}; 
    Z16: \cite{zurlo_first_2016}; 
    G18: \cite{greenbaum_gpi_2018}; 
    G19: \cite{gravity_collaboration_first_2019}; F20: \cite{flasseur_paco_2020}; 
    M20: \cite{molliere_retrieving_2020}; 
    W22: \cite{wang_atmospheric_2022_CHARIS};
    D22: \cite{doelman_l-band_2022}.
    \label{tab:data}
    \end{tablenotes}
    \end{small}
    \end{threeparttable}
\end{table*}
\section{Reprocessing SPHERE and GPI datasets}\label{ap:algocomp}
\begin{figure}[ht]
    \centering
    \includegraphics[width =\linewidth]{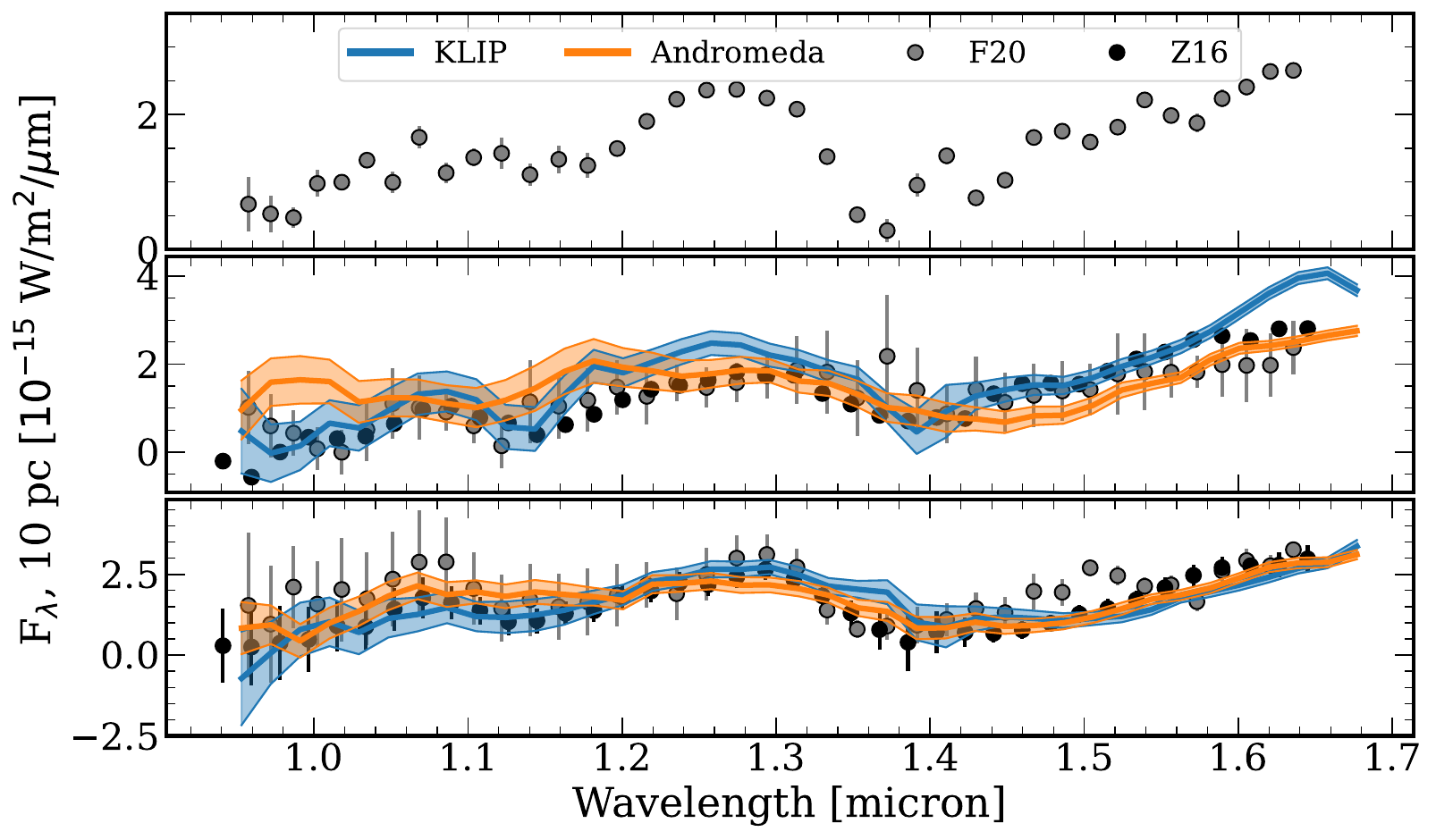}
    \caption{{\tt KLIP} and {\tt ANDROMEDA} extractions from SPHERE for HR 8799 c, d, and e compared to the spectra published in \cite{zurlo_first_2016} 
 and \cite{flasseur_exoplanet_2018}. 
 }
    \label{fig:sphereextract}
\end{figure}

\begin{figure}[ht]
    \includegraphics[width =\linewidth]{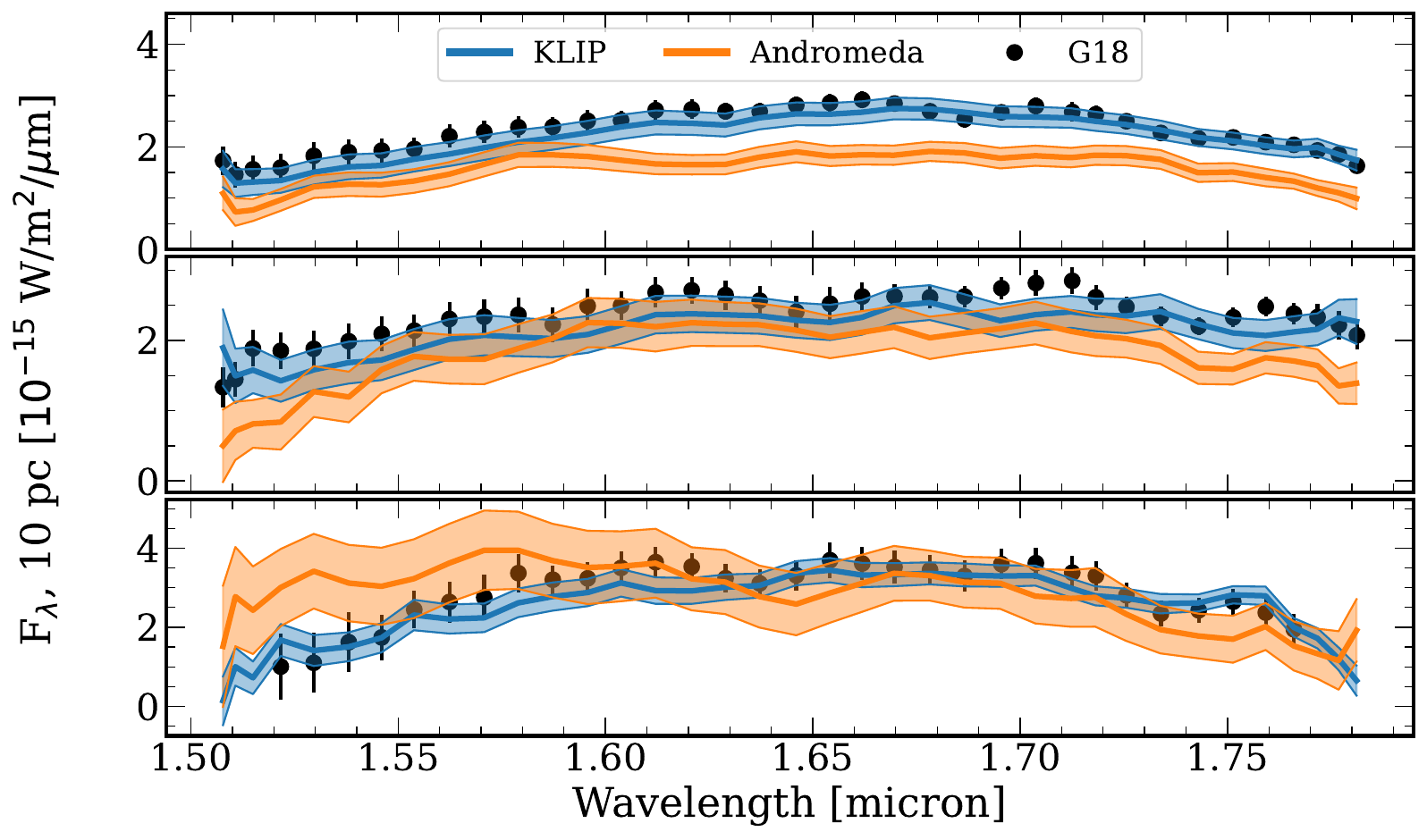}
    \caption{{\tt KLIP} and {\tt ANDROMEDA} extractions from GPI for HR 8799 c, d, and e compared to the spectra published in \cite{greenbaum_gpi_2018} }
    \label{fig:gpiextract}
\end{figure}

To resolve the known discrepancies in the H-band flux between the archival SPHERE and GPI datasets, we reprocessed each using {\tt KLIP} \citep{soummer_detection_2012,pueyo_detection_2016}, {\tt ANDROMEDA} \citep{cantalloube_direct_2015} and {\tt PynPoint} \citep{amara_pynpoint_2012,stolker_pynpoint_2019}.
We optimised the choice of algorithm parameters through a series of injection/extraction tests into each dataset, as in \cite{nasedkin_impacts_2023}.
Using two different goodness-of-fit metrics on injections representative of the true companion contrast and separation, we choose the number of principal components used in the PSF subtraction in order to extract the companion spectra with minimal bias.
Figures \ref{fig:sphereextract} and \ref{fig:gpiextract} shows the extracted spectra for each of HR 8799 c, d, and e, for SPHERE and GPI respectively, compared to published literature spectra from \cite{zurlo_first_2016}, \cite{greenbaum_gpi_2018}, and \cite{flasseur_exoplanet_2018}. 
As the goodness-of-fit metrics favoured the {\tt KLIP} extractions, we used these as the basis of our retrieval analysis.




\section{Retrieval validation}\label{ap:validation}
Extensive validation of the {\tt pRT} retrieval module was performed as part of this work. 
Following updates described in \cite{nasedkin_atmospheric_2024}, we verified that the results of \cite{molliere_retrieving_2020} could be reproduced. We independently tested updates to the c-k mixing implementation, the adaptive mesh refinement implementation, updated opacity sources for \h2o and CO, bug fixes for convergence on multiple scattering in the clouds, the inclusion of photometric data, the inclusion of scaling factors on the SPHERE and GPI datasets, including or excluding the GPI K-band spectra, updates to each of the SPHERE, GPI, and GRAVITY datasets, different prior widths and the number of live points used in the retrieval.
All of the posterior distributions were fully consistent to within 2$\sigma$ with most falling well within 1$\sigma$ of the published results, apart from the inclusion of new epochs of GRAVITY data, which led to a significantly higher retrieved metallicity ([M/H]=$1.1\pm0.32$) and an \fsed of $5\pm2.6$.

We verified several model assumptions through retrievals that only include the GRAVITY datasets, or the GRAVITY data and photometry.
We find that the GRAVITY data alone could not distinguish between clear and cloudy models ($\Delta\log_{10}\mathcal{Z} < 1$), while cloudy models were strongly favoured once the broad wavelength coverage of the photometry was included  ($\Delta\log_{10}\mathcal{Z} > 10$).

Models that use the dynamical mass estimates as priors for calculating the surface gravity were marginally favoured over those that freely retrieve $\log g$ and $R_{\rm pl}$, but that the posteriors parameter distributions were generally consistent, with \Tint and $R_{\rm pl}$ showing the greatest discrepancy.
Using the dynamical mass as a prior and setting a Gaussian prior on the radius led to more reasonable estimates of the radius of HR 8799 e ($0.97\pm0.04$) compared to the free retrieval ($0.79\pm0.05$). 
However, the composition of the planet and the degree of cloudiness did not vary significantly between the two models.

Using the full dataset for HR 8799 b, we verified that retrievals including scattering clouds are strongly favoured over those without scattering ($\Delta\log_{10}\mathcal{Z} > 10$). 
Without scattering, both the temperature and composition ([M/H] and C/O) are significantly discrepant from retrievals that include scattering clouds.

\begin{figure*}
    \centering
    \includegraphics[width=\linewidth]{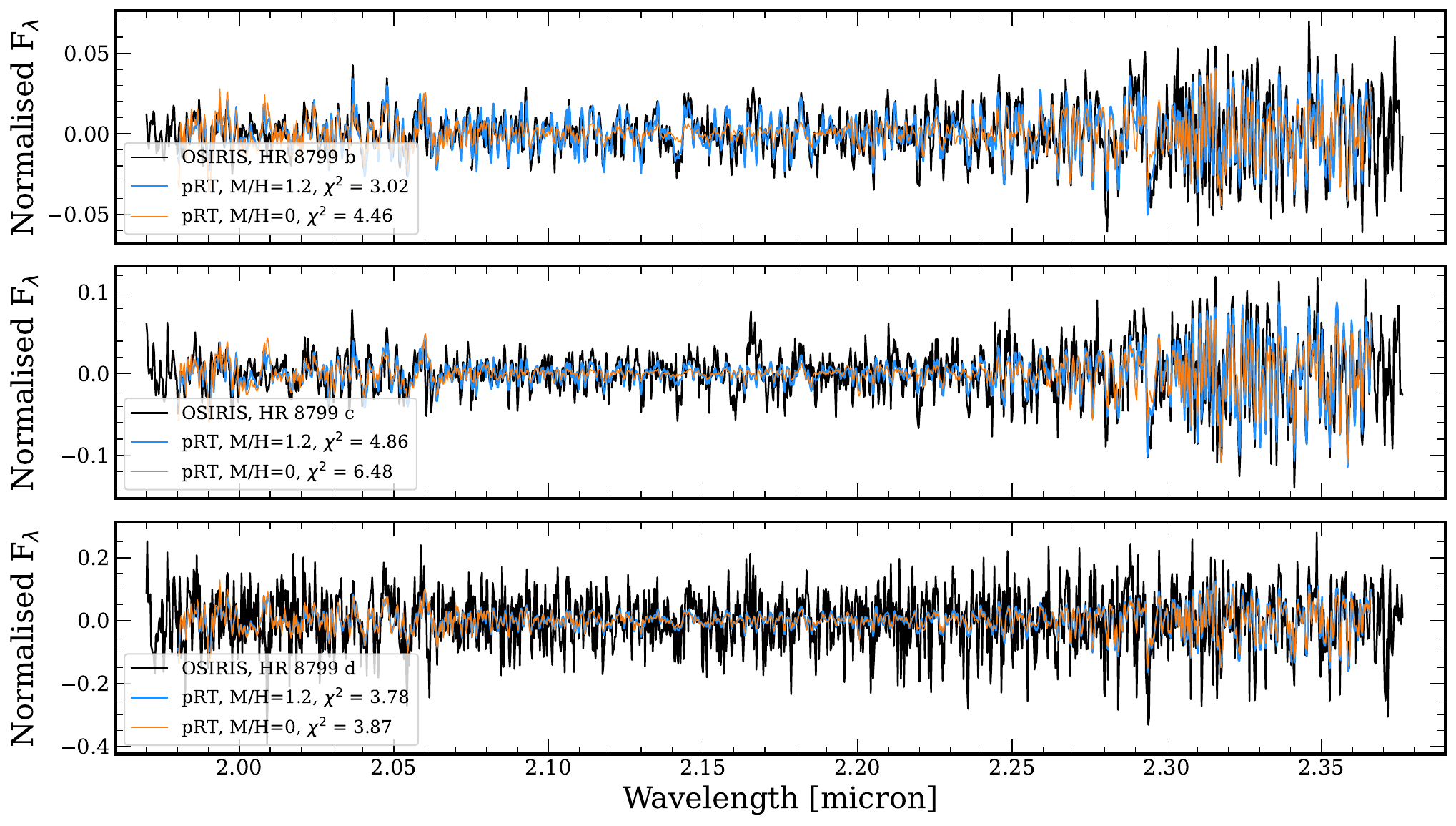}
    \caption{Comparison of best-fit disequilibrium models to OSIRIS data from \cite{ruffio_deep_2021}. From top to bottom is HR 8799 b, c, and d. In blue are the best-fit disequilibrium models, with the spectra generated using high-resolution line-by-line opacities, before being convolved, binned, and normalised for comparison. In orange is the same, but with the metallicity set to 0.}
    \label{fig:OSIRIS_Comp}
\end{figure*}
\begin{figure}
    \centering
    \includegraphics[width=\linewidth]{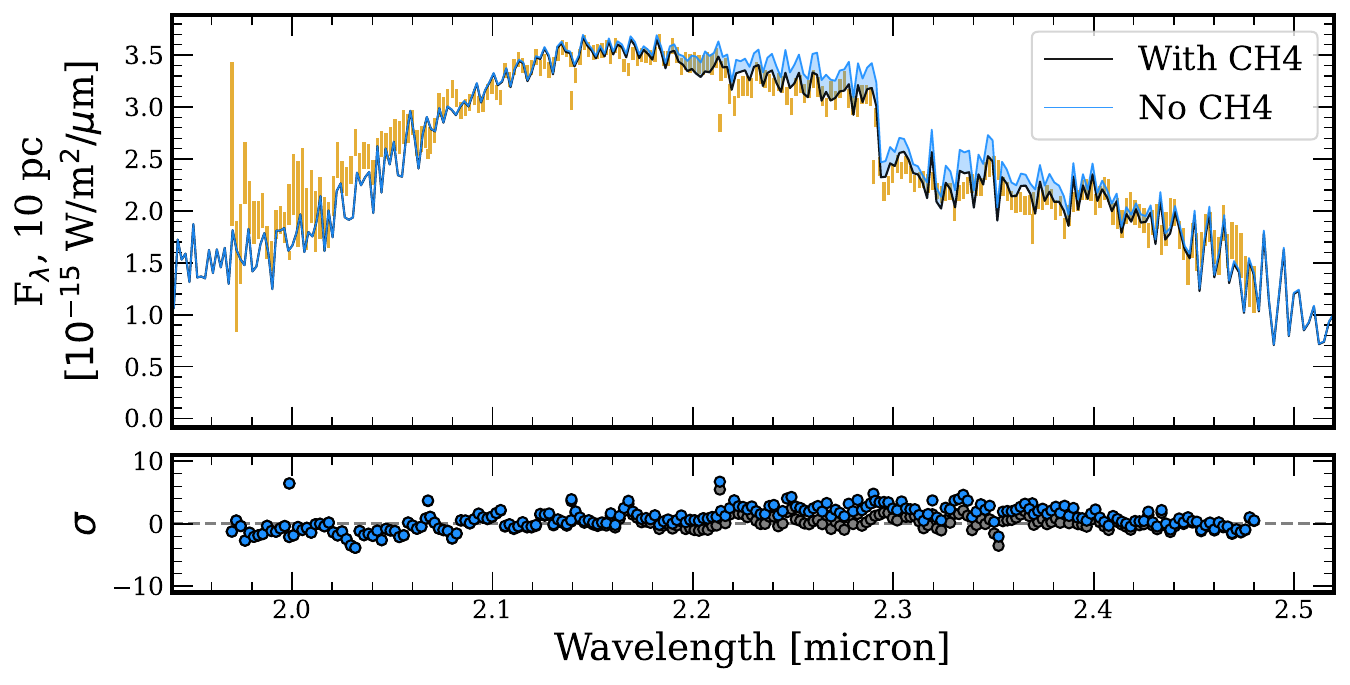}
    \caption{Comparison of best-fit disequilibrium models (black) of HR 8799 c to the data, with residuals shown in the bottom panel. In blue are the same spectra, but without opacity contributions from \ch4 .}
    \label{fig:CH4Detection}
\end{figure}
\begin{figure}
    \centering
    \includegraphics[width=\linewidth]{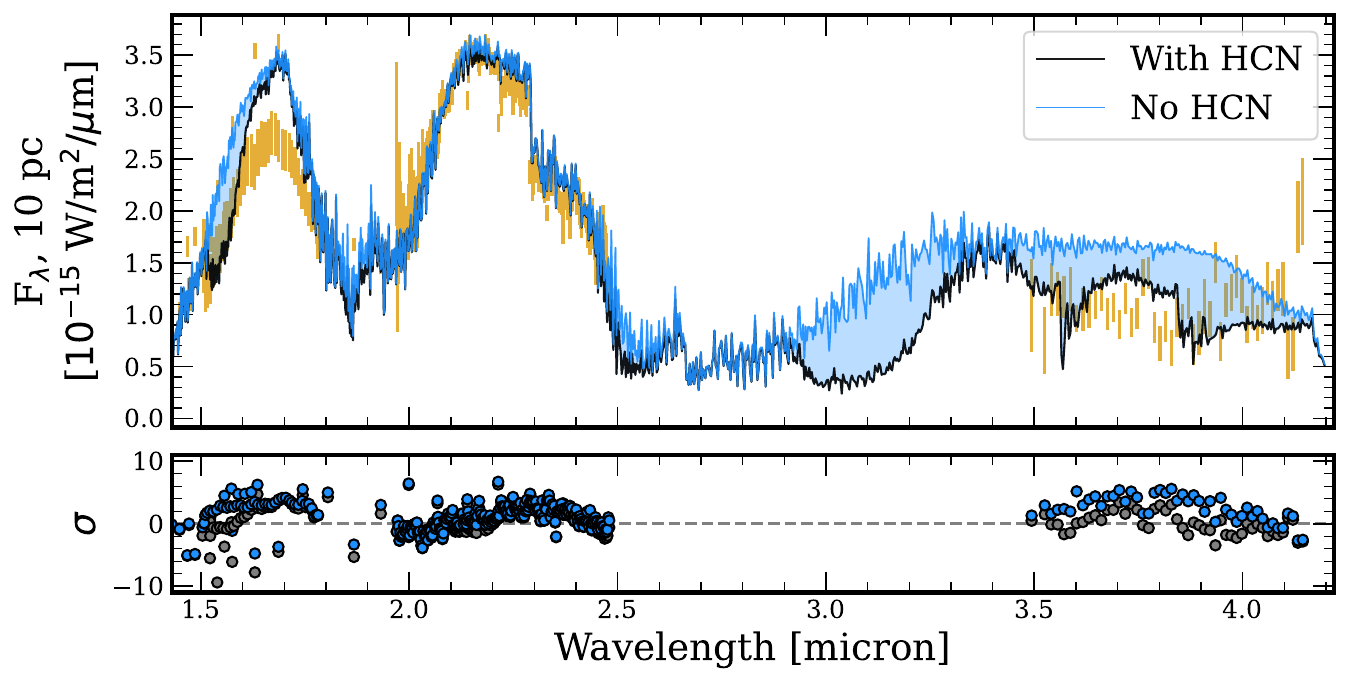}
    \caption{Comparison of best-fit disequilibrium models (black) of HR 8799 c to the data, with residuals shown in the bottom panel. In blue are the same spectra, but without opacity contributions from HCN.}
    \label{fig:HCNDetection}
\end{figure}

\section{Using the Hansen distribution with EDDYSED}\label{ap:hansen}
The \ed cloud model from \citet{ackerman_precipitating_2001} is implemented in {\tt pRT}, and is the most physically motivated model incorporated to date.
Typically, it assumes a log-normal particle size distribution, where the geometric particle radius will vary throughout the atmosphere as a function of the vertical diffusion coefficient \kzz and the sedimentation fraction \fsed.
Here, we substitute the log-normal particle size distribution with the Hansen distribution, originally introduced in \cite{hansen_multiple_1971}, and rederive the calculation for the particle radius as a function of \kzz and \fsed.

We begin with a review of the \ed model: the distribution of the number of particles as a function of particle radius, $n(r)$ is approximated as a log-normal distribution with width $\sigma_{g}$ and characteristic geometric radius $r_{g}$.
\begin{equation}
    n(r) = \frac{N}{r\sqrt{2\pi}\log\sigma_{g}}\exp\left(-\frac{\log^{2}\left(r/r_{g}\right)}{2\log^{2}\sigma_{g}}\right),
\end{equation}
$N$ is the total number of cloud particles.

The goal of the \ed model is to calculate $r_{g}$ for each layer in the atmosphere, given \kzz and \fsed. 
It balances the upwards vertical mixing, parameterised by \kzz and the particle settling velocity, $v_{f}$
\begin{equation}\label{eqn:vf}
    v_{f} = w_{*}\left(\frac{r}{r_{w}}\right)^{\alpha}.
\end{equation}
Here $w_{*}$ is the convective velocity scale. We note that $r_{w}\neq r_{g}$. $r_{w}$ is the radius at which the  particle settling velocity equals the convective velocity scale:
\begin{equation}
    w_{*} = \frac{\kzz}{L},
\end{equation}
where $L$ is the convective mixing length.
Since $w_{*}$ is known, and $v_{f}$ can be found analytically as in \citet{ackerman_precipitating_2001,podolak_contribution_2003}, a linear fit can be used to find both $\alpha$ and $r_{w}$.

With both of these quantities known, we follow \citetalias{ackerman_precipitating_2001} and define \fsed as:
\begin{equation}\label{eqn:fsed}
    \fsed = \frac{\int_{0}^{\infty}r^{3+\alpha}n(r)dr}{r_{w}^{\alpha}\int_{0}^{\infty}r^{3}n(r)dr}
\end{equation}
For the log-normal distribution, one finds:
\begin{equation}
    \int_{0}^{\infty}r^{\beta}n(r)dr = Nr_{g}^{\beta}\exp\left(\frac{1}{2}\beta^{2}\log^{2}\sigma_{g}\right)
\end{equation}
Which we can then use to solve for $r_{g}$:
\begin{equation}
    r_{g} = r_{w}\fsed^{1/\alpha}\exp\left(-\frac{\alpha + 6}{2}\log^{2}\sigma_{g}\right)
\end{equation}

In order to use the Hansen distribution, we must recalculate the total number of particles $N$, and integrate the distribution for \fsed. 
We note here that the Hansen distribution is parameterised by the effective radius, $\bar{r}$, rather than the geometric mean radius. 
In this derivation we do not correct for this difference in definition, as both act as nuisance parameters in the context of an atmospheric retrieval.

We start by giving the Hansen distribution in full:
\begin{equation}
    n_{\rm Hansen}(r) = \frac{N \left(\bar{r}v_{e}\right)^{\left(2v_{e}-1\right)/v_{e}}}{\Gamma\left[\left(1-2v_{e}\right)/v_{e}\right]} r^{(1-3v_{e})/v_{e}}\exp\left(-\frac{r}{\bar{r}v_{e}}\right)
\end{equation}
In \citet{hansen_multiple_1971}, the authors use the parameters $a$ and $b$ to denote the mean effective radius and effective variance, which we write as $\bar{r}$ and $v_{e}$ respectively.
These differ from the simple mean radius and variance by weighting them by the particle area, as the cloud particle scatters an amount of light proportional to its area. Thus:
\begin{equation}
    \bar{r} = \frac{\int_{0}^{\infty}r\pi r^{2}n(r)dr}{\int_{0}^{\infty}\pi r^{2}n(r)dr}
\end{equation}
and 
\begin{equation}
    v_{e} = \frac{\int_{0}^{\infty} \left(r-\bar{r}\right)^{2} r^{2}n(r)dr}{\bar{r}^{2}\int_{0}^{\infty}\pi r^{2}n(r)dr}
\end{equation}

As in \ed, we fit for the settling velocity, which will provide us with $\alpha$ and $r_{w}$, which we can use to find \fsed, as in \ref{eqn:fsed}.
However, we must now integrate the Hansen distribution. We find that:
\begin{equation}\label{eqn:hansint}
    \int_{0}^{\infty}r^\beta n_{\rm Hansen}(r)dr = \frac{v_{e}^{\beta} \left(v_{e}\beta + 2v_{e} + 1\right) \left(\frac{1}{\bar{r}}\right)^{-\beta} \Gamma\left(\beta + 1 + \frac{1}{v_{e}}\right)}{\left(-v_{e} + v_{e}^{\beta + 3} + 1\right) \Gamma\left(1 + \frac{1}{v_{e}}\right)}
\end{equation}
We can then use Eqns. \ref{eqn:fsed} and \ref{eqn:hansint} to solve for $\bar{r}$:
\begin{equation}
    \bar{r} = \left(\frac{ \fsed r_{w}^{\alpha}v_{e}^{-\alpha} \left(v_{e}^{3+\alpha} - v_{e} + 1\right) \Gamma\left(1 + \frac{1}{v_{e}}\right)}{\left(v_{e}\alpha + 2v_{e} + 1\right) \Gamma\left(\alpha + 1 + \frac{1}{v_{e}}\right)}\right)^{\frac{1}{\alpha}}.
\end{equation}
Thus for a given \kzz, \fsed, and $v_{e}$, we can find the effective particle radius for every layer in the atmosphere.

However, in order to compute the cloud opacity, we still require the total particle count. 
For a volume mixing ratio of a given species, $\chi_{i}$, we can integrate $n(r)$ to find $N$:
\begin{equation}
 N = \frac{\chi_{i}}{\left(\bar{r}^{3}v_{e} -1\right)\left(2v_{e} -1\right)}
\end{equation}

\section{Complete retrieval results}\label{ap:retrievalresults}
We include in the text abridged tables that present key parameters of interest.
The complete set of inferred parameters for every retrieval is available on Zenodo.

Legend: Chemistry/Profile/Clouds/Data/Info.  
\begin{itemize}
\item Chemistry: (D)isequilibrium or (F)ree.

\item Profile: (M)olliere, (Z)hang, (G)uillot or (S)pline(NNodes).

\item Clouds: Clear (CLR); (f)ree or (eq)uilibrium condensation location, (species)\_(cd/am)\_(P)atchy\_(h)ansen. `*' indicates \fsed was retrieved independently for each cloud species. 

\item Data: -(not included) or (only included). O indicates OSIRIS data was used in place of GRAVITY data.

\item Info: `-' indicates not included. `mr' indicates mass and radius were used as parameters instead of $\log g$ and radius.
\end{itemize}
In the text, models will be referred to as \textit{planet.group.index}.

\begin{sidewaystable*}[t]
\centering
\begin{threeparttable}
\centering
\begin{small}
\caption{Abridged retrieval results HR 8799 b}
\label{tab:full_results_b}
\begin{tabular}{lll|lllllllll}
\toprule
\textbf{Planet} & \textbf{Index} & \textbf{Model} & $\bm{\Delta}\log_{\mathbf{10}} \bm{\mathcal{Z}}$ & $\bm{\chi^{2}/\nu}$ & $\mathbf{T_{\rm Eff}}$ & $\bm{\log g}$ & \textbf{[M/H]} & \textbf{C/O} & \textbf{Radius} & $\bm{\log L/L_{\odot}}$ & \textbf{Mass} \\
 &  &  &  & & [K] & [cgs] &  &  & [R$_{\rm Jup}$] &  & [M$_{\rm Jup}$] \\
\midrule
b & b.A.0 & F/Z/eq*FeMg\_am/ALL/-CH4/ & 0 & 1.46 &$948_{-14}^{+11}$ & $4.11_{-0.04}^{+0.03}$ & $0.94_{-0.08}^{+0.08}$ & $0.78_{-0.04}^{+0.03}$ & $1.1_{-0.03}^{+0.03}$ & $-5.06_{-0.04}^{+0.04}$ & $6_{-0.4}^{+0.3}$ \\[2pt]
  & b.AB.1 & F/Z/eq*FeMg\_am/ALL/ & 0 & 1.47 &$942_{-13}^{+12}$ & $4.1_{-0.04}^{+0.03}$ & $0.96_{-0.08}^{+0.08}$ & $0.78_{-0.04}^{+0.03}$ & $1.11_{-0.03}^{+0.03}$ & $-5.08_{-0.04}^{+0.04}$ & $6_{-0.3}^{+0.3}$ \\[2pt]
  & b.A.2 & F/Z/eq*FeMg\_am/ALL/-H2S/ & 0 & 1.49 &$958_{-15}^{+13}$ & $4.1_{-0.04}^{+0.03}$ & $1.1_{-0.1}^{+0.2}$ & $0.83_{-0.04}^{+0.03}$ & $1.1_{-0.03}^{+0.03}$ & $-5.05_{-0.04}^{+0.04}$ & $5.9_{-0.4}^{+0.4}$ \\[2pt]
  & b.A.3 & F/S1/eqMg\_am/ALL/ & -2 & 1.48 &$977_{-10}^{+11}$ & $4.18_{-0.04}^{+0.04}$ & $1.25_{-0.1}^{+0.1}$ & $0.89_{-0.03}^{+0.02}$ & $1.01_{-0.02}^{+0.02}$ & $-5.07_{-0.03}^{+0.03}$ & $6_{-0.4}^{+0.4}$ \\[2pt]
  & b.A.4 & F/S2/eqMg\_am/ALL/ & -4 & 1.50 &$970_{-13}^{+14}$ & $4.17_{-0.04}^{+0.04}$ & $1.2_{-0.1}^{+0.1}$ & $0.87_{-0.03}^{+0.03}$ & $1.02_{-0.02}^{+0.02}$ & $-5.08_{-0.04}^{+0.04}$ & $5.9_{-0.4}^{+0.4}$ \\[2pt]
  & b.A.5 & F/S1/CLR/ALL/ & -4 & 1.53 &$985_{-11}^{+11}$ & $4.18_{-0.04}^{+0.04}$ & $1.43_{-0.09}^{+0.08}$ & $0.92_{-0.02}^{+0.02}$ & $1.01_{-0.02}^{+0.02}$ & $-5.06_{-0.04}^{+0.03}$ & $6_{-0.5}^{+0.5}$ \\[2pt]
  & b.AB.6 & D/M/eqNa\_P/ALL/ & -4 & 1.52 &$931_{-23}^{+21}$ & $4.12_{-0.05}^{+0.05}$ & $1.16_{-0.08}^{+0.08}$ & $0.73_{-0.02}^{+0.02}$ & $1.09_{-0.04}^{+0.06}$ & $-5.07_{-0.06}^{+0.06}$ & $6_{-0.3}^{+0.3}$ \\[2pt]
  & b.AB.7 & D/M/eqNa/ALL/ & -4 & 1.50 &$966_{-19}^{+17}$ & $4.19_{-0.04}^{+0.04}$ & $0.95_{-0.07}^{+0.07}$ & $0.78_{-0.03}^{+0.02}$ & $1.0_{-0.04}^{+0.05}$ & $-5.05_{-0.05}^{+0.05}$ & $6_{-0.3}^{+0.3}$ \\[2pt]
  & b.AB.8 & F/Z/f*FeMg\_am/ALL/ & -4 & 1.48 &$944_{-10}^{+10}$ & $4.15_{-0.03}^{+0.03}$ & $0.97_{-0.08}^{+0.08}$ & $0.78_{-0.04}^{+0.03}$ & $1.04_{-0.02}^{+0.02}$ & $-5.11_{-0.03}^{+0.03}$ & $5.9_{-0.3}^{+0.4}$ \\[2pt]
  & b.A.9 & F/S3/CLR/ALL/ & -4 & 1.49 &$962_{-12}^{+13}$ & $4.17_{-0.04}^{+0.04}$ & $1.1_{-0.1}^{+0.1}$ & $0.85_{-0.04}^{+0.03}$ & $1.02_{-0.02}^{+0.02}$ & $-5.09_{-0.04}^{+0.04}$ & $6_{-0.5}^{+0.4}$ \\[2pt]
  & b.A.10 & F/S2/CLR/ALL/ & -4 & 1.52 &$962_{-14}^{+15}$ & $4.17_{-0.04}^{+0.04}$ & $1.2_{-0.1}^{+0.1}$ & $0.87_{-0.04}^{+0.03}$ & $1.02_{-0.03}^{+0.03}$ & $-5.09_{-0.04}^{+0.04}$ & $5.9_{-0.5}^{+0.5}$ \\[2pt]
  & b.AB.11 & F/Z/f*FeMg\_cd/ALL/ & -4 & 1.48 &$942_{-11}^{+11}$ & $4.15_{-0.03}^{+0.03}$ & $0.94_{-0.08}^{+0.08}$ & $0.77_{-0.04}^{+0.04}$ & $1.05_{-0.02}^{+0.02}$ & $-5.11_{-0.03}^{+0.03}$ & $5.9_{-0.3}^{+0.3}$ \\[2pt]
  & b.A.12 & F/S3/eqMg\_am/ALL/ & -4 & 1.51 &$965_{-13}^{+12}$ & $4.17_{-0.04}^{+0.04}$ & $1.1_{-0.1}^{+0.1}$ & $0.85_{-0.04}^{+0.03}$ & $1.02_{-0.02}^{+0.02}$ & $-5.09_{-0.03}^{+0.04}$ & $6_{-0.4}^{+0.4}$ \\[2pt]
  & b.AB.13 & D/M/eq*FeMg\_am/ALL/ & -6 & 1.54 &$944_{-10}^{+11}$ & $4.06_{-0.03}^{+0.03}$ & $1.11_{-0.08}^{+0.08}$ & $0.72_{-0.02}^{+0.02}$ & $1.17_{-0.03}^{+0.03}$ & $-5.03_{-0.04}^{+0.04}$ & $6_{-0.3}^{+0.3}$ \\[2pt]
  & b.A.14 & F/S4/CLR/ALL/ & -6 & 1.49 &$966_{-12}^{+13}$ & $4.17_{-0.04}^{+0.04}$ & $1.15_{-0.1}^{+0.1}$ & $0.86_{-0.04}^{+0.03}$ & $1.02_{-0.02}^{+0.02}$ & $-5.09_{-0.04}^{+0.04}$ & $6_{-0.5}^{+0.4}$ \\[2pt]
  & b.A.15 & F/S4/eqMg\_am/ALL/ & -6 & 1.52 &$968_{-12}^{+13}$ & $4.17_{-0.04}^{+0.04}$ & $1.15_{-0.09}^{+0.1}$ & $0.86_{-0.03}^{+0.03}$ & $1.02_{-0.02}^{+0.02}$ & $-5.08_{-0.04}^{+0.04}$ & $6_{-0.4}^{+0.4}$ \\[2pt]
  & b.AB.16 & D/M/eq*FeMg\_cd/ALL/ & -6 & 1.55 &$911_{-13}^{+14}$ & $4.05_{-0.04}^{+0.04}$ & $1.16_{-0.08}^{+0.08}$ & $0.73_{-0.02}^{+0.02}$ & $1.18_{-0.04}^{+0.04}$ & $-5.09_{-0.04}^{+0.04}$ & $6_{-0.3}^{+0.3}$ \\[2pt]
  & b.AB.17 & D/M/eq*FeMg\_am\_P/ALL/ & -7 & 1.55 &$941_{-12}^{+12}$ & $4.07_{-0.03}^{+0.03}$ & $1.08_{-0.09}^{+0.08}$ & $0.73_{-0.02}^{+0.02}$ & $1.16_{-0.03}^{+0.03}$ & $-5.05_{-0.03}^{+0.04}$ & $6_{-0.3}^{+0.3}$ \\[2pt]
  & b.AB.18 & D/Z/eq*FeMg\_am/ALL/ & -7 & 1.57 &$934_{-10}^{+10}$ & $4.06_{-0.03}^{+0.03}$ & $1.06_{-0.07}^{+0.06}$ & $0.73_{-0.02}^{+0.02}$ & $1.17_{-0.03}^{+0.03}$ & $-5.05_{-0.03}^{+0.03}$ & $6_{-0.3}^{+0.3}$ \\[2pt]
  & b.A.19 & F/S5/CLR/ALL/ & -7 & 1.50 &$955_{-14}^{+13}$ & $4.16_{-0.04}^{+0.04}$ & $1.1_{-0.1}^{+0.1}$ & $0.84_{-0.05}^{+0.04}$ & $1.03_{-0.02}^{+0.02}$ & $-5.09_{-0.04}^{+0.04}$ & $6_{-0.5}^{+0.4}$ \\[2pt]
  & b.A.20 & F/S5/eqMg\_am/ALL/ & -7 & 1.51 &$958_{-14}^{+13}$ & $4.17_{-0.04}^{+0.03}$ & $1.09_{-0.1}^{+0.1}$ & $0.84_{-0.04}^{+0.03}$ & $1.03_{-0.02}^{+0.02}$ & $-5.09_{-0.03}^{+0.04}$ & $6_{-0.4}^{+0.4}$ \\[2pt]
  & b.AB.21 & D/M/eqKCl/ALL/ & -8 & 1.56 &$906_{-13}^{+12}$ & $4.08_{-0.03}^{+0.03}$ & $1.22_{-0.08}^{+0.08}$ & $0.7_{-0.02}^{+0.02}$ & $1.13_{-0.03}^{+0.03}$ & $-5.13_{-0.04}^{+0.04}$ & $6_{-0.3}^{+0.2}$ \\[2pt]
  & b.A.22 & F/S6/CLR/ALL/ & -9 & 1.50 &$960_{-13}^{+13}$ & $4.17_{-0.04}^{+0.04}$ & $1_{-0.1}^{+0.1}$ & $0.83_{-0.05}^{+0.04}$ & $1.03_{-0.02}^{+0.02}$ & $-5.09_{-0.04}^{+0.04}$ & $6_{-0.4}^{+0.4}$ \\[2pt]
  & b.AB.23 & D/Z/f*FeMg\_am/ALL/ & -9 & 1.53 &$954_{-59}^{+45}$ & $4.17_{-0.03}^{+0.03}$ & $1.22_{-0.06}^{+0.06}$ & $0.76_{-0.02}^{+0.01}$ & $1.02_{-0.02}^{+0.02}$ & $-5.1_{-0.1}^{+0.08}$ & $5.9_{-0.3}^{+0.3}$ \\[2pt]
  & b.A.24 & F/S6/eqMg\_am/ALL/ & -9 & 1.49 &$928_{-15}^{+17}$ & $4.01_{-0.04}^{+0.04}$ & $0.82_{-0.07}^{+0.07}$ & $0.63_{-0.05}^{+0.05}$ & $1.21_{-0.04}^{+0.04}$ & $-5.03_{-0.04}^{+0.04}$ & $5.8_{-0.4}^{+0.4}$ \\[2pt]
  & b.AB.25 & D/S4/eq*FeMg\_am/ALL/ & -10 & 1.53 &$867_{-45}^{+75}$ & $4.03_{-0.03}^{+0.03}$ & $1.13_{-0.07}^{+0.07}$ & $0.67_{-0.04}^{+0.03}$ & $1.21_{-0.03}^{+0.03}$ & $-5.15_{-0.1}^{+0.1}$ & $6_{-0.3}^{+0.3}$ \\[2pt]
  & b.A.26 & F/S7/CLR/ALL/ & -10 & 1.52 &$964_{-15}^{+13}$ & $4.17_{-0.04}^{+0.04}$ & $1.1_{-0.1}^{+0.1}$ & $0.85_{-0.04}^{+0.04}$ & $1.02_{-0.02}^{+0.02}$ & $-5.09_{-0.04}^{+0.04}$ & $6_{-0.4}^{+0.4}$ \\[2pt]
  & b.AB.27 & D/M/fNa/ALL/ & -11 & 1.56 &$918_{-57}^{+46}$ & $4.13_{-0.03}^{+0.03}$ & $1.18_{-0.06}^{+0.06}$ & $0.7_{-0.02}^{+0.02}$ & $1.07_{-0.02}^{+0.03}$ & $-5.1_{-0.1}^{+0.09}$ & $6_{-0.3}^{+0.2}$ \\[2pt]
  & b.A.28 & F/S8/CLR/ALL/ & -11 & 1.52 &$955_{-14}^{+14}$ & $4.16_{-0.04}^{+0.04}$ & $0.98_{-0.1}^{+0.1}$ & $0.79_{-0.1}^{+0.05}$ & $1.03_{-0.03}^{+0.03}$ & $-5.1_{-0.04}^{+0.04}$ & $6_{-0.4}^{+0.4}$ \\[2pt]
  & b.A.29 & F/S9/CLR/ALL/ & -12 & 1.52 &$965_{-12}^{+13}$ & $4.18_{-0.04}^{+0.04}$ & $0.95_{-0.09}^{+0.1}$ & $0.57_{-0.1}^{+0.2}$ & $1.02_{-0.02}^{+0.02}$ & $-5.09_{-0.04}^{+0.04}$ & $6.1_{-0.4}^{+0.4}$ \\[2pt]
  & b.A.30 & F/S0/CLR/ALL/ & -21 & 1.84 &$1020_{-12}^{+12}$ & $4.24_{-0.04}^{+0.04}$ & $1.92_{-0.06}^{+0.04}$ & $0.97_{-0.005}^{+0.005}$ & $0.93_{-0.03}^{+0.02}$ & $-5.08_{-0.04}^{+0.04}$ & $5.8_{-0.5}^{+0.5}$ \\[2pt]
  & b.A.31 & F/S0/eqMg\_am/ALL/ & -21 & 1.83 &$1022_{-11}^{+13}$ & $4.24_{-0.04}^{+0.04}$ & $1.93_{-0.06}^{+0.04}$ & $0.97_{-0.005}^{+0.004}$ & $0.93_{-0.02}^{+0.02}$ & $-5.08_{-0.04}^{+0.04}$ & $5.8_{-0.5}^{+0.5}$ \\[2pt]
  & b.A.32 & D/Z/eq*FeMg\_am/Gr/ & -948 & 0.74 &$891_{-27}^{+30}$ & $4.18_{-0.07}^{+0.07}$ & $1.4_{-0.4}^{+0.3}$ & $0.42_{-0.2}^{+0.1}$ & $0.99_{-0.08}^{+0.08}$ & $-5.3_{-0.1}^{+0.09}$ & $5.8_{-0.3}^{+0.3}$ \\[2pt]
\bottomrule
\end{tabular}
\end{small}
\begin{tablenotes}
\small
\item\textbf{Notes}
\item All values presented are the median values from the fits, with uncertainties given as the $\pm 34.1$\% percentiles.
\end{tablenotes}
\end{threeparttable}
\end{sidewaystable*}
\begin{sidewaystable*}[t]
\centering
\begin{threeparttable}
\centering
\begin{small}
\caption{Abridged retrieval results HR 8799 c}
\label{tab:full_results_c}
\begin{tabular}{lll|lllllllll}
\toprule
\textbf{Planet} & \textbf{Index} & \textbf{Model} & $\bm{\Delta}\log_{\mathbf{10}} \bm{\mathcal{Z}}$ & $\bm{\chi^{2}/\nu}$ & $\mathbf{T_{\rm Eff}}$ & $\bm{\log g}$ & \textbf{[M/H]} & \textbf{C/O} & \textbf{Radius} & $\bm{\log L/L_{\odot}}$ & \textbf{Mass} \\
 &  &  &  & & [K] & [cgs] &  &  & [R$_{\rm Jup}$] &  & [M$_{\rm Jup}$] \\
\midrule
c & c.A.0 & F/Z/eq*FeMg\_am/O-Gr/mr & 0 & 1.72 &$1145_{-15}^{+14}$ & $4.27_{-0.04}^{+0.03}$ & $0.95_{-0.1}^{+0.1}$ & $0.6_{-0.1}^{+0.09}$ & $1.06_{-0.03}^{+0.03}$ & $-4.75_{-0.03}^{+0.03}$ & $8_{-0.5}^{+0.5}$ \\[2pt]
  & c.A.1 & D/M/eq*FeMg\_cd/O-Gr/mr & -1 & 1.83 &$1207_{-10}^{+11}$ & $4.34_{-0.03}^{+0.03}$ & $1.89_{-0.06}^{+0.06}$ & $0.87_{-0.006}^{+0.005}$ & $0.99_{-0.02}^{+0.02}$ & $-4.73_{-0.03}^{+0.03}$ & $8.2_{-0.5}^{+0.5}$ \\[2pt]
  & c.AB.2 & F/Z/eq*FeMg\_am/ALL/mr & -430 & 2.88 &$1158_{-12}^{+12}$ & $4.26_{-0.03}^{+0.02}$ & $1.27_{-0.06}^{+0.06}$ & $0.66_{-0.01}^{+0.01}$ & $1.1_{-0.01}^{+0.01}$ & $-4.71_{-0.02}^{+0.02}$ & $8.5_{-0.4}^{+0.4}$ \\[2pt]
  & c.AB.3 & F/G/f*FeMg\_am/ALL/mr & -432 & 2.87 &$1173_{-8}^{+18}$ & $4.26_{-0.02}^{+0.02}$ & $1.3_{-0.06}^{+0.06}$ & $0.67_{-0.01}^{+0.01}$ & $1.09_{-0.01}^{+0.01}$ & $-4.71_{-0.02}^{+0.03}$ & $8.3_{-0.4}^{+0.4}$ \\[2pt]
  & c.AB.4 & F/Z/f*FeMg\_am/ALL/mr & -435 & 2.91 &$1173_{-6}^{+5}$ & $4.28_{-0.02}^{+0.02}$ & $1.27_{-0.07}^{+0.05}$ & $0.67_{-0.01}^{+0.01}$ & $1.07_{-0.01}^{+0.01}$ & $-4.71_{-0.01}^{+0.01}$ & $8.5_{-0.4}^{+0.4}$ \\[2pt]
  & c.AB.5 & D/M/eq*FeMg\_cd/ALL/mr & -443 & 2.93 &$1057_{-13}^{+14}$ & $4.05_{-0.02}^{+0.02}$ & $1.05_{-0.04}^{+0.04}$ & $0.62_{-0.01}^{+0.01}$ & $1.4_{-0.03}^{+0.03}$ & $-4.69_{-0.03}^{+0.03}$ & $8.6_{-0.4}^{+0.3}$ \\[2pt]
  & c.A.6 & F/G/f*FeMg\_am/ALL/-CH4/mr & -444 & 3.02 &$1191_{-12}^{+40}$ & $4.28_{-0.02}^{+0.02}$ & $1.2_{-0.06}^{+0.06}$ & $0.7_{-0.01}^{+0.01}$ & $1.07_{-0.01}^{+0.01}$ & $-4.69_{-0.02}^{+0.05}$ & $8.4_{-0.3}^{+0.3}$ \\[2pt]
  & c.AB.7 & D/S4/eq*FeMg\_am/ALL/mr & -464 & 3.10 &$1198_{-23}^{+21}$ & $4.29_{-0.02}^{+0.02}$ & $1.02_{-0.03}^{+0.03}$ & $0.6_{-0.01}^{+0.01}$ & $1.08_{-0.01}^{+0.01}$ & $-4.67_{-0.04}^{+0.03}$ & $8.8_{-0.3}^{+0.3}$ \\[2pt]
  & c.A.8 & F/G/f*FeMg\_am/ALL/-HCN/mr & -465 & 3.14 &$1197_{-33}^{+45}$ & $4.27_{-0.04}^{+0.03}$ & $1.36_{-0.06}^{+0.05}$ & $0.69_{-0.01}^{+0.01}$ & $1.1_{-0.03}^{+0.05}$ & $-4.62_{-0.05}^{+0.06}$ & $8.7_{-0.2}^{+0.2}$ \\[2pt]
  & c.AB.9 & D/M/eq*FeMg\_am\_P/ALL/mr & -468 & 3.25 &$1099_{-16}^{+15}$ & $4.14_{-0.03}^{+0.03}$ & $1.1_{-0.05}^{+0.05}$ & $0.61_{-0.01}^{+0.01}$ & $1.28_{-0.05}^{+0.04}$ & $-4.72_{-0.03}^{+0.03}$ & $8.9_{-0.4}^{+0.4}$ \\[2pt]
  & c.AB.10 & D/Z/f*FeMg\_am/ALL/mr & -478 & 3.38 &$1145_{-7}^{+9}$ & $4.23_{-0.02}^{+0.02}$ & $1.18_{-0.03}^{+0.03}$ & $0.62_{-0.009}^{+0.01}$ & $1.16_{-0.01}^{+0.01}$ & $-4.69_{-0.01}^{+0.02}$ & $8.8_{-0.3}^{+0.3}$ \\[2pt]
  & c.A.11 & D/Z/eq*FeMg\_am/Gr/mr & -2678 & 0.84 &$1234_{-17}^{+18}$ & $4.29_{-0.03}^{+0.04}$ & $0.79_{-0.2}^{+0.1}$ & $0.69_{-0.02}^{+0.02}$ & $1.01_{-0.03}^{+0.03}$ & $-4.66_{-0.03}^{+0.04}$ & $7.6_{-0.4}^{+0.4}$ \\[2pt]
\bottomrule
\end{tabular}
\end{small}
\begin{tablenotes}
\small
\item\textbf{Notes}
\item All values presented are the median values from the fits, with uncertainties given as the $\pm 34.1$\% percentiles.
\end{tablenotes}
\end{threeparttable}
\end{sidewaystable*}
\begin{sidewaystable*}[t]
\centering
\begin{threeparttable}
\centering
\begin{small}
\caption{Abridged retrieval results HR 8799 d}
\label{tab:full_results_d}
\begin{tabular}{lll|lllllllll}
\toprule
\textbf{Planet} & \textbf{Index} & \textbf{Model} & $\bm{\Delta}\log_{\mathbf{10}} \bm{\mathcal{Z}}$ & $\bm{\chi^{2}/\nu}$ & $\mathbf{T_{\rm Eff}}$ & $\bm{\log g}$ & \textbf{[M/H]} & \textbf{C/O} & \textbf{Radius} & $\bm{\log L/L_{\odot}}$ & \textbf{Mass} \\
 &  &  &  & & [K] & [cgs] &  &  & [R$_{\rm Jup}$] &  & [M$_{\rm Jup}$] \\
\midrule
d & d.AB.0 & D/M/eq*FeMg\_am/ALL/mr & 0 & 1.42 &$1177_{-21}^{+21}$ & $4.18_{-0.03}^{+0.04}$ & $1.2_{-0.1}^{+0.2}$ & $0.61_{-0.04}^{+0.03}$ & $1.26_{-0.06}^{+0.05}$ & $-4.63_{-0.04}^{+0.04}$ & $9.19_{-0.07}^{+0.08}$ \\[2pt]
  & d.AB.1 & D/Z/f*FeMg\_am/ALL/mr & 0 & 1.39 &$1139_{-19}^{+38}$ & $4.13_{-0.02}^{+0.02}$ & $1.3_{-0.1}^{+0.1}$ & $0.6_{-0.04}^{+0.03}$ & $1.34_{-0.03}^{+0.03}$ & $-4.6_{-0.03}^{+0.06}$ & $9.2_{-0.07}^{+0.07}$ \\[2pt]
  & d.AB.2 & D/M/eq*FeMg\_am\_P/ALL/mr & -1 & 1.47 &$1220_{-10}^{+10}$ & $4.25_{-0.02}^{+0.02}$ & $1.1_{-0.1}^{+0.1}$ & $0.49_{-0.04}^{+0.04}$ & $1.16_{-0.02}^{+0.02}$ & $-4.58_{-0.03}^{+0.03}$ & $9.2_{-0.09}^{+0.09}$ \\[2pt]
  & d.AB.3 & D/M/eq*FeMg\_cd/ALL/mr & -1 & 1.47 &$1220_{-9}^{+10}$ & $4.25_{-0.02}^{+0.02}$ & $1.1_{-0.1}^{+0.1}$ & $0.49_{-0.04}^{+0.04}$ & $1.16_{-0.02}^{+0.02}$ & $-4.58_{-0.03}^{+0.03}$ & $9.21_{-0.09}^{+0.09}$ \\[2pt]
  & d.AB.4 & F/G/f*FeMg\_cd/ALL/mr & -3 & 1.43 &$1194_{-15}^{+18}$ & $4.18_{-0.02}^{+0.02}$ & $1.5_{-0.2}^{+0.2}$ & $0.68_{-0.04}^{+0.03}$ & $1.26_{-0.03}^{+0.02}$ & $-4.58_{-0.03}^{+0.03}$ & $9.2_{-0.07}^{+0.07}$ \\[2pt]
  & d.AB.5 & F/G/f*FeMg\_am/ALL/mr & -3 & 1.42 &$1196_{-17}^{+20}$ & $4.18_{-0.02}^{+0.03}$ & $1.5_{-0.2}^{+0.2}$ & $0.67_{-0.04}^{+0.03}$ & $1.26_{-0.04}^{+0.03}$ & $-4.58_{-0.03}^{+0.03}$ & $9.2_{-0.07}^{+0.07}$ \\[2pt]
  & d.AB.6 & F/Z/eq*FeMg\_am/ALL/mr & -4 & 1.46 &$1146_{-13}^{+13}$ & $4.17_{-0.02}^{+0.02}$ & $1.6_{-0.1}^{+0.2}$ & $0.67_{-0.05}^{+0.04}$ & $1.27_{-0.03}^{+0.03}$ & $-4.62_{-0.03}^{+0.03}$ & $9.19_{-0.07}^{+0.07}$ \\[2pt]
  & d.AB.7 & D/Z/eqMg\_am/ALL/mr & -4 & 1.47 &$1150_{-16}^{+21}$ & $4.17_{-0.03}^{+0.04}$ & $0.95_{-0.1}^{+0.1}$ & $0.5_{-0.04}^{+0.04}$ & $1.27_{-0.06}^{+0.05}$ & $-4.67_{-0.04}^{+0.04}$ & $9.18_{-0.07}^{+0.07}$ \\[2pt]
  & d.AB.8 & D/Z/eq*FeMg\_am/ALL/mr & -5 & 1.49 &$1166_{-16}^{+19}$ & $4.21_{-0.03}^{+0.03}$ & $0.98_{-0.1}^{+0.1}$ & $0.48_{-0.04}^{+0.04}$ & $1.22_{-0.05}^{+0.05}$ & $-4.66_{-0.04}^{+0.04}$ & $9.19_{-0.06}^{+0.07}$ \\[2pt]
  & d.AB.9 & D/Z/eq*FeMg\_am/ALL/ & -6 & 1.50 &$1158_{-30}^{+22}$ & $4.5_{-0.3}^{+0.3}$ & $1.3_{-0.3}^{+0.3}$ & $0.48_{-0.04}^{+0.04}$ & $1.23_{-0.06}^{+0.08}$ & $-4.61_{-0.06}^{+0.05}$ & $20.70_{-10}^{+19}$ \\[2pt]
  & d.AB.10 & F/Z/f*FeMg\_am/ALL/mr & -6 & 1.44 &$1157_{-16}^{+30}$ & $4.16_{-0.01}^{+0.01}$ & $1.7_{-0.1}^{+0.1}$ & $0.64_{-0.04}^{+0.04}$ & $1.29_{-0.02}^{+0.02}$ & $-4.57_{-0.03}^{+0.04}$ & $9.22_{-0.05}^{+0.05}$ \\[2pt]
  & d.AB.11 & D/S4/eq*FeMg\_am/ALL/mr & -12 & 1.51 &$1198_{-20}^{+21}$ & $4.24_{-0.04}^{+0.03}$ & $0.71_{-0.1}^{+0.2}$ & $0.34_{-0.06}^{+0.07}$ & $1.17_{-0.04}^{+0.05}$ & $-4.62_{-0.04}^{+0.04}$ & $9.19_{-0.07}^{+0.07}$ \\[2pt]
  & d.A.12 & D/Z/eq*FeMg\_am/Gr/mr & -2276 & 0.79 &$1172_{-18}^{+19}$ & $4.18_{-0.03}^{+0.03}$ & $2_{-0.2}^{+0.2}$ & $0.78_{-0.03}^{+0.02}$ & $1.25_{-0.04}^{+0.04}$ & $-4.61_{-0.04}^{+0.04}$ & $9.19_{-0.08}^{+0.08}$ \\[2pt]
  & d.A.13 & D/Z/eq*FeMg\_am/-Gr/mr & -3498 & 1.03 &$1162_{-28}^{+28}$ & $4.22_{-0.04}^{+0.04}$ & $0.46_{-0.7}^{+0.7}$ & $0.31_{-0.1}^{+0.1}$ & $1.19_{-0.06}^{+0.06}$ & $-4.68_{-0.06}^{+0.06}$ & $9.2_{-0.08}^{+0.08}$ \\[2pt]
\bottomrule
\end{tabular}
\end{small}
\begin{tablenotes}
\small
\item\textbf{Notes}
\item All values presented are the median values from the fits, with uncertainties given as the $\pm 34.1$\% percentiles.
\end{tablenotes}
\end{threeparttable}
\end{sidewaystable*}
\begin{sidewaystable*}[t]
\centering
\begin{threeparttable}
\centering
\begin{small}
\caption{Abridged retrieval results HR 8799 e}
\label{tab:full_results_e}
\begin{tabular}{lll|lllllllll}
\toprule
\textbf{Planet} & \textbf{Index} & \textbf{Model} & $\bm{\Delta}\log_{\mathbf{10}} \bm{\mathcal{Z}}$ & $\bm{\chi^{2}/\nu}$ & $\mathbf{T_{\rm Eff}}$ & $\bm{\log g}$ & \textbf{[M/H]} & \textbf{C/O} & \textbf{Radius} & $\bm{\log L/L_{\odot}}$ & \textbf{Mass} \\
 &  &  &  & & [K] & [cgs] &  &  & [R$_{\rm Jup}$] &  & [M$_{\rm Jup}$] \\
\midrule
e & e.A.0 & F/Z/eq*FeMg\_am/ALL/-CH4/mr & 0 & 1.23 &$1139_{-20}^{+21}$ & $4.18_{-0.05}^{+0.05}$ & $1.9_{-0.1}^{+0.1}$ & $0.88_{-0.02}^{+0.01}$ & $1.13_{-0.05}^{+0.05}$ & $-4.73_{-0.05}^{+0.05}$ & $7.5_{-0.6}^{+0.6}$ \\[2pt]
  & e.AB.1 & F/G/eq*FeMg\_am/ALL/mr & 0 & 1.22 &$1172_{-29}^{+27}$ & $4.2_{-0.06}^{+0.06}$ & $1.8_{-0.2}^{+0.1}$ & $0.88_{-0.02}^{+0.02}$ & $1.1_{-0.05}^{+0.05}$ & $-4.71_{-0.05}^{+0.06}$ & $7.5_{-0.7}^{+0.7}$ \\[2pt]
  & e.AB.2 & F/Z/eq*FeMg\_am/ALL/mr & 0 & 1.23 &$1134_{-20}^{+24}$ & $4.18_{-0.05}^{+0.05}$ & $1.9_{-0.1}^{+0.1}$ & $0.87_{-0.02}^{+0.02}$ & $1.14_{-0.05}^{+0.05}$ & $-4.74_{-0.05}^{+0.06}$ & $7.5_{-0.6}^{+0.6}$ \\[2pt]
  & e.AB.3 & F/G/fMg\_am/ALL/mr & -1 & 1.22 &$1206_{-23}^{+27}$ & $4.24_{-0.05}^{+0.05}$ & $1.8_{-0.2}^{+0.1}$ & $0.88_{-0.02}^{+0.02}$ & $1.06_{-0.03}^{+0.03}$ & $-4.67_{-0.05}^{+0.05}$ & $7.6_{-0.7}^{+0.7}$ \\[2pt]
  & e.AB.4 & F/G/fFe/ALL/mr & -2 & 1.22 &$1203_{-28}^{+30}$ & $4.24_{-0.05}^{+0.05}$ & $1.8_{-0.2}^{+0.1}$ & $0.89_{-0.02}^{+0.02}$ & $1.07_{-0.04}^{+0.03}$ & $-4.69_{-0.05}^{+0.06}$ & $7.6_{-0.6}^{+0.7}$ \\[2pt]
  & e.AB.5 & F/G/grey\_P/ALL/mr & -3 & 1.23 &$1261_{-18}^{+18}$ & $4.35_{-0.06}^{+0.05}$ & $1.7_{-0.3}^{+0.2}$ & $0.89_{-0.03}^{+0.02}$ & $0.94_{-0.03}^{+0.04}$ & $-4.64_{-0.05}^{+0.04}$ & $7.6_{-0.7}^{+0.7}$ \\[2pt]
  & e.AB.6 & F/G/f*FeMg\_am/ALL/mr & -3 & 1.23 &$1214_{-23}^{+22}$ & $4.25_{-0.04}^{+0.04}$ & $1.8_{-0.2}^{+0.1}$ & $0.89_{-0.02}^{+0.02}$ & $1.05_{-0.03}^{+0.03}$ & $-4.69_{-0.04}^{+0.04}$ & $7.5_{-0.6}^{+0.6}$ \\[2pt]
  & e.AB.7 & F/S4/eq*FeMg\_am/ALL/mr & -6 & 1.26 &$1202_{-20}^{+19}$ & $4.28_{-0.05}^{+0.05}$ & $1.8_{-0.2}^{+0.1}$ & $0.9_{-0.03}^{+0.02}$ & $1.01_{-0.03}^{+0.03}$ & $-4.7_{-0.05}^{+0.05}$ & $7.6_{-0.7}^{+0.8}$ \\[2pt]
  & e.AB.8 & D/M/eqFeMg\_cd\_P/ALL/mr & -6 & 1.32 &$1215_{-18}^{+19}$ & $4.31_{-0.05}^{+0.05}$ & $2.1_{-0.3}^{+0.2}$ & $0.83_{-0.02}^{+0.01}$ & $1.0_{-0.03}^{+0.03}$ & $-4.68_{-0.05}^{+0.04}$ & $7.9_{-0.7}^{+0.7}$ \\[2pt]
  & e.AB.9 & D/Z/eq*FeMg\_cd/ALL/mr & -7 & 1.27 &$1084_{-23}^{+27}$ & $4.09_{-0.06}^{+0.06}$ & $1.3_{-0.1}^{+0.1}$ & $0.77_{-0.03}^{+0.03}$ & $1.29_{-0.07}^{+0.06}$ & $-4.82_{-0.05}^{+0.06}$ & $7.9_{-0.7}^{+0.7}$ \\[2pt]
  & e.AB.10 & D/M/eq*FeMg\_cd\_P/ALL/mr & -7 & 1.32 &$1215_{-18}^{+17}$ & $4.31_{-0.05}^{+0.05}$ & $2.1_{-0.3}^{+0.2}$ & $0.83_{-0.02}^{+0.01}$ & $1.0_{-0.03}^{+0.03}$ & $-4.68_{-0.05}^{+0.05}$ & $7.9_{-0.7}^{+0.6}$ \\[2pt]
  & e.AB.11 & D/M/eq*FeMg\_cd/ALL/mr & -7 & 1.28 &$1106_{-28}^{+33}$ & $4.11_{-0.06}^{+0.08}$ & $1.3_{-0.1}^{+0.2}$ & $0.78_{-0.03}^{+0.03}$ & $1.26_{-0.09}^{+0.07}$ & $-4.82_{-0.06}^{+0.07}$ & $7.9_{-0.7}^{+0.7}$ \\[2pt]
  & e.A.12 & F/Z/eq*FeMg\_am/ALL/-HCN/mr & -8 & 1.30 &$1148_{-23}^{+24}$ & $4.15_{-0.05}^{+0.06}$ & $1.91_{-0.1}^{+0.07}$ & $0.87_{-0.02}^{+0.02}$ & $1.17_{-0.06}^{+0.06}$ & $-4.73_{-0.06}^{+0.06}$ & $7.5_{-0.6}^{+0.7}$ \\[2pt]
  & e.AB.13 & D/Z/f*FeMg\_am/ALL/mr & -9 & 1.33 &$1187_{-30}^{+50}$ & $4.26_{-0.04}^{+0.04}$ & $2_{-0.3}^{+0.3}$ & $0.83_{-0.02}^{+0.02}$ & $1.08_{-0.03}^{+0.03}$ & $-4.69_{-0.06}^{+0.07}$ & $8.1_{-0.6}^{+0.6}$ \\[2pt]
  & e.AB.14 & D/Z/eq*FeMg2/Gr/mr & -9 & 1.35 &$1170_{-27}^{+26}$ & $4.26_{-0.06}^{+0.06}$ & $1.3_{-0.2}^{+0.2}$ & $0.76_{-0.03}^{+0.02}$ & $1.06_{-0.05}^{+0.05}$ & $-4.77_{-0.06}^{+0.05}$ & $7.8_{-0.7}^{+0.7}$ \\[2pt]
  & e.AB.15 & F/Z/f*FeMg\_am/ALL/mr & -10 & 1.29 &$1143_{-18}^{+12}$ & $4.19_{-0.05}^{+0.05}$ & $1.99_{-0.05}^{+0.05}$ & $0.86_{-0.02}^{+0.02}$ & $1.09_{-0.03}^{+0.03}$ & $-4.73_{-0.04}^{+0.04}$ & $7_{-0.7}^{+0.6}$ \\[2pt]
  & e.A.16 & D/Z/eqFeMg\_am/ALL/Fe-H15mr & -10 & 1.37 &$1202_{-17}^{+20}$ & $4.32_{-0.05}^{+0.05}$ & $1.5_{-0}^{+0}$ & $0.76_{-0.02}^{+0.02}$ & $0.99_{-0.03}^{+0.03}$ & $-4.73_{-0.05}^{+0.05}$ & $7.8_{-0.7}^{+0.7}$ \\[2pt]
  & e.A.17 & D/Z/eqFeMg\_am/ALL/Fe-H20mr & -10 & 1.33 &$1179_{-28}^{+29}$ & $4.25_{-0.06}^{+0.06}$ & $2.0_{-0}^{+0}$ & $0.82_{-0.01}^{+0.01}$ & $1.08_{-0.07}^{+0.07}$ & $-4.77_{-0.05}^{+0.06}$ & $8_{-0.6}^{+0.6}$ \\[2pt]
  & e.A.18 & D/M/grey\_P/ALL/mr & -10 & 1.38 &$1245_{-16}^{+16}$ & $4.36_{-0.05}^{+0.05}$ & $1.5_{-0.1}^{+0.1}$ & $0.63_{-0.02}^{+0.02}$ & $0.95_{-0.03}^{+0.03}$ & $-4.68_{-0.04}^{+0.04}$ & $8_{-0.8}^{+0.8}$ \\[2pt]
  & e.AB.19 & D/M/eq*FeMg\_am/ALL/mr & -11 & 1.33 &$1206_{-22}^{+21}$ & $4.3_{-0.05}^{+0.05}$ & $2.1_{-0.4}^{+0.2}$ & $0.83_{-0.02}^{+0.01}$ & $1.02_{-0.04}^{+0.05}$ & $-4.69_{-0.06}^{+0.05}$ & $8.1_{-0.6}^{+0.6}$ \\[2pt]
  & e.AB.20 & D/Z/eq*FeMg\_am\_h/ALL/mr & -11 & 1.33 &$1129_{-27}^{+29}$ & $4.15_{-0.05}^{+0.06}$ & $1.4_{-0.1}^{+0.2}$ & $0.78_{-0.03}^{+0.03}$ & $1.21_{-0.06}^{+0.06}$ & $-4.77_{-0.07}^{+0.07}$ & $7.9_{-0.6}^{+0.6}$ \\[2pt]
  & e.AB.21 & D/M/grey\_P/ALL/mr & -11 & 1.37 &$1245_{-15}^{+15}$ & $4.34_{-0.07}^{+0.05}$ & $1.5_{-0.1}^{+0.1}$ & $0.63_{-0.02}^{+0.02}$ & $0.97_{-0.03}^{+0.06}$ & $-4.64_{-0.06}^{+0.08}$ & $8_{-0.8}^{+0.8}$ \\[2pt]
  & e.A.22 & D/Z/eqFeMg\_am/ALL/Fe-H10mr & -11 & 1.41 &$1208_{-19}^{+21}$ & $4.29_{-0.05}^{+0.05}$ & $1.0_{-0}^{+0}$ & $0.69_{-0.02}^{+0.02}$ & $0.98_{-0.03}^{+0.03}$ & $-4.72_{-0.05}^{+0.05}$ & $7.3_{-0.6}^{+0.7}$ \\[2pt]
  & e.A.23 & D/Z/eqFeMg\_am/ALL/mr & -11 & 1.36 &$1209_{-17}^{+20}$ & $4.32_{-0.05}^{+0.04}$ & $1.7_{-0.3}^{+0.7}$ & $0.79_{-0.05}^{+0.03}$ & $0.99_{-0.03}^{+0.03}$ & $-4.72_{-0.05}^{+0.05}$ & $7.9_{-0.7}^{+0.6}$ \\[2pt]
  & e.AB.24 & D/Z/eqAl2O3/ALL/mr & -11 & 1.34 &$1235_{-14}^{+17}$ & $4.36_{-0.05}^{+0.04}$ & $2.3_{-0.3}^{+0.1}$ & $0.82_{-0.01}^{+0.01}$ & $0.94_{-0.03}^{+0.03}$ & $-4.73_{-0.04}^{+0.04}$ & $7.9_{-0.7}^{+0.7}$ \\[2pt]
  & e.AB.25 & D/Z/eq*FeMg\_am/ALL/mr & -11 & 1.36 &$1207_{-17}^{+19}$ & $4.32_{-0.04}^{+0.04}$ & $2.1_{-0.6}^{+0.3}$ & $0.81_{-0.06}^{+0.01}$ & $0.99_{-0.03}^{+0.03}$ & $-4.72_{-0.05}^{+0.04}$ & $8_{-0.6}^{+0.6}$ \\[2pt]
  & e.AB.26 & D/Z/eq*FeMg\_am/ALL/ & -12 & 1.36 &$1224_{-22}^{+21}$ & $4.8_{-0.2}^{+0.2}$ & $2_{-0.3}^{+0.3}$ & $0.76_{-0.03}^{+0.05}$ & $0.96_{-0.03}^{+0.03}$ & $-4.78_{-0.06}^{+0.06}$ & $19.91_{-7}^{+13}$ \\[2pt]
  & e.AB.27 & F/G/grey/ALL/mr & -12 & 1.37 &$1245_{-31}^{+58}$ & $4.24_{-0.06}^{+0.06}$ & $2.05_{-0.08}^{+0.09}$ & $0.91_{-0.02}^{+0.02}$ & $1.04_{-0.04}^{+0.04}$ & $-4.6_{-0.07}^{+0.09}$ & $7.4_{-0.8}^{+0.8}$ \\[2pt]
  & e.AB.28 & D/S4/eq*FeMg\_cd/ALL/mr & -12 & 1.27 &$1093_{-24}^{+27}$ & $4.08_{-0.05}^{+0.05}$ & $1.4_{-0.1}^{+0.1}$ & $0.78_{-0.03}^{+0.02}$ & $1.3_{-0.05}^{+0.05}$ & $-4.73_{-0.06}^{+0.06}$ & $7.8_{-0.7}^{+0.7}$ \\[2pt]
  & e.A.29 & D/Z/eqFeMg\_am/ALL/Fe-H05mr & -17 & 1.49 &$1217_{-18}^{+20}$ & $4.29_{-0.05}^{+0.05}$ & $0.5_{-0}^{+0}$ & $0.52_{-0.04}^{+0.03}$ & $0.97_{-0.03}^{+0.03}$ & $-4.71_{-0.05}^{+0.04}$ & $7.1_{-0.6}^{+0.7}$ \\[2pt]
  & e.A.30 & D/Z/eqFeMg\_am/ALL/Fe-H00mr & -26 & 1.58 &$1227_{-16}^{+17}$ & $4.29_{-0.05}^{+0.05}$ & $0.0_{-0}^{+0}$ & $0.21_{-0.02}^{+0.02}$ & $0.95_{-0.03}^{+0.03}$ & $-4.72_{-0.05}^{+0.04}$ & $6.9_{-0.6}^{+0.6}$ \\[2pt]
  & e.A.31 & D/Z/eq*FeMg\_am/Gr/mr & -2294 & 0.72 &$1173_{-28}^{+31}$ & $4.24_{-0.06}^{+0.06}$ & $1.8_{-0.5}^{+0.4}$ & $0.85_{-0.04}^{+0.02}$ & $1.06_{-0.05}^{+0.06}$ & $-4.71_{-0.07}^{+0.07}$ & $7.5_{-0.7}^{+0.7}$ \\[2pt]
  & e.A.32 & D/Z/CLR/Gr/mr & -2301 & 0.79 &$1339_{-38}^{+36}$ & $4.3_{-0.07}^{+0.07}$ & $0.65_{-0.2}^{+0.2}$ & $0.71_{-0.07}^{+0.05}$ & $0.94_{-0.05}^{+0.06}$ & $-4.53_{-0.08}^{+0.08}$ & $6.9_{-0.6}^{+0.8}$ \\[2pt]
  & e.A.33 & D/Z/eq*FeMg\_am/-Gr/mr & -3496 & 1.88 &$1143_{-32}^{+38}$ & $4.17_{-0.08}^{+0.08}$ & $2_{-0.4}^{+0.3}$ & $0.71_{-0.2}^{+0.08}$ & $1.15_{-0.09}^{+0.09}$ & $-4.79_{-0.09}^{+0.08}$ & $7.5_{-0.7}^{+0.7}$ \\[2pt]
\bottomrule
\end{tabular}
\end{small}
\begin{tablenotes}
\small
\item\textbf{Notes}
\item All values presented are the median values from the fits, with uncertainties given as the $\pm 34.1$\% percentiles.
\end{tablenotes}
\end{threeparttable}
\end{sidewaystable*}


\end{appendix}

\end{document}